\documentclass[onecolumn,12pt]{elsarticle}

\usepackage{amssymb, bm, color}
\usepackage{amsthm, amsmath}

\usepackage{graphicx}
\usepackage{caption}
\usepackage{subcaption}
\usepackage{multirow}
\usepackage{mathtools,xparse}

\makeatletter
\newenvironment{smallermatrix}[1][c]
{\null\,\vcenter\bgroup
  \Let@\restore@math@cr\default@tag
  \baselineskip0pt \lineskip0.4pt \lineskiplimit0pt
  \ialign\bgroup\if#1l\else\hfil\fi$\m@th\scriptstyle##$\if#1r\else\hfil\fi&&\thickspace\hfil
  $\m@th\scriptstyle##$\hfil\crcr
}{%
  \crcr\egroup\egroup\,%
}
\makeatother

\NewDocumentCommand{\ts}{O{c} e{^?_}}{
  \begin{smallermatrix}[#1]
  \mathstrut\IfValueT{#2}{#2} \\
  \mathstrut\IfValueT{#3}{#3} \\
  \mathstrut\IfValueT{#4}{#4}
  \end{smallermatrix}%
}

\newcommand{\revise}[1]{{\color{black}{#1}}}

\journal{arXiv}

\begin{document}

\begin{frontmatter}



\title{An adaptive viscosity regularization approach for the numerical solution of conservation laws: Application to  finite element methods}


\author[inst1]{Ngoc Cuong Nguyen}

\author[inst1]{Jordi Vila-P\'erez}


\author[inst1]{Jaime Peraire}

\affiliation[inst1]{organization={Center for Computational Engineering, Department of Aeronautics and Astronautics, Massachusetts Institute of Technology},
            addressline={77 Massachusetts
Avenue}, 
            city={Cambridge},
            state={MA},
            postcode={02139}, 
            country={USA}}

\begin{abstract}
We introduce an adaptive viscosity regularization approach for the numerical solution of systems of nonlinear conservation laws with shock waves.
The approach seeks to solve a sequence of regularized problems consisting of the system of conservation laws and an additional Helmholtz equation for the artificial viscosity.
We propose a homotopy continuation  of the regularization parameters to minimize the amount of artificial viscosity subject to positivity-preserving and smoothness constraints on the numerical solution.
The regularization methodology is combined with a mesh adaptation strategy that identifies the shock location and generates shock-aligned meshes, which allows to further reduce the amount of artificial dissipation and capture shocks with increased accuracy.
We use the hybridizable discontinuous Galerkin method to numerically solve the regularized system of conservation laws and the continuous Galerkin method to solve the Helmholtz equation for the artificial viscosity. We show that the approach can produce  approximate solutions that converge to the exact solution of the Burgers' equation. Finally, we demonstrate the performance of the method on  inviscid transonic, supersonic,  hypersonic  flows in two dimensions. The approach is found to be accurate, robust and efficient, and yields very sharp yet smooth solutions in a few homotopy iterations.  
\end{abstract}









\end{frontmatter}


\section{Introduction}
\label{sec:intro}

The formation of shock waves is one of the most challenging problems in numerical approximation of nonlinear conservation laws. Difficulties in capturing shock waves  are that (1) at the very moment a shock is formed it poses a source of instability in the shock region, which then leads to numerical instabilities if no treatment of shock waves is introduced; (2) it is hard to predict when and where new shocks arise, and track them as they propagate through the physical domain and interact with each other and with boundary layers and vortices; and (3) numerical treatment of shock waves should not cause deterioration in resolution and reduction of accuracy in domains where the solution is smooth. For high-order numerical methods, insufficient resolution or an inadequate treatment of shocks can result in Gibbs oscillations, which grow rapidly and contribute to numerical instabilities. These challenges have been a driving force behind the development of shock capturing methods designed to detect and stabilize shocks. 

 
A number of shock detection methods rely on the non-smoothness of the numerical solution to detect shocks as well as other sharp features \cite{Cook2004,Cook2005,Fiorina2007,Kawai2008,Kawai2010,Klockner2011,MR2056921,Mani2009,Olson2013,persson06:_shock_capturing,Persson2013}. Among them, the sensor by \cite{MR2056921}, devised in the context of DG methods, takes advantage of the theoretical convergence rate of DG schemes for smooth solutions in order to detect discontinuities. The shock sensor by \cite{persson06:_shock_capturing,Persson2013} is based on the decay rate of the coefficients of the DG polynomial approximation. Other methods that rely on high-order derivatives of the solution include \cite{Cook2004,Cook2005,Fiorina2007,Kawai2008,Kawai2010,Klockner2011,Mani2009,Olson2013,Premasuthan2013,Premasuthan2014}, and apply to numerical schemes for which such derivatives can be computed, such as spectral methods and finite difference methods. The most simple shock-detection method is to take advantage of the strong compression that a fluid undergoes across a shock wave and use the divergence of the velocity field as a shock sensor \cite{Fernandez2018,Moro2016,Nguyen2011a}.  

Shock stabilization methods lie within one of the following two categories: limiters and artificial viscosity. Limiters, in the form of flux limiters \cite{Burbeau2001,Cockburn1989,Krivodonova2007}, slope limiters \cite{Cockburn1998a,MR2056921,Lv2015,Sonntag2017}, and WENO-type schemes \cite{Luo2007,Qiu2005,Zhu2008,Zhu2013} pose implementation difficulties for implicit time integration schemes and high-order methods on complex geometries. As for artificial viscosity methods, Laplacian-based \cite{Barter2010,Hartmann2013,Lv2016,Moro2016,Nguyen2011a,persson06:_shock_capturing,Persson2013} and physics-based \cite{Abbassi2014,Chaudhuri2017,Cook2004,Cook2005,Fernandez2018,Fiorina2007,Kawai2008,Kawai2010,Mani2009,Olson2013,persson06:_shock_capturing} approaches have been proposed. An assessment of artificial viscosity methods for LES is presented in \cite{Johnsen2010}.

Shock capturing using artificial viscosity may date back as early as 1950~\cite{VonNeumann1950}. The main idea is to add an artificial viscous term into the governing equations to stabilize shock waves without affecting the solution away from the shock region. When the amount of viscosity is properly added in a neighborhood of shocks, the solution can converge uniformly except in the region around shocks, where it is smoothed and spread out over some length scale. On the other hand, excessive addition of artificial viscosity may negatively affect  the computed solution not only in the shock region but also in other parts of the domain where the solution is smooth. Artificial viscosity has been widely used in finite volume methods~\cite{Jameson1995}, streamline upwind Petrov-Galerkin (SUPG) methods~\cite{HughesMalletMisukamiII86},  spectral methods~\cite{madtad93,Tadmor89}, as well as DG methods~\cite{Barter2010,Ching2019,HartmannHoustonCompressible02,persson06:_shock_capturing,Bai2022a,Vila-Perez2021}. Both Laplacian-based \cite{Hartmann2013,Lv2016,Nguyen2011a,Moro2016,persson06:_shock_capturing,Persson2013} and physics-based \cite{Abbassi2014,Bhagatwala2009,Mani2009,Cook2005,Cook2007,Fiorina2007,Kawai2008,Kawai2010,Olson2013,Premasuthan2010b} artificial viscosity methods have been used for shock capturing.
 


The recent work \cite{Zahr2018,Zahr2020,Shi2022} introduces an optimization-based method for resolving discontinuous solutions of conservation laws with high-order numerical discretizations that support inter-element solution discontinuities, such as discontinuous Galerkin or finite volume methods \cite{Zahr2020}. The method aims to align inter-element
boundaries with discontinuities in the solution by deforming the computational mesh in order to avoid Gibbs’ phenomena. It requires solution of a PDE-constrained optimization problem for both the computational mesh and the numerical solution using sequential quadratic programming
solver. Recently, a moving discontinuous Galerkin finite element method with interface condition enforcement (MDG-ICE) \cite{Corrigan2019,Kercher2021,Kercher2021a} is formulated for shock flows by enforcing the interface condition separately from the conservation laws. In the MDG-ICE method, the discrete grid geometry is treated as an additional variable to detect interfaces and satisfy the interface condition, thereby directly fitting shocks and preserving high-order accurate solutions. The Levenberg-Marquardt method is used to solve the regularized coupled system of the conservation laws and the interface condition to obtain the approximate solution and the shock-aligned mesh.

In this paper, we introduce an adaptive viscosity regularization approach for the numerical solution of nonlinear conservation laws with shock waves. The approach aims to numerically solve a sequence of viscosity-regularized problems by making the amount of viscosity as small as possible while simultaneously enforcing relevant physics and smoothness constraints on the numerical solution.
The methodology is based on two main ingredients: the viscosity regularization of nonlinear conservation laws and the homotopy continuation of regularization parameters.
In particular, the viscosity regularization consists of a PDE-based artificial viscosity method \cite{Barter2010,Ching2019} which couples the regularized conservation laws with an additional Helmholtz equation for the artificial viscosity. On the other hand, homotopy continuation is used to minimize the amount of artificial viscosity subject to positivity-preserving and smoothness constraints on the numerical solution.
We propose a relaxation variant of our method to solve the regularized conservation laws and the Helmholtz equation separately. 

\revise{Dissipation-based continuation was developed in \cite{Hicken2011} as a form of globalization suitable for inexact--Newton flow solvers and an alternative to pseudo-transient continuation.} Our  method minimizes the amount of artificial viscosity needed for stabilizing the numerical solution under the presence of shocks on a given mesh. 
Nevertheless, this strategy can be coupled to mesh adaptation algorithms that identify the shock location and generate a shock-aligned grid in order to further reduce the amount of artificial dissipation.
To this end, we also introduce an algorithm to locate shock waves and generate shock-aligned meshes so as to reduce the amount of artificial viscosity and approximate the exact solution of the original conservation laws with increased accuracy. 


While different numerical schemes, such as finite volume methods and finite difference methods, can be used to discretize the governing equations, this paper employs finite element methods.
In particular, we use the hybridizable discontinuous Galerkin (HDG) method \cite{Nguyen2012,Fernandez2018a,Moro2011a,Peraire2010,Vila-Perez2021,Woopen2014c,Fidkowski2016,Fernandez2017a,williams2018entropy} to numerically solve the viscosity-regularized conservation laws and the continuous Galerkin (CG) method to solve the Helmholtz equation for the artificial viscosity. The HDG method is considered here due to its efficiency and high-order accuracy, while the CG method is employed to provide a continuous viscosity field. The continuity and smoothness of the artificial viscosity is highly desirable \cite{Barter2010,Moro2016,Bai2022a,Ching2019}. We demonstrate the approach on inviscid transonic, supersonic, hypersonic  flows in two dimensions. 


The paper is organized as follows. We present the adaptive viscosity regularization approach in Section~\ref{sec:viscosity}, which is accompanied by the description of the homotopy continuation procedure, the specification of the physical and numerical constraints, and the proposed shock identification and mesh alignment algorithms. In Section~\ref{sec:results}, we present numerical results to assess the performance of the proposed approach on the inviscid Burgers' equation and inviscid transonic, supersonic, and hypersonic flows. Finally, in Section~\ref{sec:conclusions}, we conclude the paper with some remarks and future work.

\section{Adaptive Viscosity Regularization Approach} \label{sec:viscosity}

\subsection{Viscosity regularization of nonlinear hyperbolic systems}

We consider a hyperbolic system of $m$  conservation laws, defined on the physical domain $\Omega \in \mathbb{R}^d$ and subject to appropriate initial and boundary conditions, as follows
\begin{equation}
\label{eq1}
\frac{\partial \bm u}{\partial t} + \nabla \cdot \bm F(\bm u) = 0  \quad \mbox{in }\Omega,    
\end{equation}
where $\bm u(\bm x) \in \mathbb{R}^m$ is the solution of the system of conservation laws at $\bm x \in \Omega$ and the  physical fluxes $\bm F = (\bm f_1(\bm u), \ldots, \bm f_d(\bm u)) \in \mathbb{R}^{m \times d}$ include $d$ vector-valued functions of the solution. We assume that the fluxes $\bm F$ are smooth functions of  $\bm u$ and that the system is  hyperbolic in the sense that the Jacobian matrix $\bm A(\bm u) = \sum_{i=1}^d \alpha_i \partial \bm f_i(\bm u) / \partial \bm u$ has real eigenvalues and is diagonalizable for all $\alpha_1,\ldots, \alpha_d \in \mathbb{R}$. Furthermore, we assume that (\ref{eq1}) is a non-dimensional system.  \revise{Let $\Gamma_{\rm wall} \subset \partial \Omega$ be the wall boundary. For the Euler equations, the boundary condition at the wall boundary $\Gamma_{\rm wall}$ is $\bm v \cdot \bm n = 0$, where $\bm v$ is the velocity field and $\bm n$ is the unit normal vector outward the boundary.}

A distinguished feature of nonlinear hyperbolic systems is the possible development of shock waves in the exact solution even if the initial data is smooth. Shock waves have always been a considerable source of difficulties toward a rigorous mathematical treatment of  nonlinear hyperbolic systems. In the presence of shock waves, the system of conservation laws (\ref{eq1}) admits a class of weak solutions in a distributional sense \cite{Bianchini2005}. In order to single out the unique ``physically relevant'' solution among all possible weak solutions, additional entropy conditions must be imposed along shocks \cite{Tadmor89}.   \revise{The  entropic solutions of the hyperbolic system (\ref{eq1}) coincide with the vanishing viscosity solutions of the following parabolic system
\begin{equation}
\label{eq2}
\frac{\partial \bm u}{\partial t} + \nabla \cdot \bm F(\bm u) -  \nabla \cdot \varepsilon  \nabla \bm u = 0  \quad \mbox{in }\Omega,  
\end{equation}
in the limit of the vanishing viscosity $\varepsilon \to 0$ in one dimension \cite{Bianchini2005}. A recent work \cite{Chen2015} establishes the strong convergence of the viscosity approximate solutions to finite-energy entropy solutions of the multidimensional Euler equations with spherical symmetry.} The parabolic system (\ref{eq2}) is a viscosity regularization of the original hyperbolic system (\ref{eq1}) and inspires the artificial viscosity method for capturing shock waves. 


Herein we follow \cite{Barter2010,Ching2019} to consider a more sophisticated viscosity regularization of the original hyperbolic system (\ref{eq1}) as follows
\begin{subequations}
\label{eq3}
\begin{alignat}{2}
\frac{\partial \bm u}{\partial t} + \nabla \cdot \bm F(\bm u) - \lambda_1 \nabla \cdot \bm G(\bm u,   \nabla \bm u,  \eta) = 0  \quad \mbox{in }\Omega, \\
\eta - \lambda_2^2 \nabla  \cdot \left( 
\ell^2 \nabla  \eta  \right)- s(\bm u, \nabla \bm u) = 0 \quad \mbox{in }\Omega ,
\end{alignat}
\end{subequations}
where $\eta(\bm x)$ is the solution of the Helmholtz  equation (\ref{eq3}b) with homogeneous  Neumann boundary conditions 
\begin{equation}
\eta = 0 \quad \mbox{on } \Gamma_{\rm wall}, \qquad 
\ell^2 \nabla  \eta \cdot \bm n = 0 \quad \mbox{on }\partial \Omega \backslash \Gamma_{\rm wall} \ .
\end{equation}
\revise{Here $\lambda_1$ is the first regularization parameter that controls the amplitude of artificial viscosity, and $\lambda_2$ is the second regularization parameter that controls the thickness of artificial viscosity. Furthermore, $\ell$ is an appropriate length scale. For notational convenience, we denote $\bm \lambda = (\lambda_1, \lambda_2)$.}



\begin{figure}[htbp]
\centering
\includegraphics[scale=0.18]{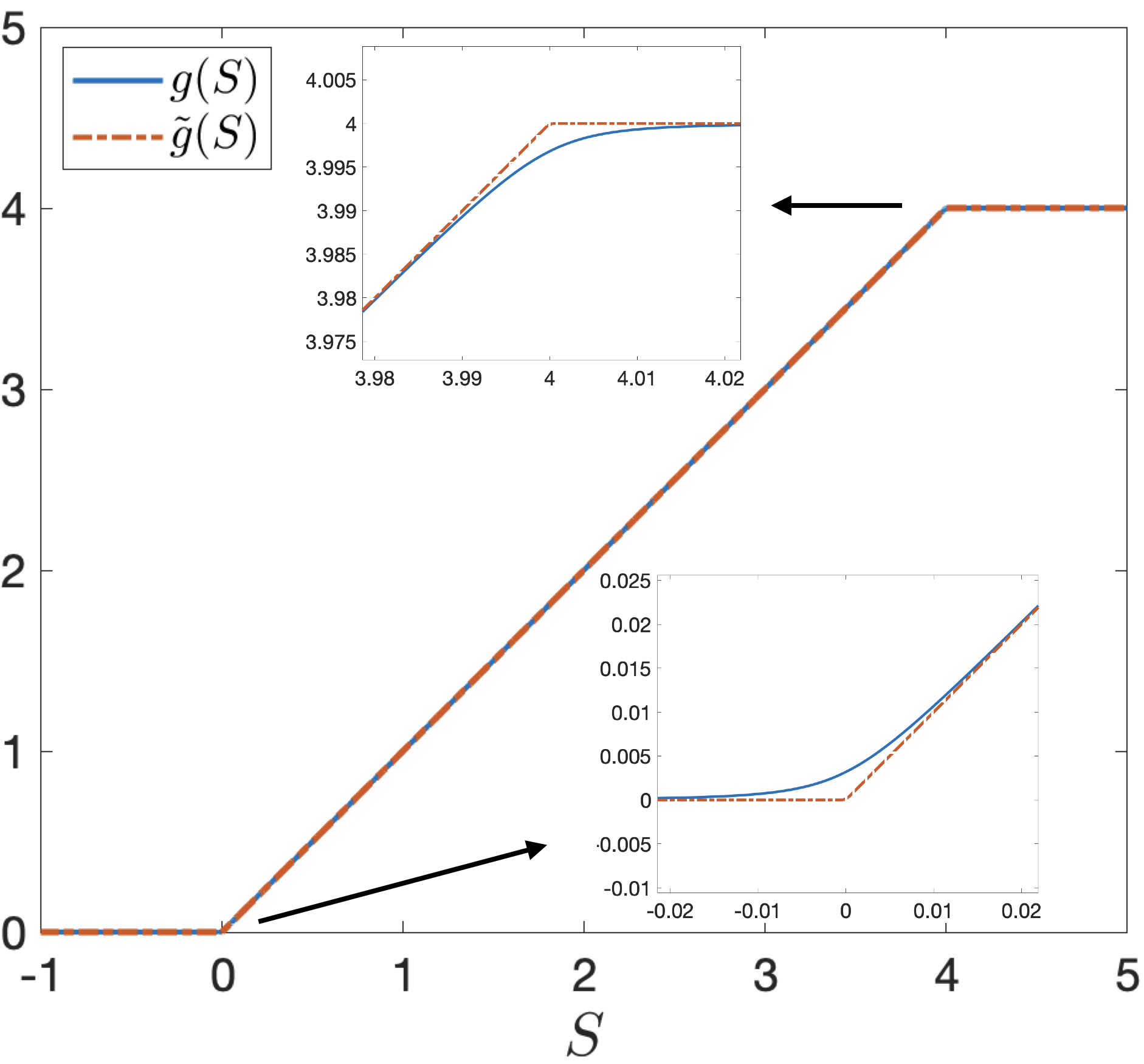} \hfill 
\includegraphics[scale=0.375]{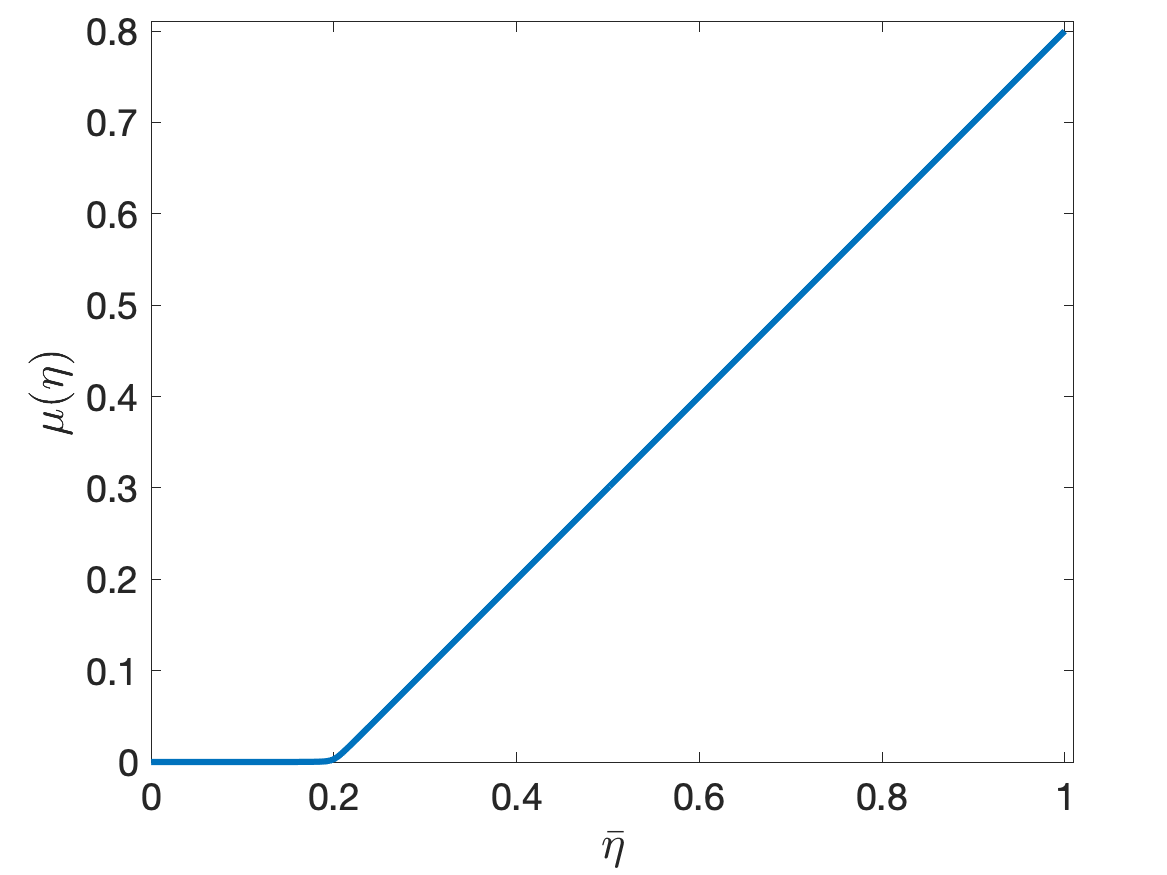} \\
\caption {\label{fig1} The source term $s$ as a function of the shock strength $S$ for $s_{\max} = 4$ (left) and the artificial viscosity $\mu(\eta)$ as a function of $\bar{\eta}$ for $\bar{\eta}_{\rm T} = 0.2$ (right). Note that the derivative of $g(S)$ is continuous at $S=0$ and $S= s_{\rm max}$, whereas that of $\tilde{g}(S)$ is discontinuous there.}
\end{figure}

The source term $s$ in (\ref{eq3}b) is required to determine $\eta$. It must meet several requirements in order to yield an effective artificial viscosity. First and foremost, it must be a non-negative function. Second, it takes positive value in the shock region and smoothly vanishes to zero away from the shock region. Third, the positive value of the source term is proportional to the  shock strength. And last, it must be a smooth function of $\bm u$ and $\nabla \bm u$. The source term should depend on the solution gradient because gradient information is needed to determine the shock strength. A particular source term that satisfies the above requirements is defined as follows
\begin{equation}
s(\bm u, \nabla \bm u) =  {g}(S(\bm u, \nabla \bm u)) 
\end{equation}
where  $g(S)$ is a smooth approximation of the following step function 
\begin{equation}
\tilde{g}(S) = \left\{
\begin{array}{cl}
   0  & \mbox{if } S < 0, \\
   S  & \mbox{if } 0 \le S \le s_{\rm max}, \\
   s_{\rm max} & \mbox{if } S > s_{\rm max} . 
\end{array}
\right.
\end{equation}
\revise{The quantity $S(\bm u, \nabla \bm u)$ is a measure of the shock strength whose definition depends on the governing equations.  
For systems of hyperbolic conservation laws such as the Euler equations, we consider  
\begin{equation}
S(\bm u, \nabla \bm u) = -\nabla \cdot \bm v \ ,
\end{equation}
where $\bm v$ is the non-dimensional velocity field that is determined from the state vector $\bm u$. For scalar conservation laws, $S$ can be defined as the magnitude of the gradient of the scalar variable.} The parameter $s_{\max}$ is used to put an upper bound on the source term when the divergence of the velocity becomes too negatively large. Herein we choose $s_{\max} = 0.5 \|S\|_\infty$, where $\|S\|_{\infty} = \max_{\bm x \in \Omega}  |S(\bm x)|$ is the $L_\infty$ norm. \revise{Since $S$ depends on the solution, so its norm may not be known prior. In practice, we employ a homotopy continuation scheme to  iteratively solve the problem (\ref{eq3}).  Hence, $s_{\max}$ is computed by using the numerical solution at the previous iteration of the homotopy continuation.} Figure \ref{fig1} shows $g(S)$ and $\tilde{g}(S)$ as a function of $S$.  The source term is only active in the region of the flow where the divergence of the velocity field is negative. The use of the velocity divergence as shock strength for defining an artificial viscosity field has appeared in \cite{Fernandez2018,Moro2016,Nguyen2011a}.


The artificial fluxes $\bm G$ provide a viscosity regularization  to smooth out the discontinuities in the exact solution of the original hyperbolic system. With a proper choice of the artificial fluxes, the solution $\bm u$ of the regularized system (\ref{eq3}a) must be continuous for  positive values of $\lambda_1$ and $\eta$. There are a number of different options for the artificial fluxes $\bm G$. In this paper, we use the Laplacian fluxes of the form
\begin{equation}
\bm G(\bm u,   \nabla \bm u,  \eta) = \mu(\eta) \nabla \bm u  ,  
\end{equation} 
where 
\begin{equation}
\mu(\eta) = (\bar{\eta}-\bar{\eta}_{\rm T})\left(\frac{\arctan(100(\bar{\eta} -\bar{\eta}_{\rm T}))}{\pi} + \frac{1}{2} \right) - \frac{\arctan(100)}{\pi} + \frac{1}{2}  
\end{equation} 
is a smooth approximation of a ramp function as shown in Figure \ref{fig1}. Here $\bar{\eta} = \eta/\|\eta\|_\infty$ is the normalized function with $\|\eta\|_{\infty} = \max_{\bm x \in \Omega}  |\eta(\bm x)|$ being the $L_\infty$ norm. Note that $\bar{\eta}_{\rm T}$ is the artificial viscosity threshold that makes $\mu(\eta)$ vanish to zero when $\bar{\eta} \le \bar{\eta}_{\rm T}$. In other words, artificial viscosity is only added to the shock region where $\bar{\eta}$ exceeds $\bar{\eta}_{\rm T}$. Therefore, the threshold $\bar{\eta}_{\rm T}$ will help remove excessive artificial viscosity. Since $\|\bar{\eta}\|_\infty = 1$, $\bar{\eta}_{\rm T} = 0.2$ is a sensible choice. Note that the artificial viscosity field $\varepsilon(\bm x)$ is equal to $\lambda_1 \mu(\bm x)$, where  $\mu(\bm x)$ is bounded by $\mu(\bm x) \in [0, 1 - \bar{\eta}_{\rm T}]$ for any $\bm x \in \Omega$. We can also consider a more general form $\bm G = \mu(\eta) \nabla \bm u^*$ \cite{Barter2010,Nguyen2011a}, where $\bm u^*$ is a modified state vector. Both are known as Laplacian-based artificial viscosity.  Another option is physics-based artificial viscosity  by taking $\bm G$ to be the viscous stress tensor and the heat flux of the Navier-Sokes equation and adding the artificial viscosity to the physical viscosities and thermal conductivity \cite{CuongNguyen2022,Nguyen2023a}. 

We still need to determine $\lambda_1$ and $\lambda_2$ in order to close the system (\ref{eq3}). If $\lambda_1$ is too small, then the artificial viscosity will not be large enough to stabilize the numerical method used to solve the system (\ref{eq3}). But if $\lambda_1$ is too large, then  the solution of the system (\ref{eq3}) will be no longer an accurate approximation to the solution of the original system (\ref{eq1}). Likewise, if $\lambda_2$ is too small, then the artificial viscosity will not be sufficiently smooth to stabilize the numerical method used to solve the system (\ref{eq3}). But if $\lambda_2$ is too large, then the solution of the system (\ref{eq3}) will be no longer an accurate approximation to the solution of the original system (\ref{eq1}) since the artificial viscosity spreads out beyond the shock region. 

 
\subsection{Homotopy continuation of the regularization parameters}

In this paper we will focus on steady-state problems. We describe the adaptive regularization approach for numerically solving the steady-state version of the nonlinear hyperbolic system. In this case, the coupled system (\ref{eq3}) reduces to 
\begin{subequations}
\label{eq4}
\begin{alignat}{2}
\nabla \cdot \bm F(\bm u) - \lambda_1 \nabla \cdot \bm G(\bm u,   \nabla \bm u,  \eta) = 0  \quad \mbox{in }\Omega, \\
\eta - \lambda_2^2 \nabla \cdot \left( \ell^2 \nabla \eta \right) - s(\bm u, \nabla \bm u) = 0 \quad \mbox{in }\Omega .
\end{alignat}
\end{subequations}
For time-dependent problems, the same approach can be applied to the fully discrete system at every time step. 


We denote by $\mathcal{T}_h$ a collection of  curved elements that partition the physical domain $\Omega$. For any element $K \in \mathcal{T}_h$, we say that it belongs to the shock region if the following inequality holds
\begin{equation}
\frac{\int_K  \bar{\eta} d \bm x}{\int_K d \bm x} \ge \bar{\eta}_{\rm T} \ 
\end{equation}
where $\bar{\eta} = \eta/ \|\eta\|_\infty$. The left quantity is the cell average of the shock strength measure. The shock region is defined by 
\begin{equation}
\mathcal{T}_h^{\rm shock} = \{K \in \mathcal{T}_h \ : \ \int_K  \bar{\eta}  d \bm x  \ge \bar{\eta}_{\rm T} |K|  \} . 
\end{equation}
\revise{The artificial viscosity field $\eta$ is a smooth approximation of the source term $s$. We note that $\eta$ approaches $s$ in the limit $\lambda_2 \ell \to 0$, and that $\eta$ becomes smoother than $s$ as $\lambda_2 \ell$ increases. Let $h_{\rm min}$ be the smallest edge over all elements in the shock region. For $\lambda_2 \ell < h_{\min}$, $\eta$ may not be smooth enough as it is not much different from $s$. Hence, we  choose $\lambda_2 \ell \ge h_{\min}$ because we would like the artificial viscosity to be sufficiently smooth. Furthermore, we will consider $\ell = h_{\min}$ for the numerical examples reported herein. In this case, the parameter $\lambda_2$ should be greater than or equal to 1 so that $\lambda_2 \ell \ge h_{\min}$. More generally, the length scale $\ell$ can be set to a suitable tensor-valued function which allows for anisotropic smoothing of the artificial viscosity.} 



\revise{
The pair of regularization  parameters $\bm \lambda = (\lambda_1, \lambda_2)$ controls the magnitude and thickness of the artificial viscosity in order to obtain accurate solutions. On the one hand, if $\bm \lambda$ is too small then the numerical solution can develop oscillations across the shock waves. On the other hand, if $\bm \lambda$ is too large the solution becomes less accurate in the shock region, which in turn  affects the accuracy of the solution in the remaining region. Therefore, we propose a homotopy continuation method to determine $\bm \lambda$. The key idea is to solve the regularized system with a large value of $\bm \lambda$ first and then gradually decrease $\bm \lambda$ until any of the physics or smoothness constraints on the numerical solution are violated. At this point, we take the value of $\bm \lambda$ from the previous iteration where the numerical solution still satisfies all of the physics and smoothness constraints. This procedure is summarized in the following algorithm:
}

\revise{
\begin{itemize}
 \item Given  initial value $\bm \lambda_0 = (\lambda_{0,1}, \lambda_{0,2})$, numerically solve the coupled system (\ref{eq4}) with $\bm \lambda = \bm \lambda_0$ to obtain the initial solution $(\bm u_0, \eta_0)$.
 \item  Set $\bm \lambda_n = (\zeta^n \lambda_{0,1}, 1 + \zeta^n (\lambda_{0,2} - 1))$ for some constant $\zeta \in (0,1)$ and solve the coupled system (\ref{eq4}) with $\bm \lambda = \bm \lambda_n$ to obtain the iterative solution $(\bm u_n, \eta_n)$ for $n = 1, 2, \ldots$ until $\bm u_n$ violates any of the constraints.
 \item Finally, we accept $\bm u_{n-1}$ as the numerical solution of the original system of conservation laws.
\end{itemize}
}

The adaptive viscosity regularization approach can be seen as a method to solve the following minimization problem
\begin{subequations}
\label{eq5}
\begin{alignat}{2}
\min_{\lambda_1 \in \mathbb{R}^+, \lambda_2 \ge 1, \bm u, \eta} & \quad \lambda_1 \lambda_2  \\
\mbox{s.t.} & \quad \mathcal{L}(\bm u, \eta, \bm \lambda) = 0 \\
 & \quad \bm u \in \mathcal{C} .
\end{alignat}
\end{subequations}
Here $\mathcal{L}$ represents the spatial discretization of the coupled system (\ref{eq4}) by a numerical method and $\mathcal{C}$ represents a set of constraints on the numerical solution. The objective function is to minimize the amount of artificial viscosity which is proportional to $\lambda_1 \lambda_2$. The constraints rule out unwanted solutions of the discrete system (\ref{eq5}b) and play an important role in yielding a high-quality numerical solution. Hence, the optimization problem (\ref{eq5}) is to minimize the amount of artificial viscosity while ensuring the physicality and accuracy of the numerical solution. We will later introduce the constraints that are used to obtain such numerical solution.


\subsection{Relaxation variant of the adaptive viscosity regularization}

The above homotopy continuation method requires us to solve the coupled system (\ref{eq4}). In order to  be able to decouple the system (\ref{eq4}), we propose the following variant of the homotopy continuation: 

\revise{
\begin{itemize}
 \item Given an initial choice of $\bm \lambda_0 = (\lambda_{0,1}, \lambda_{0,2})$ and $\eta_0$ such that $\|\eta_0\|_{\infty} = 1$, solve the regularized system (\ref{eq4}a) with $\lambda_1 = \lambda_{0,1}, \eta = \eta_0$ to obtain the initial solution $\bm u_0$. 
 \item   Set $\lambda_{n,1} = \zeta^{n-1} \lambda_{{n-1},1}$ and $\lambda_{n,2} = 1 + \zeta^{n-1}(\lambda_{{n-1},2} -1)$ for some constant $\zeta \in (0,1)$; solve the Helmholtz equation (\ref{eq4}b) with $\lambda_2 = \lambda_{n,2}$ and the source term from $\bm u_{n-1}$ to obtain $\eta_{n}$; and solve the regularized system (\ref{eq4}a) with $\lambda_1 = \lambda_{n,1}, \eta = \eta_{n}$ to obtain the iterative solution $\bm u_{n}$ for $n = 1, 2, \ldots$ until $\bm u_{n}$ violates any of the constraints.
 \item Finally, we accept $\bm u_{n-1}$ as the numerical solution of the system of conservation laws.
\end{itemize}

The initial function $\eta_0$ can be set to 1 on most of the physical domain $\Omega$ except near the wall boundary where it vanishes smoothly to zero at the wall. The initial value $\lambda_{0,1}$ is conservatively large to make the initial solution $\bm u_0$ very smooth. The initial value $\lambda_{0,2}$ depends on the type of meshes used to compute the numerical solution. For regular meshes that have the elements of the same size in the shock region, $\lambda_{0,2} = 1.5$ is a sensible choice. For adaptive meshes that are refined toward the shock region, we choose $\lambda_{0,2} = 5$ since $\ell = h_{\min}$ is extremely small for shock-adaptive meshes. In any case, $\lambda_{n,2}$ will decrease from $\lambda_{0,2}$ toward 1 during the homotopy iteration. Hence, the choice of $\lambda_{0,2}$ can be flexible.


This homotopy procedure solves the Helmholtz equation (\ref{eq4}b) separately from the regularized system (\ref{eq4}a). Hence, different  numerical methods can be used to solve (\ref{eq4}a) and (\ref{eq4}b) separately.   The method is robust enough that the number of homotopy iterations required to reach the convergence is usually around 10.
}

\subsection{Solution constraints}

The physical constraints are that pressure and density must be positive. In order to establish a smoothness constraint on the numerical solution, we express an approximate scalar variable $\xi$ of degree $k$ within each element in terms of an orthogonal basis and its truncated expansion of degree $k-1$ as
\begin{equation}
\xi = \sum_{i=1}^{N(k)} \xi_i \psi_i, \qquad \xi^* = \sum_{i=1}^{N(k-1)} \xi_i \psi_i    
\end{equation}
where $N(k)$ is the total number of terms in the $k$-degree expansion and $\psi_i$ are the basis functions \cite{persson06:_shock_capturing}. In this paper, $\xi$ is chosen to be either density or pressure when we use the method to solve the Euler equations. For scalar conservation laws, $\xi$ is the numerical solution.  We introduce the following quantity
\begin{equation}
\label{eq10}
\sigma(\bm \lambda_n) = \max_{K \in \mathcal{T}_h^{\rm shock}} \sigma_K(\bm \lambda_n), \qquad  \sigma_K(\bm \lambda_n) \equiv   \frac{\int_K  |\xi_n/\xi_n^* - 1| d \bm x}{\int_K d \bm x} .
\end{equation}
Some shock capturing methods \cite{persson06:_shock_capturing,Ching2019} use a piecewise-constant function similar to $\sigma_K$ in (\ref{eq10}) to compute the artificial viscosity field. Herein we employ this type of functions to devise a smoothness constraint on the numerical solution as follows. 

\begin{figure}[htbp]
\centering
	\begin{subfigure}[b]{0.32\textwidth}
		\centering		\includegraphics[width=\textwidth]{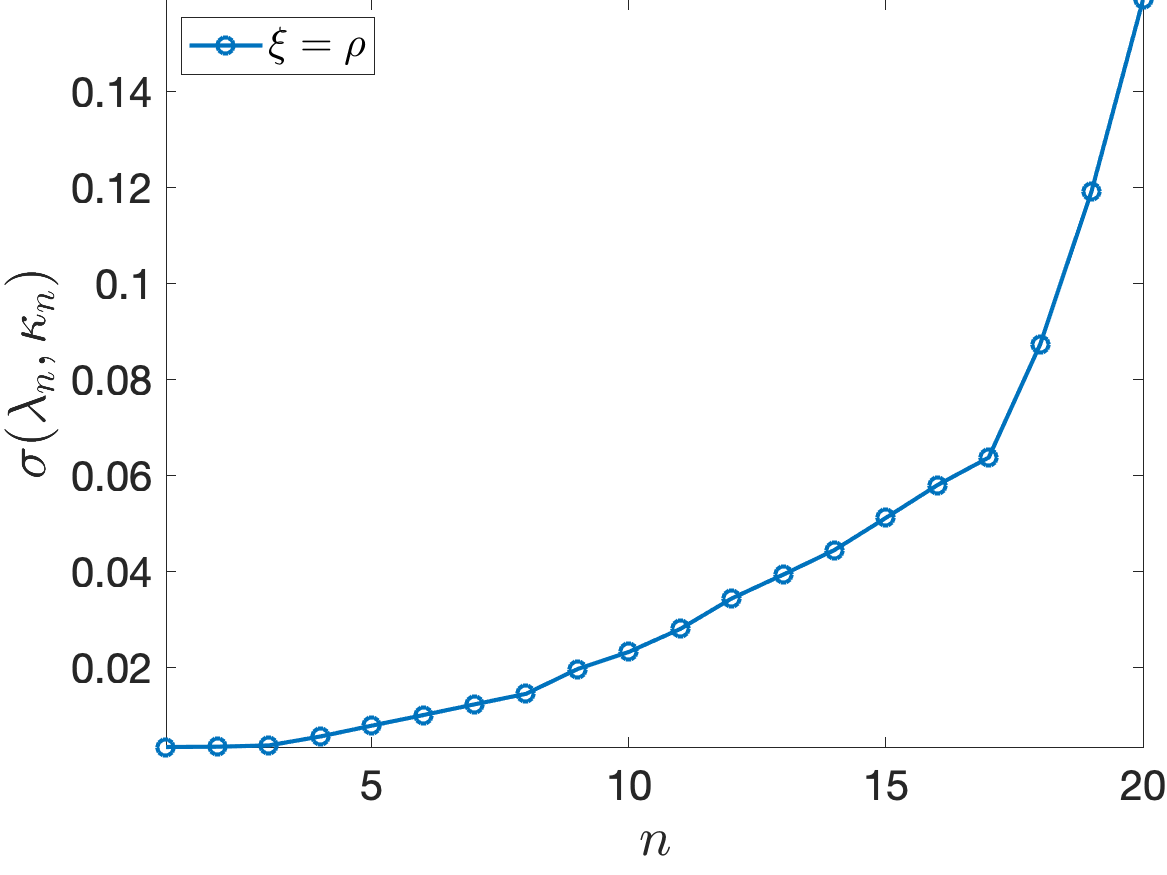}
		\caption{density}
	\end{subfigure}
	\hfill
	\begin{subfigure}[b]{0.34\textwidth}
		\centering		\includegraphics[width=\textwidth]{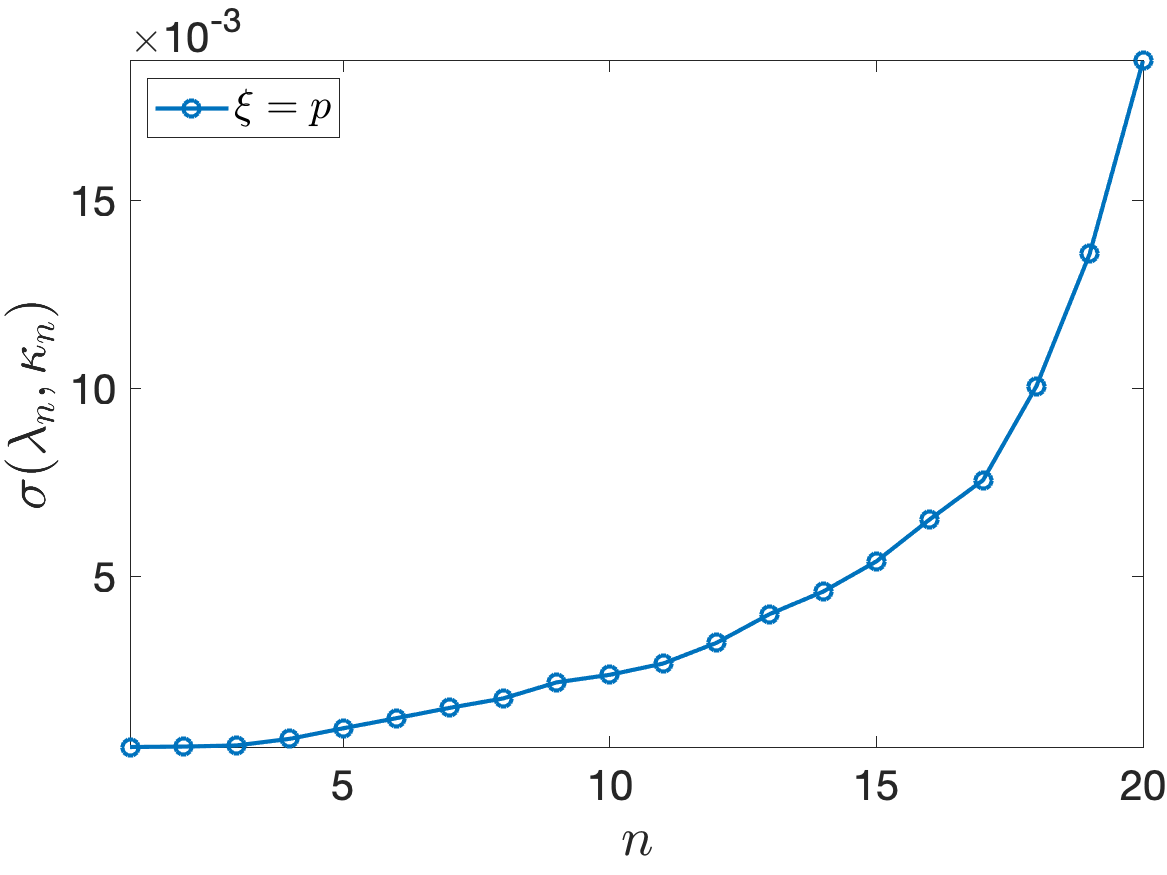}
		\caption{pressure}
	\end{subfigure}
	\begin{subfigure}[b]{0.32\textwidth}
		\centering		\includegraphics[width=\textwidth]{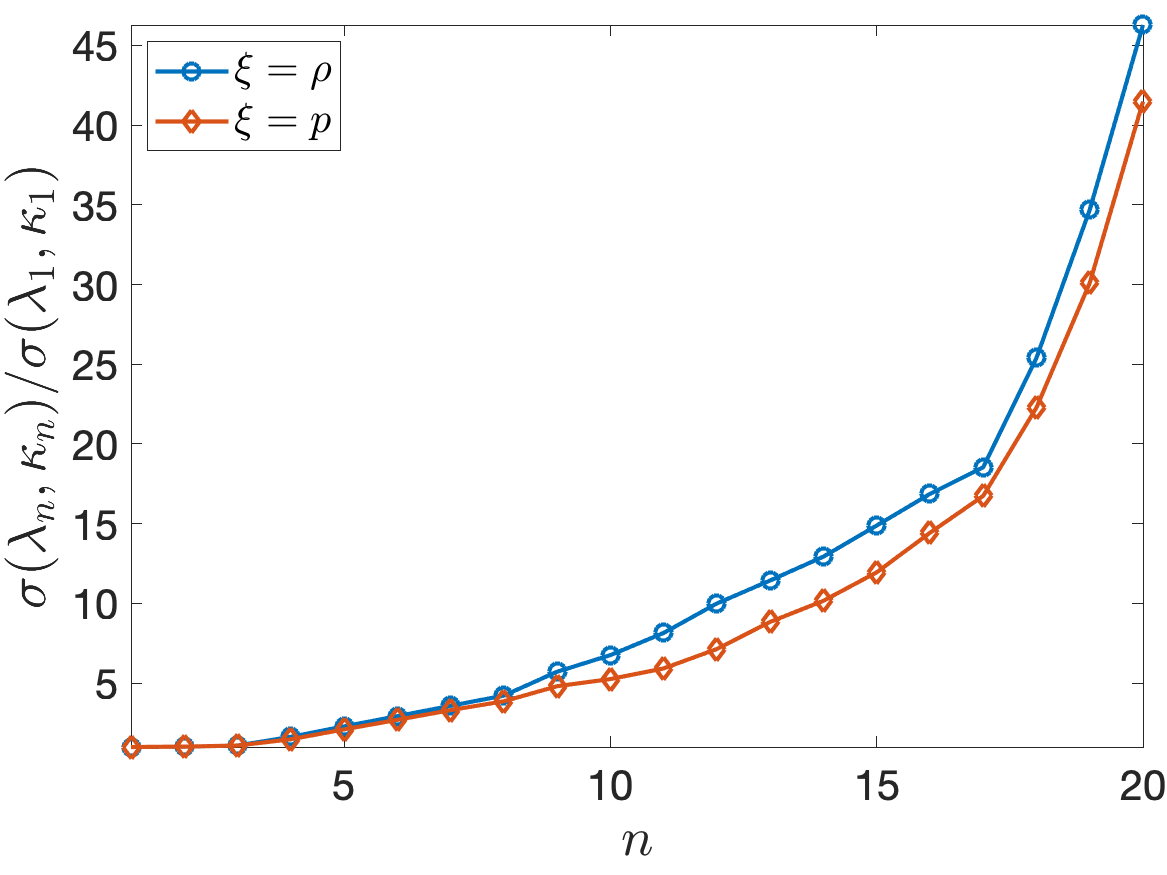}
		\caption{Density and pressure}
	\end{subfigure}
\caption{The computed value of $\sigma(\bm \lambda_n)$ as we gradually decrease $(\bm \lambda_n)$ in the homotopy continuation for the inviscid hypersonic flow past a circular cylinder at $M_\infty = 7$ on a regular mesh: (a) $\xi$ is chosen to be density, (b) $\xi$ is chosen to be pressure, (c) the normalized function $\sigma(\bm \lambda_n)/\sigma(\bm \lambda_1)$ for both cases.}
\label{figure2}	
\vspace{-0.5cm}	
\end{figure}
\begin{figure}[htbp]
	\centering
	\begin{subfigure}[b]{0.49\textwidth}
		\centering		\includegraphics[width=\textwidth]{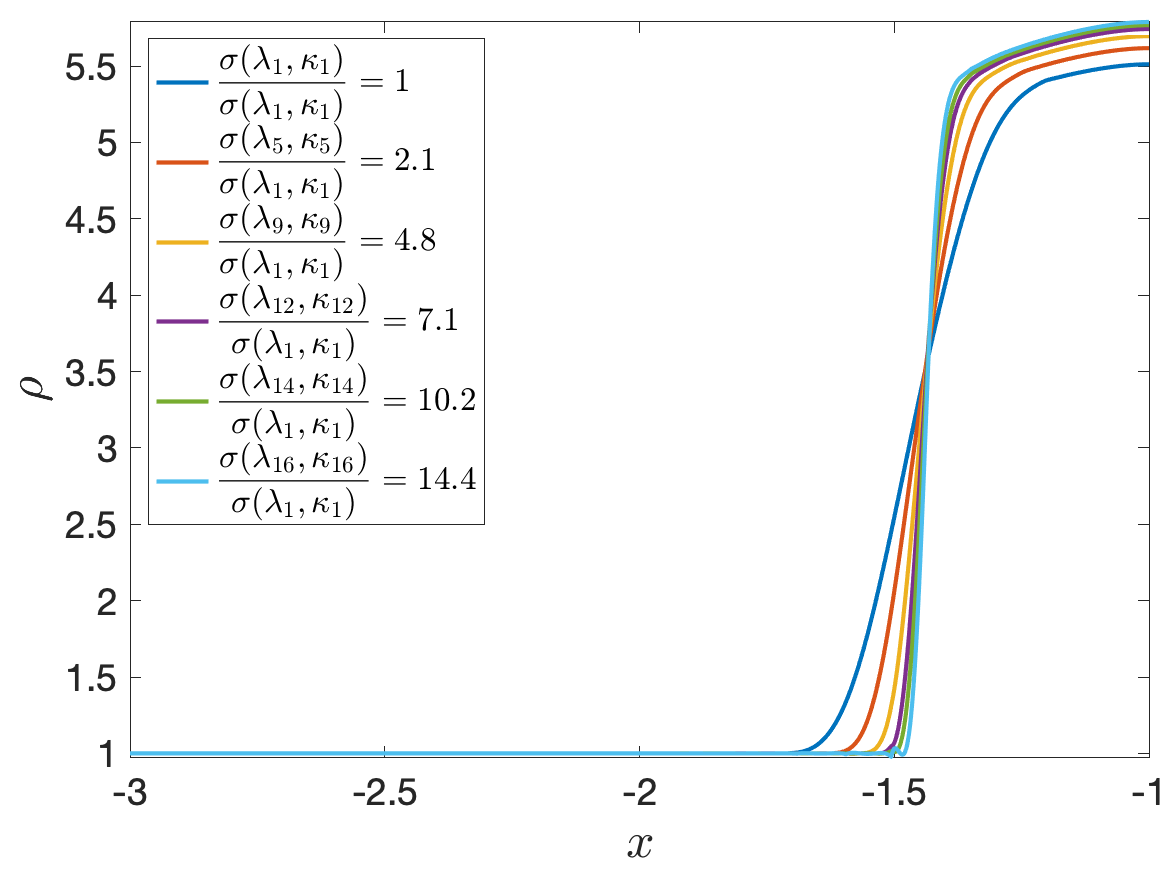}
		\caption{Approximate density along $y=0$}
	\end{subfigure}
	\hfill
	\begin{subfigure}[b]{0.49\textwidth}
		\centering
		\includegraphics[width=\textwidth]{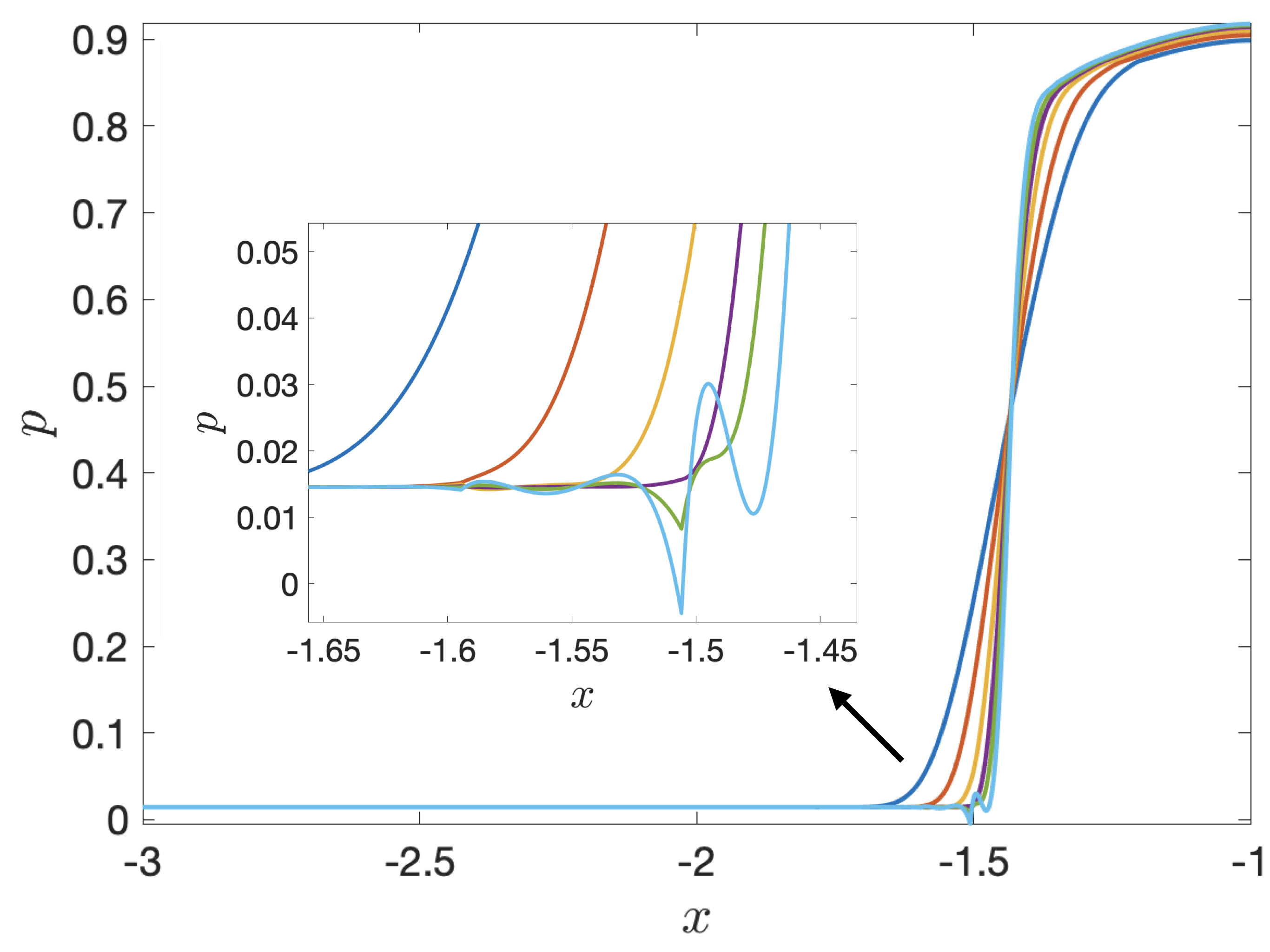}
		\caption{Approximate pressure along $y=0$}
	\end{subfigure} 
	\caption{The plot of the approximate density and pressure along the horizontal line $y=0$ for several values of $\sigma(\bm \lambda_n)/\sigma(\bm \lambda_1)$ for the inviscid hypersonic flow past a cylinder at $M_\infty = 7$ on a regular mesh. The approximate density is sharper and more accurate  as $\sigma(\bm \lambda_n)/\sigma(\bm \lambda_1)$ increases during the homotopy continuation. The approximate pressure is smooth for $\sigma(\bm \lambda_n)/\sigma(\bm \lambda_1) \le 7.1$,  mildly oscillatory for $\sigma(\bm \lambda_n)/\sigma(\bm \lambda_1) = 10.2$, and oscillatory and negative for $\sigma(\bm \lambda_n)/\sigma(\bm \lambda_1) = 14.4$.}
\label{fig2bb} 
\end{figure}

When $\bm \lambda_n$ are sufficiently large at the beginning of the homotopy loop, we expect $\sigma(\bm \lambda_n)$ to be small. As we gradually decrease $\bm \lambda_n$ during the homotopy continuation, we expect $\sigma(\bm \lambda_n)$ to increase. Figure \ref{figure2} illustrates this behavior of $\sigma(\bm \lambda_n)$ for the inviscid hypersonic flow past a circular cylinder at $M_\infty = 7$. It is important to note that $\sigma(\bm \lambda_n)$ depends on $\xi$, being this term either density or pressure.
Indeed, $\sigma(\bm \lambda_n)$ is about 10 times larger when $\xi$ is chosen to be density instead of pressure. However, note that the ratio $\sigma(\bm \lambda_n)/\sigma(\bm \lambda_1)$ behaves very similarly in both cases.
Finally, since we would like to bound $\sigma(\bm \lambda_n)$ because the numerical solution will be oscillatory if $\sigma(\bm \lambda_n)$ exceeds a certain value, we impose a bound on the ratio $\sigma(\bm \lambda_n)/\sigma(\bm \lambda_1)$. Hence, we introduce the following smoothness constraint 
\begin{equation}
\label{eq12a}
 \frac{ \sigma(\bm \lambda_n)}{\sigma(\bm \lambda_1)} \le C_\sigma,
\end{equation}
where $C_\sigma > 1$ is a smoothness parameter that controls the smoothness of the numerical solution. Figure \ref{fig2bb} shows the approximate density and pressure along $y=0$ for the inviscid hypersonic flow past a cylinder. While the approximate solution is sharper and more accurate as $\sigma(\bm \lambda_n)/\sigma(\bm \lambda_1)$ increases, it becomes oscillatory when $\sigma(\bm \lambda_n)/\sigma(\bm \lambda_1)$ exceeds 10. A reasonably conservative choice for the smoothness parameter is $C_{\sigma} = 5$. 



It is important to choose a sufficiently large value for $\bm \lambda_0$, so that $\sigma(\bm \lambda_n)$ slowly increases when $(\bm \lambda_n)$ gradually decreases for the first few homotopy iterations. A rapid increase of $\sigma(\bm \lambda_n)$ for the first few homotopy iterations should not be allowed, because it is likely to result in oscillatory solutions. If this is encountered, it is likely that $\bm \lambda_0$ is not large enough and a larger value for $\bm \lambda_0$ must be used. It is completely acceptable for $\sigma(\bm \lambda_n)$ to decrease during the first few homotopy iterations. If this occurs, we replace the smoothness constraint (\ref{eq12a}) with 
\begin{equation}
 \frac{ \sigma(\bm \lambda_n)}{\sigma(\bm \lambda_m)} \le C_\sigma ,  
\end{equation}
where $m = \arg \min_{k \in [1,2,\ldots,n-1]} \sigma(\bm \lambda_k)$. In other words, $\sigma(\bm \lambda_m)$ is the smallest value of $\sigma(\bm \lambda_k)$ for $k=1,\ldots,n-1,$ during the homotopy continuation loop. 

The physical constraints ensure that the integral in (\ref{eq10}) is bounded since both $\xi_n$ and $\xi_n^*$ are positive. For scalar conservation laws, there may be no such constraints and it is possible for $\xi_n$ and $\xi_n^*$ to be zero. In that case, we suggest to replace $\xi_n/\xi_n^*$ with $1$ whenever $\xi_n^*$ is zero or very close to zero. For the Euler equations, the constraint set $\mathcal{C}$ in (\ref{eq5}) consists of the following contraints    
\begin{equation}
 \rho(\bm x) > 0, \quad  p(\bm x) > 0,   \quad  \frac{ \sigma(\bm \lambda_n)}{\sigma(\bm \lambda_m)} \le C_\sigma .
\end{equation}
\revise{The first two constraints enforce the positivity of density and pressure, while the last constraint guarantees the smoothness of the numerical solution. The smoothness constraint  imposes a degree of regularity on the numerical solution and plays a vital role in yielding sharp and smooth solutions. If any of the constraints (18) is violated at iteration $n$, we end the homotopy continuation and accept $\bm u_{n-1}$ as the numerical solution of the problem. Because the numerical solution $\bm u_{n-1}$ satisfies the constraints (18), the associated density and pressure must be positive.}


\subsection{Finite element discretizations} \label{sec:FEmethods}

\revise{In the adaptive viscosity regularization approach described earlier, one can use any appropriate numerical method to solve the viscosity-regularized conservation laws and the Helmholtz equation. In this paper, we employ the hybridizable discontinuous Galerkin (HDG) method to solve the former and the continuous Galerkin (CG) method to solve the latter. We use the CG method since it allows us to obtain a continuous artificial viscosity field. The HDG method \cite{Nguyen2012,Fernandez2018a,Moro2011a,Peraire2010,Vila-Perez2021,Woopen2014c,Fidkowski2016,Fernandez2017a,williams2018entropy} is suitable for solving the regularized conservation laws because of its efficiency and high-order accuracy.}

\subsection{Mesh sensitivity: shock-aligned grid generation} \label{sec:shocks}

The proposed adaptive regularization approach is a general procedure that is able to provide the optimal artificial viscosity for any kind of mesh.
Nevertheless, it is important to remark that the artificial dissipation field depends on the grid $\mathcal{T}_h$.
Meshes that are aligned and refined along the shock allow for  a significant reduction of the artificial viscosity, which leads to sharper and more accurate approximations.
In this study, we will compare the performance of the viscosity regularization procedure for both uniform and shock-aligned grids.
In particular, we introduce a mesh adaptation procedure that identifies the shock location based on the solution on the uniform grid and constructs a new mesh that is aligned and refined along the approximated shock location, as described throughout this section.

\revise{The process can be iterated successively until the amount of artificial viscosity is sufficiently small, leading to a robust shock-alignment strategy that renders accurate approximation of the original system of conservation laws. Nevertheless, since the aim of this work focuses on introducing an adaptive regularization approach that can be used with any numerical method, the shock-aligned mesh generation employed in the numerical examples will consist of a single shock adaptation iteration.}


\subsubsection{Identifying the shock location}
\begin{figure}[h]
	\centering
	\begin{subfigure}[b]{0.49\textwidth}
		\centering		\includegraphics[width=\textwidth]{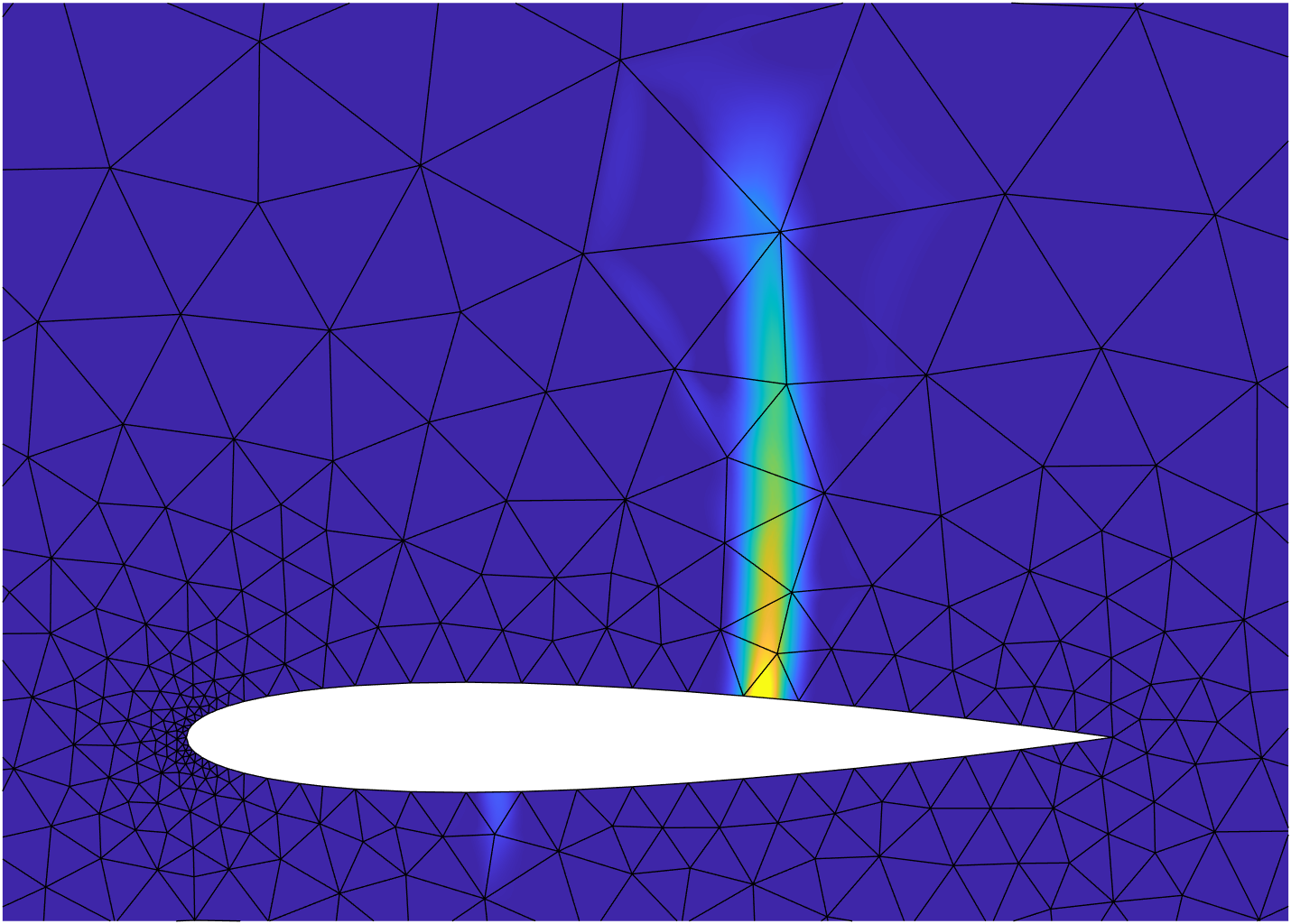}
        \caption{Artificial viscosity}
	\end{subfigure}
	\hfill
	\begin{subfigure}[b]{0.49\textwidth}
		\centering
		\includegraphics[width=\textwidth]{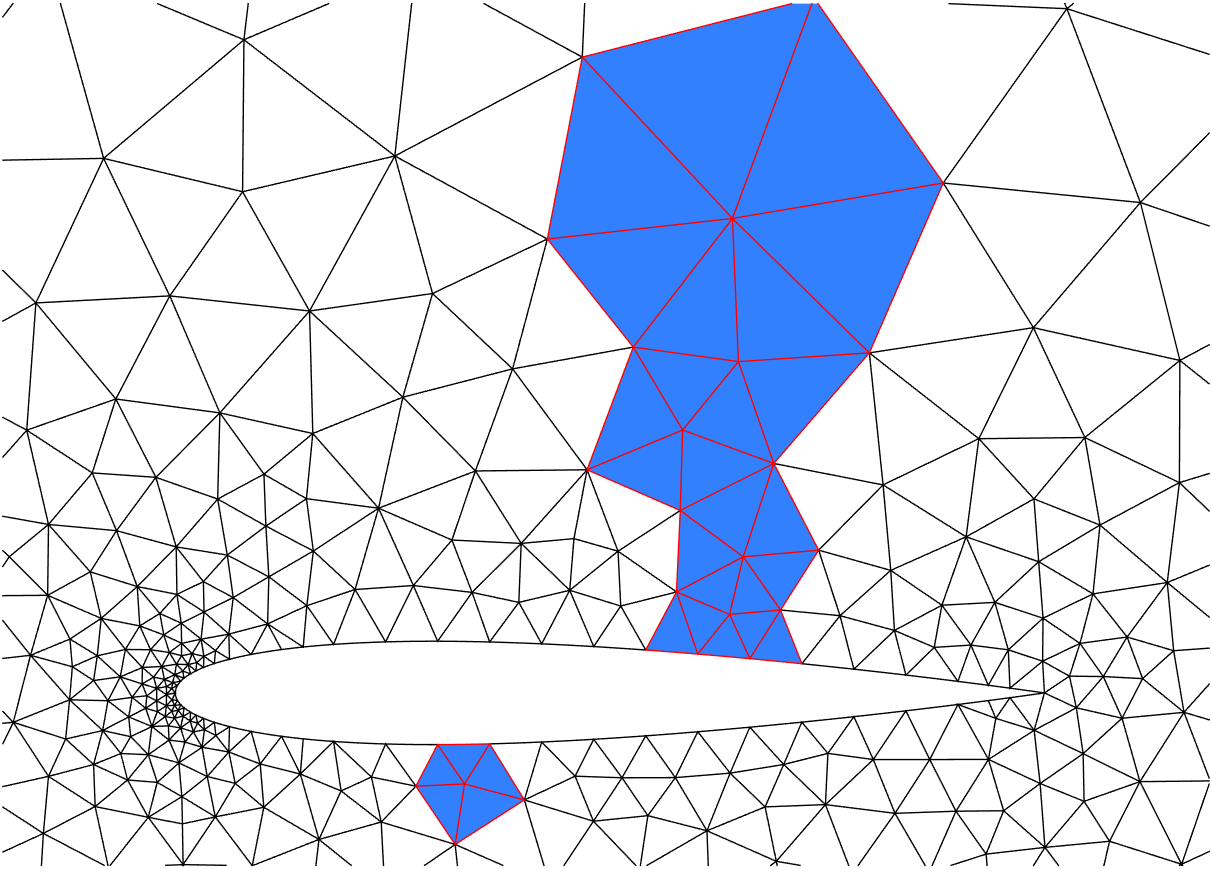}
        \caption{Elements in the shock region}
	\end{subfigure} 
 	\begin{subfigure}[b]{0.49\textwidth}
		\centering		\includegraphics[width=\textwidth]{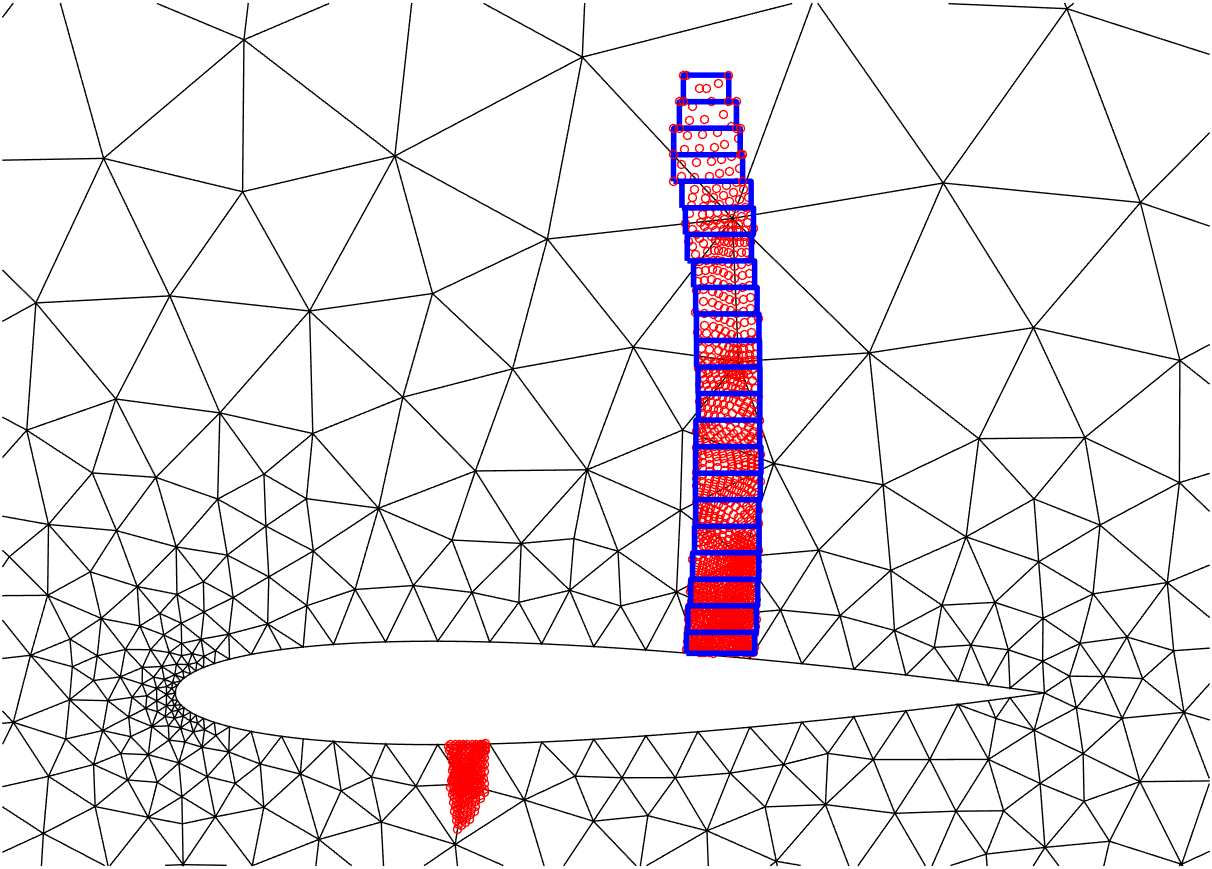}
		\caption{Selected points and associated bounding boxes}
	\end{subfigure}
	\hfill
	\begin{subfigure}[b]{0.49\textwidth}
		\centering
		\includegraphics[width=\textwidth]{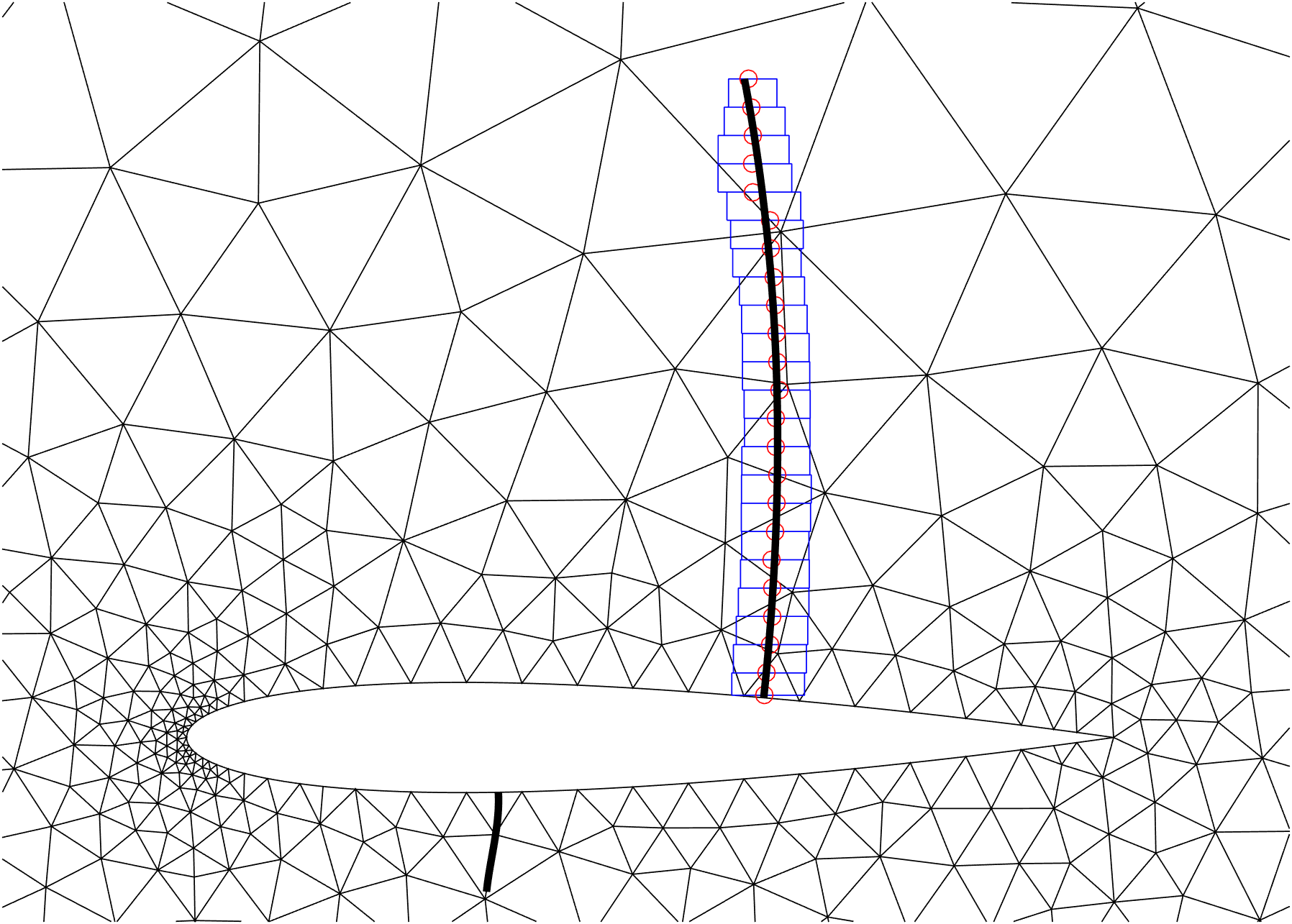}
		\caption{Shocks are determined by the set of faces.}
	\end{subfigure} 
	\caption{\label{fignaca3b} Identification of shock locations for the inviscid transonic flow past a NACA 0012 airfoil.}
\end{figure}

A shock can be represented by a curve in two dimensions or a surface in three dimensions. There can be several shocks that exist in the physical domain. We propose a method to construct curves or surfaces  to represent possible shocks in the physical domain. The idea is to use the normalized shock indicator $\bar{\eta}(\bm x)$ to find a set of faces and construct shock curves/surfaces from those faces. The method can be described as follows:

\revise{
\begin{itemize}
    \item Find the shock region $\mathcal{T}_h^{\rm shock} = \{K \in \mathcal{T}_h \ : \ \int_K  \bar{\eta}(\bm x) d \bm x  \ge \bar{\eta}_{\rm T}  |K| \}$, as illustrated in Figure \ref{fignaca3b}(b).
    \item Map a set of points on the reference element $K_{\rm ref}$ to a set of points on a physical  element $K$ for all $K \in \mathcal{T}_h^{\rm shock}$. The total number of points in the shock region is equal to $N_{\rm s} \times N_{\rm p}$, where $N_{\rm s}$ is the number of elements in the shock region and $N_{\rm p}$ is the number of points on the reference element. 
    \item Select only the points in the shock region  that satisfy $\bar{\eta}(\bm x) \ge 2 \bar{\eta}_{\rm T}$, as illustrated in  Figure \ref{fignaca3b}(c).
    \item Determine $M$ bounding boxes $\{\mathcal{B}_m\}_{m=1}^M$ that contain the selected points. $M$ should be chosen large enough to map the shock curves/surfaces accurately,  as illustrated in  Figure \ref{fignaca3b}(c).
    \item For every edge $L_{m j}$ of each bounding box $\mathcal{B}_m$ along the freestream velocity direction, find  $\bm x_{mj} = \arg \max_{\bm x \in L_{m j}} \bar{\eta}(\bm x)$ for $j = 1, \ldots 2^{d-1}$. Each bounding box $\mathcal{B}_m$ contains two points $\bm x_{m1},\bm x_{m2}$ in two dimensions, or four points $\bm x_{m1}$, $\bm x_{m2}$, $\bm x_{m3}$, $\bm x_{m4}$ in three dimensions. Let $\mathcal{F}_m$ be a face formed by $\{\bm x_{mj}\}_{j=1}^{2^{d-1}}$ for each bounding box $\mathcal{B}_m$.  See  Figure \ref{fignaca3b}(d).
    \item The shock location is determined by the set of $M$ faces $\{\mathcal{F}_m\}_{m=1}^M$. This set may contain a number of disjoint subsets of connected faces. Each disjoint subset of connected faces represents one shock curve/surface. Hence, the number of shocks in the physical domain is equal to the number of disjoint subsets of  connected faces.  See Figure \ref{fignaca3b}(d).    
\end{itemize}
}

The shock identification procedure is illustrated in Figure \ref{fignaca3b} for  inviscid transonic flows past NACA 0012 airfoil, where there are an upper (strong) shock and a lower (weak) shock. As a result, the procedure yields two disjoint subsets of connected faces.


\subsubsection{Generating shock-aligned meshes}
\begin{figure}[h]
	\centering
	\begin{subfigure}[b]{0.49\textwidth}
		\centering		\includegraphics[width=\textwidth]{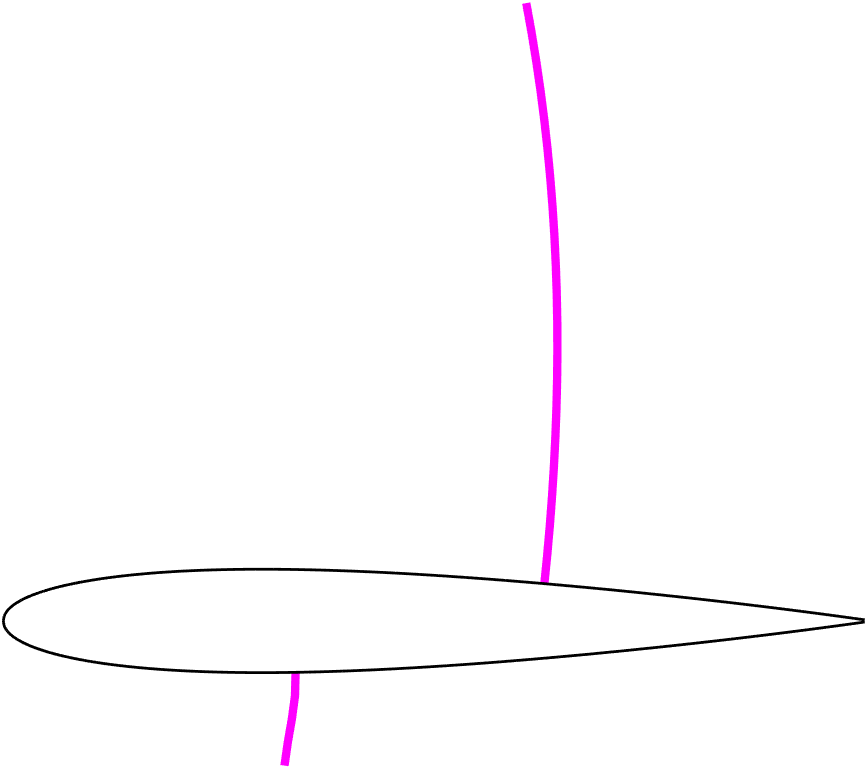}
  \caption{Shock curves}
	\end{subfigure}
	\hfill
	\begin{subfigure}[b]{0.49\textwidth}
		\centering
		\includegraphics[width=\textwidth]{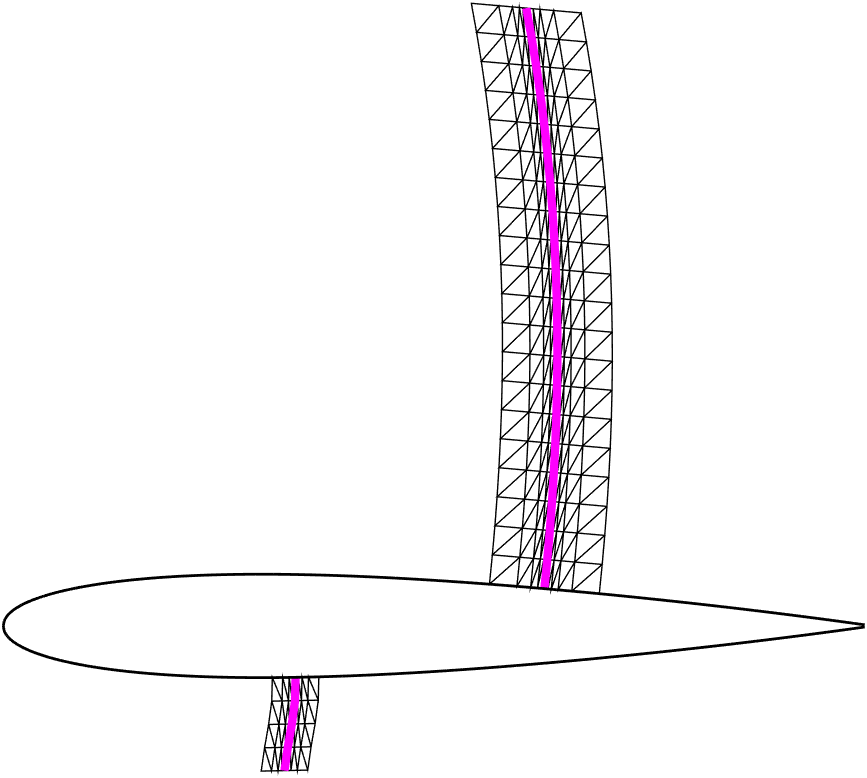}
  \caption{Meshes in the shock region}
	\end{subfigure} 
 	\begin{subfigure}[b]{0.49\textwidth}
		\centering		\includegraphics[width=\textwidth]{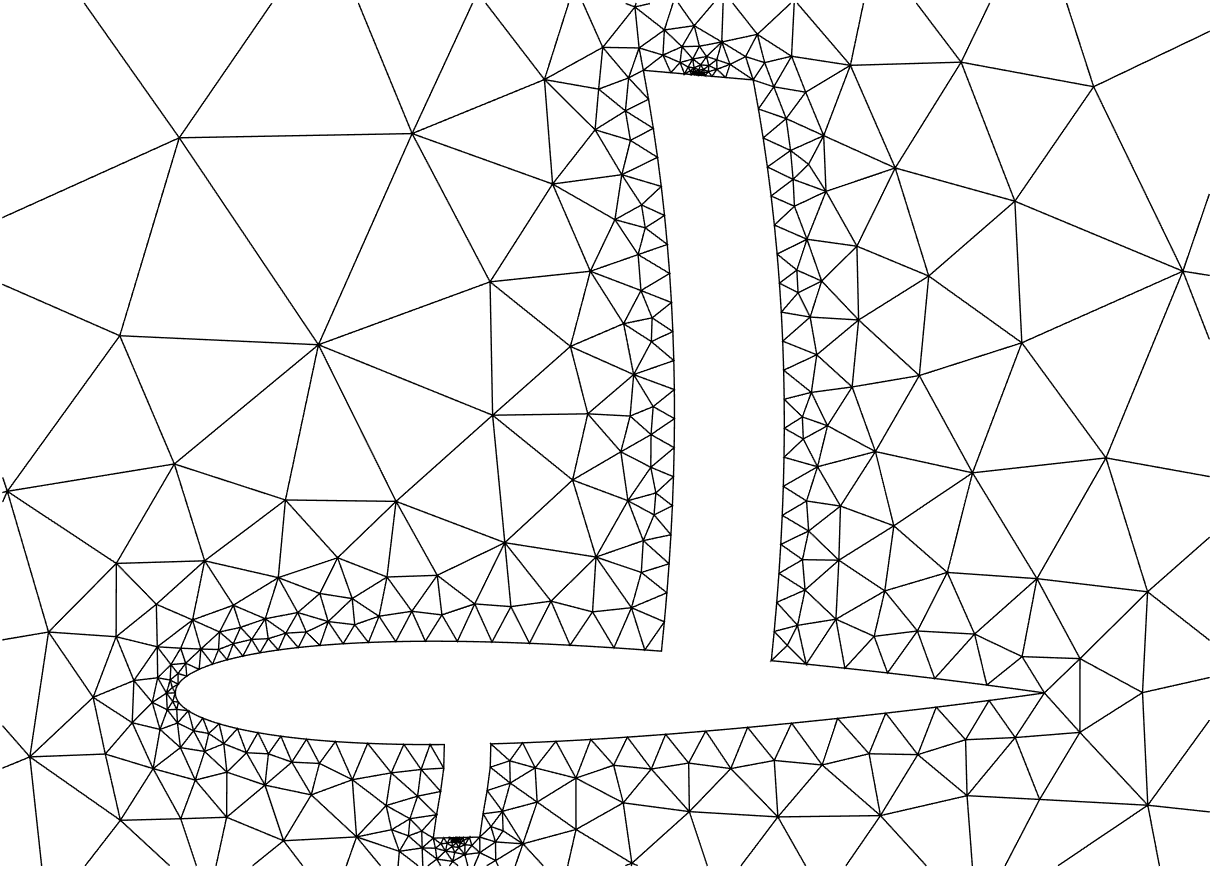}
		\caption{Mesh in the smooth region}
	\end{subfigure}
	\hfill
	\begin{subfigure}[b]{0.49\textwidth}
		\centering
		\includegraphics[width=\textwidth]{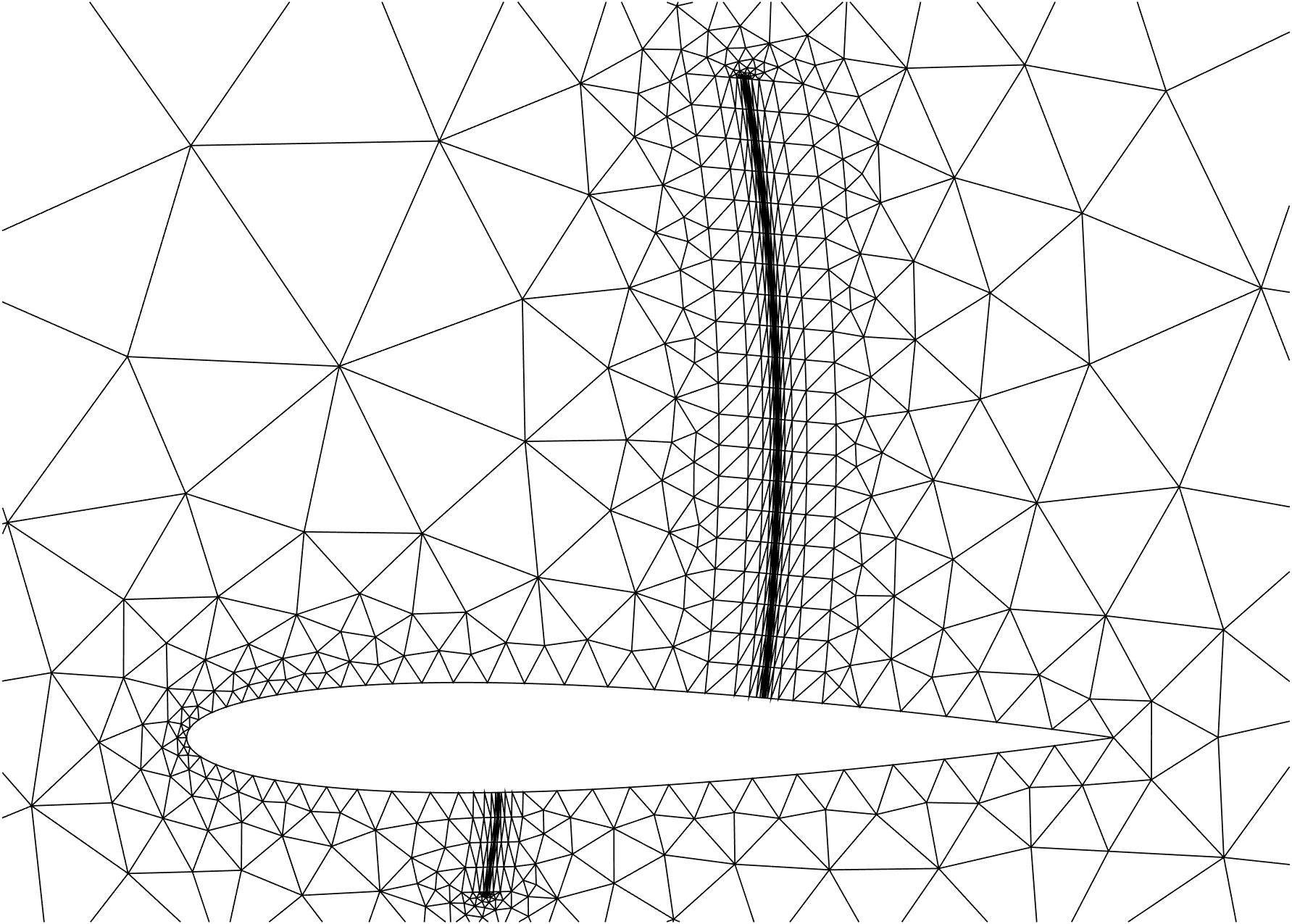}
		\caption{Shock-aligned mesh on the whole domain}
	\end{subfigure} 
	\caption{\label{fignaca4b} Generation of a shock-aligned mesh for the NACA 0012 example.}
\end{figure}

Having identified the shock location, we generate shock-aligned meshes to substantially reduce the regularization parameter and improve the numerical solution in the shock region. The mesh generation method can be described as follows: 

\revise{
\begin{itemize}
    \item  Polynomial interpolation is used to fit each disjoint subset of connected faces to  obtain smooth shock curves, as illustrated in Figure \ref{fignaca4b}(a).
    \item Next, create a grid of $N$ nodes and $M$ faces for the shock curve. For high-order methods, the grid should be made high-order to represent the shock curve accurately as illustrated in Figure \ref{fignaca4b}(b).
    \item Extrude each node on the shock curve along a specified direction to create a number of new nodes. Typically, the direction is the normal vector at the node. However, if the shock curve intersects with the domain boundary, different directions can be chosen to make the mesh conform to the domain boundary. Furthermore, the directions behind the shock curve can also be different from the directions before the shock curve.   See Figure \ref{fignaca4b}(b).
    \item Connect the nodes on the shock curve and the newly created nodes to form a shock-aligned mesh based on the grid of the shock curve. 
    \item Repeat the above steps for the remaining disjoint subsets of connected faces to generate their  shock-aligned meshes, as shown in Figure \ref{fignaca4b}(b).
    \item Generate a mesh for the remaining region of the physical domain, as illustrated in Figure \ref{fignaca4b}(c).
    \item Finally, connect these meshes to obtain a full shock-aligned mesh for the entire physical domain, as illustrated in Figure \ref{fignaca4b}(d).
\end{itemize}
}

The mesh generation is illustrated in Figure \ref{fignaca4b}. We use Gmsh \cite{Geuzaine2009} to generate meshes for the smooth region of the physical domain. This is done by collecting the mesh points on the boundary of the shock-aligned meshes together with the points on the boundary of the physical domain to define a geometry description of the smooth region.

\section{Numerical Results} \label{sec:results}

In this section, we present numerical results for a number of inviscid steady-state problems to demonstrate the adaptive viscosity regularization approach. The initial artificial viscosity is set to $\eta_0(\bm x) = \tanh(20 d(\bm x))$, where $d(\bm x)$ is a distance from $\bm x \in \Omega$ to the wall boundary $\Gamma_{\rm wall}$. Hence, $\eta_0$ is equal to 1 on most of the physical domain and vanishes smoothly to zero at the wall boundary. This choice is made because adding viscosity on the entire wall boundary can affect the wall boundary conditions, which may negatively impact the accuracy and convergence of the numerical solution near the wall. The initial regularization parameters $\bm \lambda_0$ are chosen large enough so that the initial numerical solution $\bm u_0$ is smooth, but not necessarily accurate. Unless otherwise specified, the homotopy parameter $\zeta $ is set to $0.8$, the artificial viscosity threshold $\bar{\eta}_{\rm T}$ is set to 0.2, and polynomial degree $k = 4$ is used to represent the approximate solutions. 

\subsection{Inviscid Burgers' equation}

We consider the two-dimensional space-time inviscid Burgers' equation \cite{Zahr2020}:
\begin{equation}
\frac{\partial u}{\partial t} + \frac{1}{2} \frac{\partial u^2}{\partial x} = 0  \quad \mbox{in } \Omega \equiv  (-1, 1) \times (0,1),  
\end{equation}
\revise{with boundary condition $u(x,t) = 2(x+1)^2(1 - H(x))$ on $\Gamma_D \equiv \{x=-1\} \cup \{t=0\}$, where $H(x)$ is the Heaviside function. The exact solution is found by the method of characteristics as follows
\begin{equation}
u(x,t) = \left\{
\begin{array}{cc}
  U(x,t)   &  x < x_s(t) \\
   0  & x > x_s(t) 
\end{array}
\right.
\end{equation}
where $U(x,t)$ is obtained by solving the quadratic equation $U = 2(x - U t + 1)^2$. The exact shock location $x_s(t)$ is the solution of the following ordinary differential equation
\begin{equation}
\frac{d x_s}{d t} = \frac{U(x_s,t)}{2} , \quad x_s(0) = 0 ,   
\end{equation}
which stems from the Rankine-Hugoniot condition at the shock.} We use an initial uniform mesh of $288$ $k=4$ elements to obtain an approximate solution which is shown in Figure \ref{figbg2}(a). 
This solution is compared to the numerical approximation on a shock-aligned mesh of 523 $k=4$ elements which is generated by means of the procedure described in Section~\ref{sec:shocks}.

\begin{figure}[htbp]
\centering
	\centering
	\begin{subfigure}[b]{0.42\textwidth}
		\centering		\includegraphics[width=\textwidth]{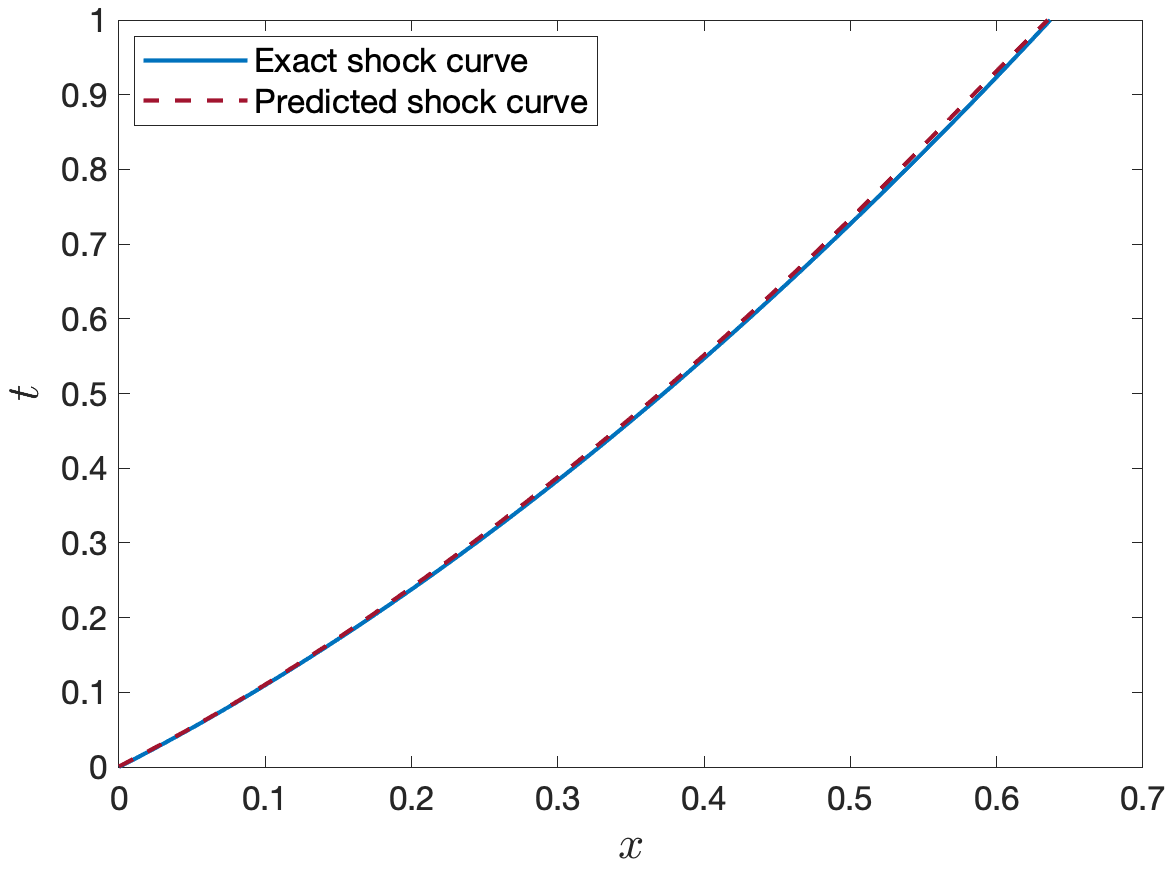}
		\caption{Shock location}
	\end{subfigure}
	\hfill
	\begin{subfigure}[b]{0.57\textwidth}
		\centering
		\includegraphics[width=\textwidth]{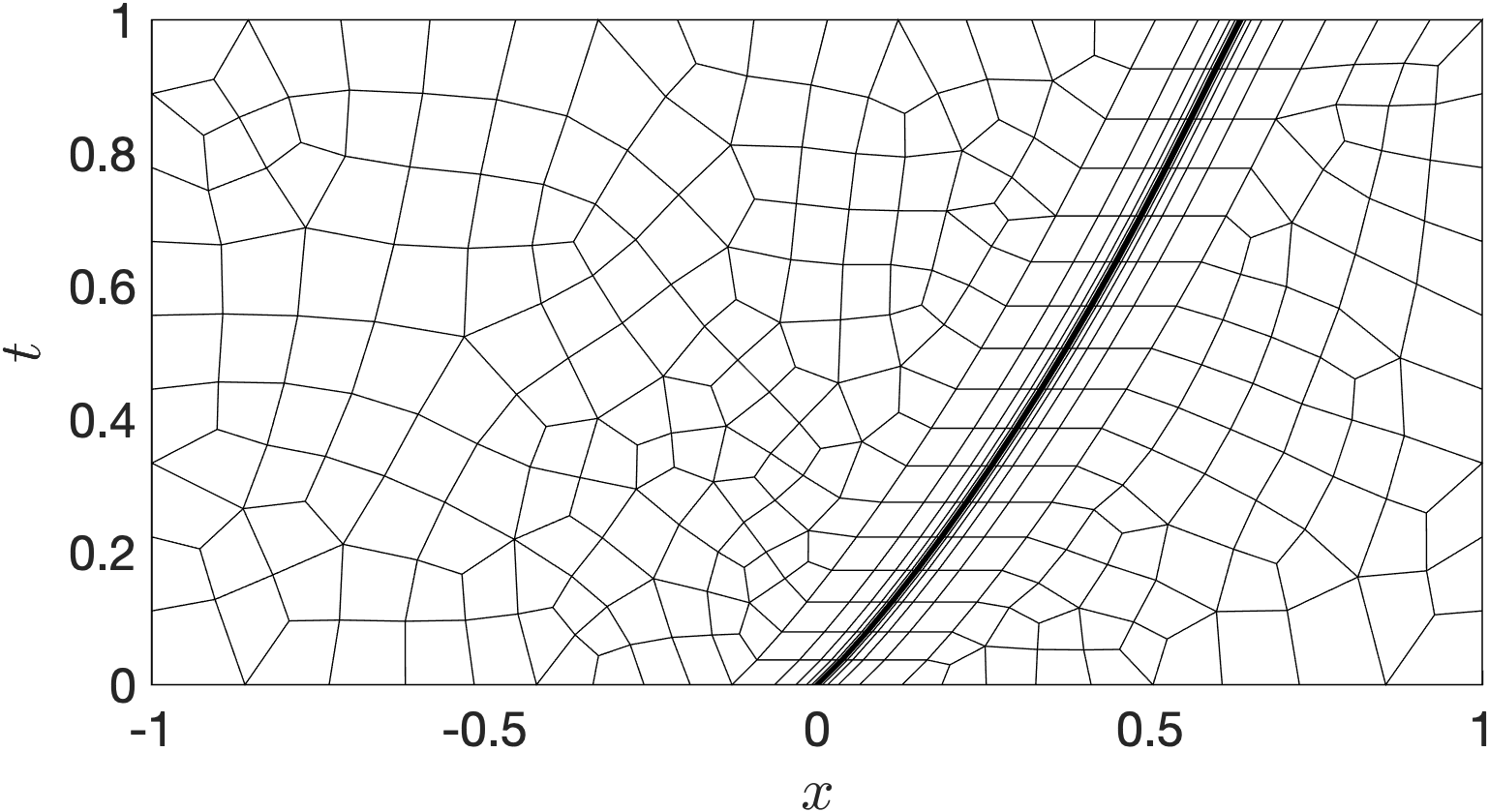}
		\caption{Shock-aligned mesh of 523 $k=4$ elements}
	\end{subfigure} 
\caption {\label{figbg1} Shock location and shock-aligned mesh for the inviscid Burgers' equation.}
\end{figure}

\begin{figure}[htbp]
	\centering
	\begin{subfigure}[b]{0.49\textwidth}
		\centering		\includegraphics[width=\textwidth]{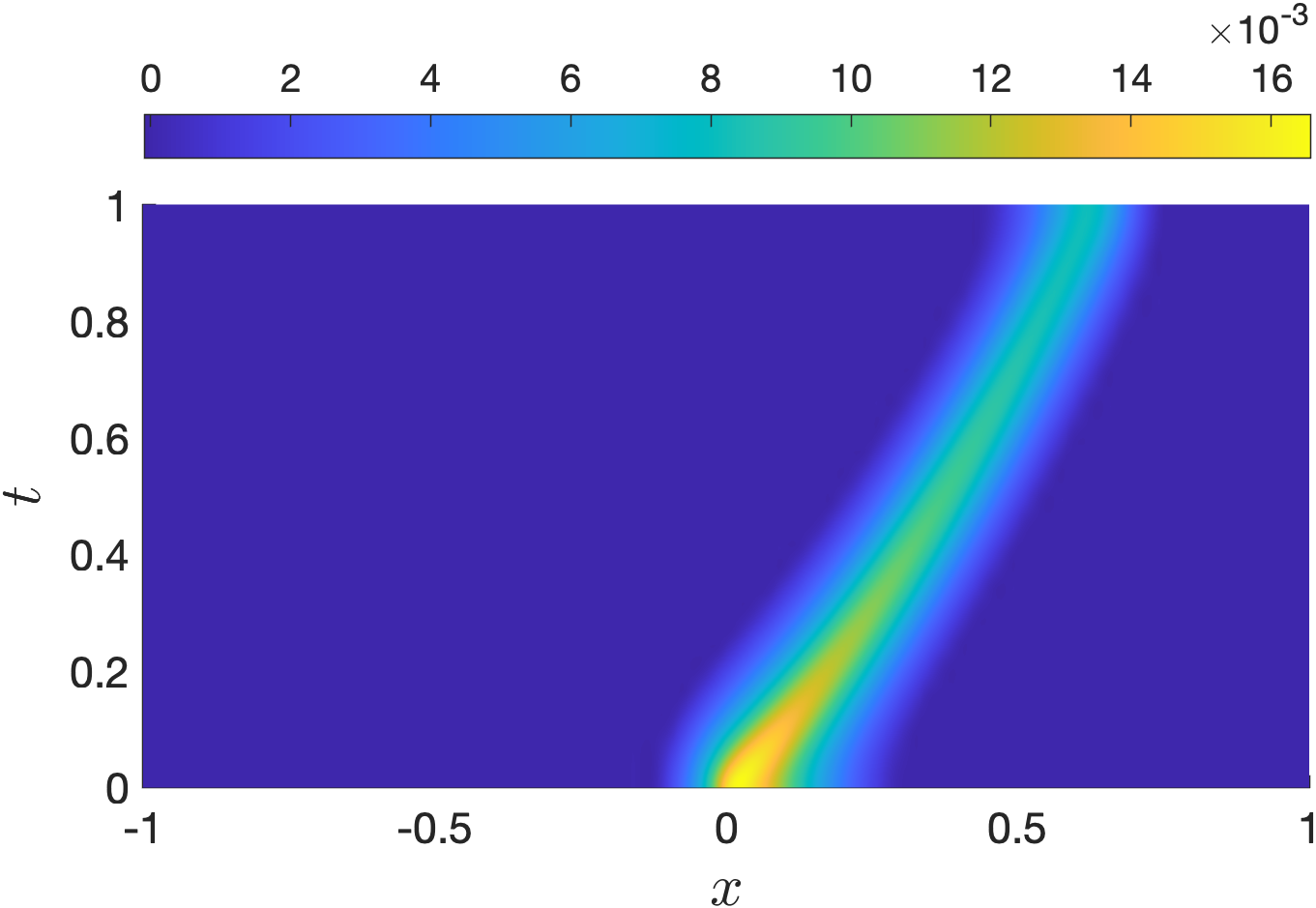}
	\end{subfigure}
	\hfill
	\begin{subfigure}[b]{0.49\textwidth}
		\centering
		\includegraphics[width=\textwidth]{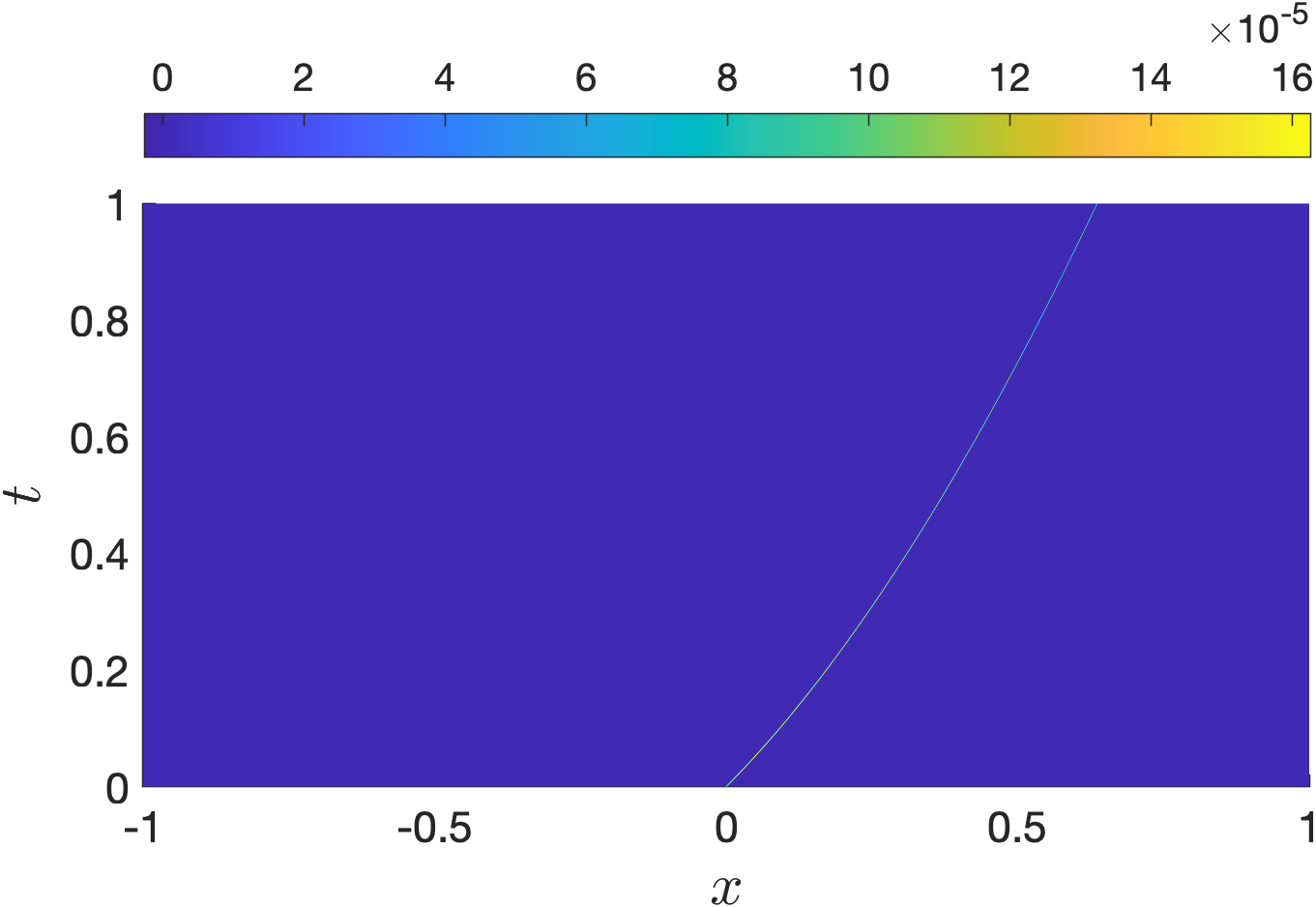}
	\end{subfigure} 
 	\begin{subfigure}[b]{0.49\textwidth}
		\centering		\includegraphics[width=\textwidth]{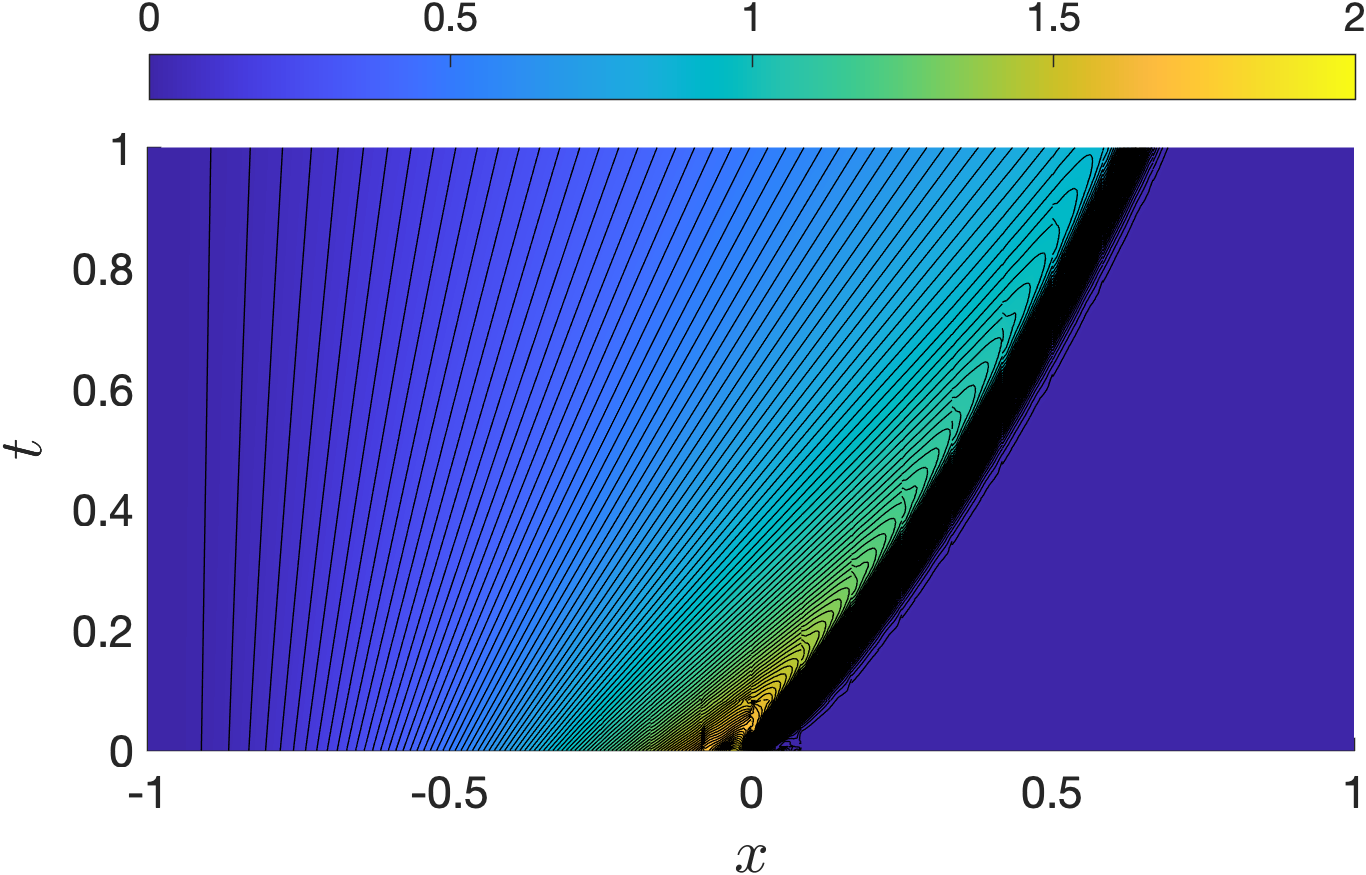}
		\caption{Regular mesh}
	\end{subfigure}
	\hfill
	\begin{subfigure}[b]{0.49\textwidth}
		\centering
		\includegraphics[width=\textwidth]{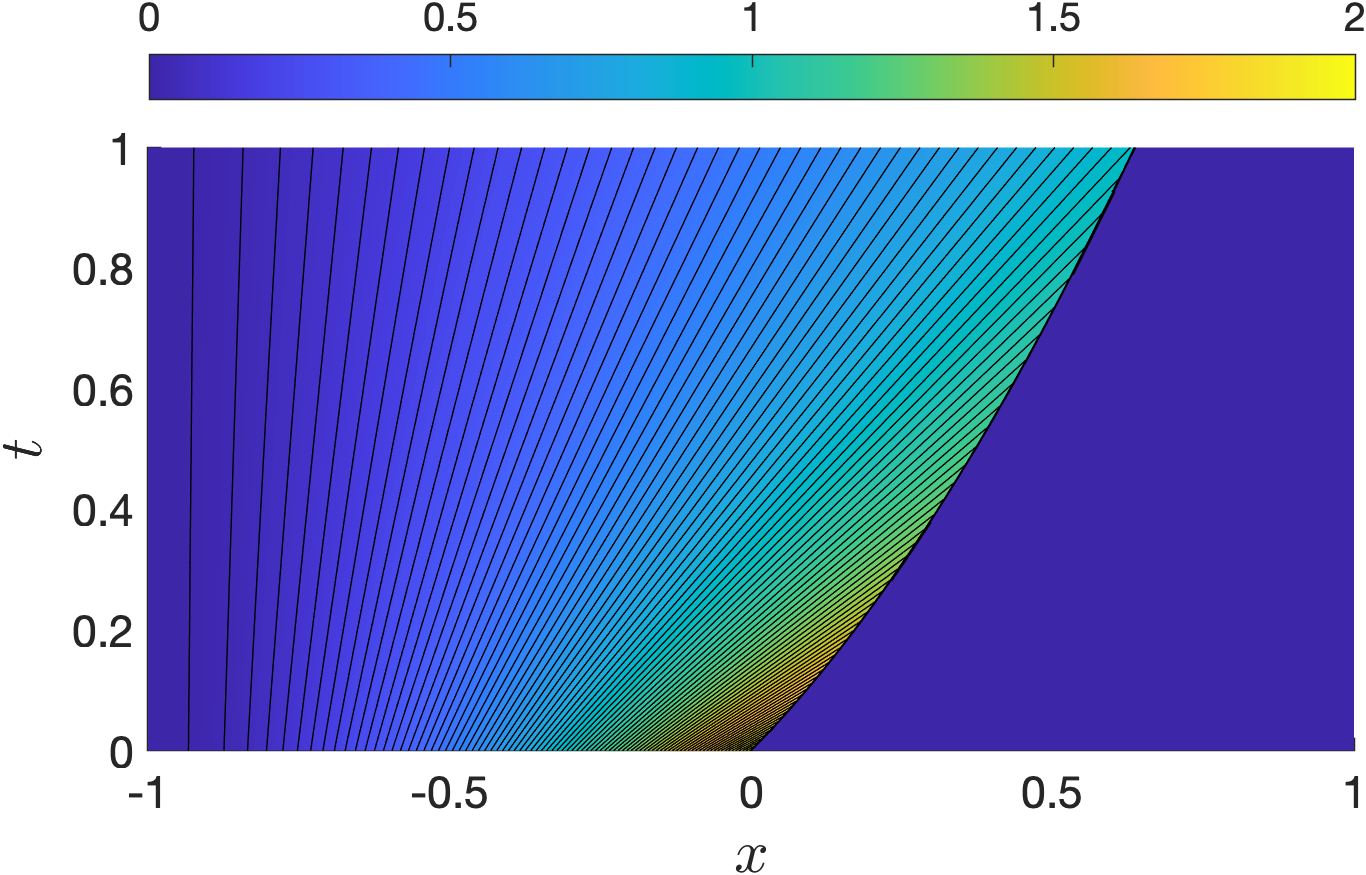}
		\caption{Shock-aligned mesh}
	\end{subfigure} 
	\caption{\label{figbg2} Artificial viscosity (top) and approximate  solution (bottom) for the inviscid Burgers' equation.} 
\end{figure}

\begin{figure}[htbp]
\centering
	\centering
	\begin{subfigure}[b]{0.51\textwidth}
		\centering		\includegraphics[width=\textwidth]{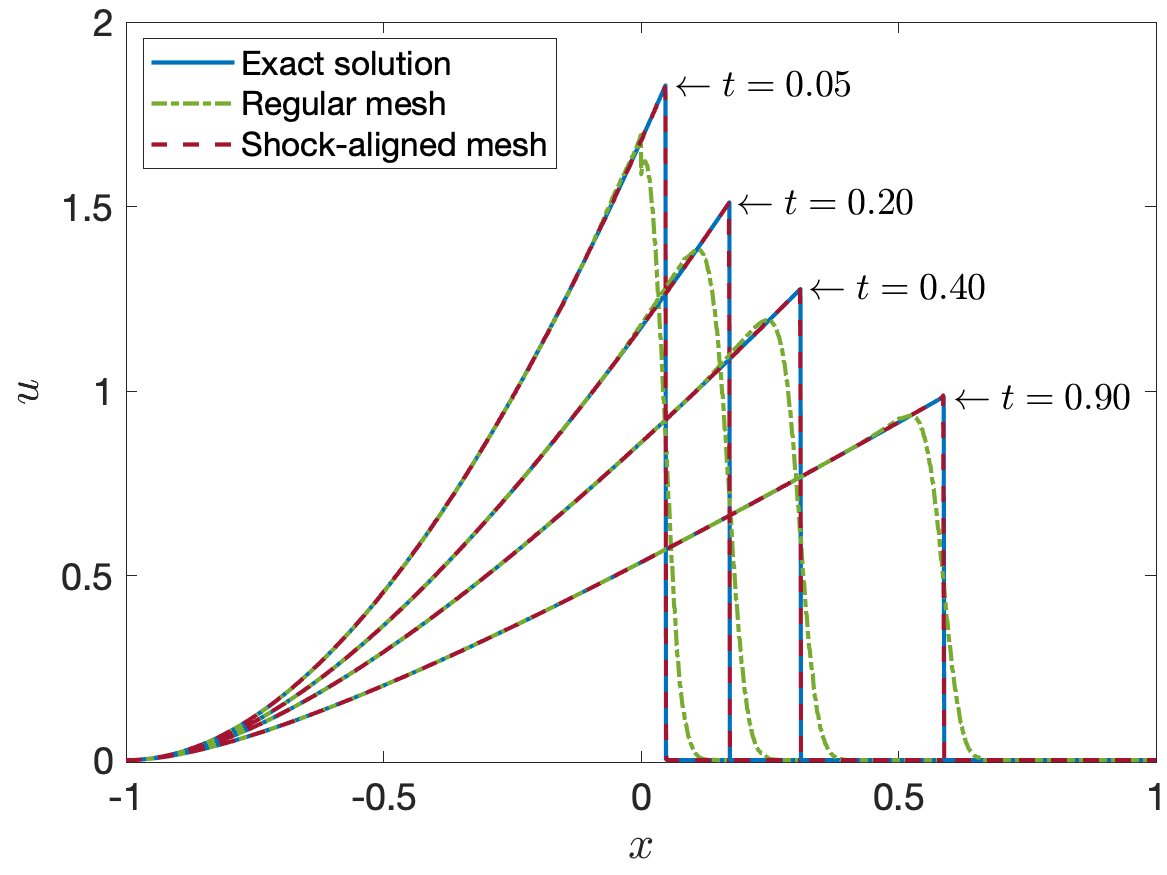}
		\caption{}
	\end{subfigure}
	\hfill
	\begin{subfigure}[b]{0.48\textwidth}
		\centering
		\includegraphics[width=\textwidth]{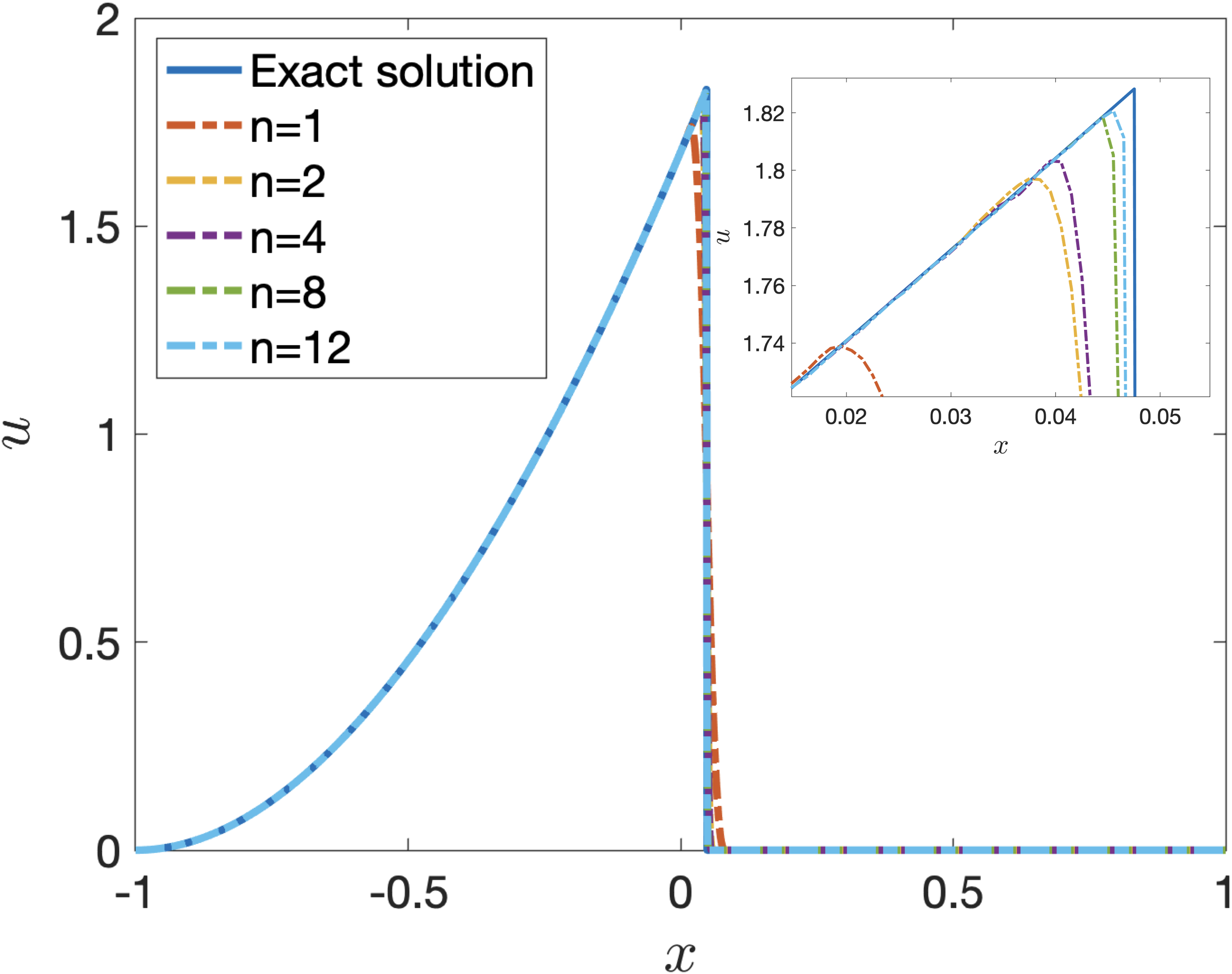}
		\caption{}
	\end{subfigure} 
\caption {\label{figbg3} Comparison of the approximate solution and the exact solution for the inviscid Burgers' equation. (a) Profiles of the exact and approximate solutions at different times. (b) Profiles of the approximate solution for different homotopy iterations $n$ at $t=0.05$.}
\end{figure}

The resulting shock location and shock-aligned mesh are shown in Figure \ref{figbg1}. We observe that the shock location, which is identified based on the approximate solution on the uniform mesh, accurately approximates the exact shock location. Figure \ref{figbg2}(b) shows the artificial viscosity and the approximate solution on the shock-aligned mesh. We see that the artificial viscosity on the shock-aligned mesh is about 100 times smaller than that on the regular mesh. The approximate solution on the shock-aligned mesh is  sharper and more accurate than that on the regular mesh. This can also be seen in Figure \ref{figbg3}, which displays both the exact and approximate solutions. We see that the approximate solution on the shock-aligned mesh is almost indistinguishable from the exact solution.

\begin{figure}[htbp]
\centering
	\centering
	\begin{subfigure}[b]{0.51\textwidth}
		\centering		\includegraphics[width=\textwidth]{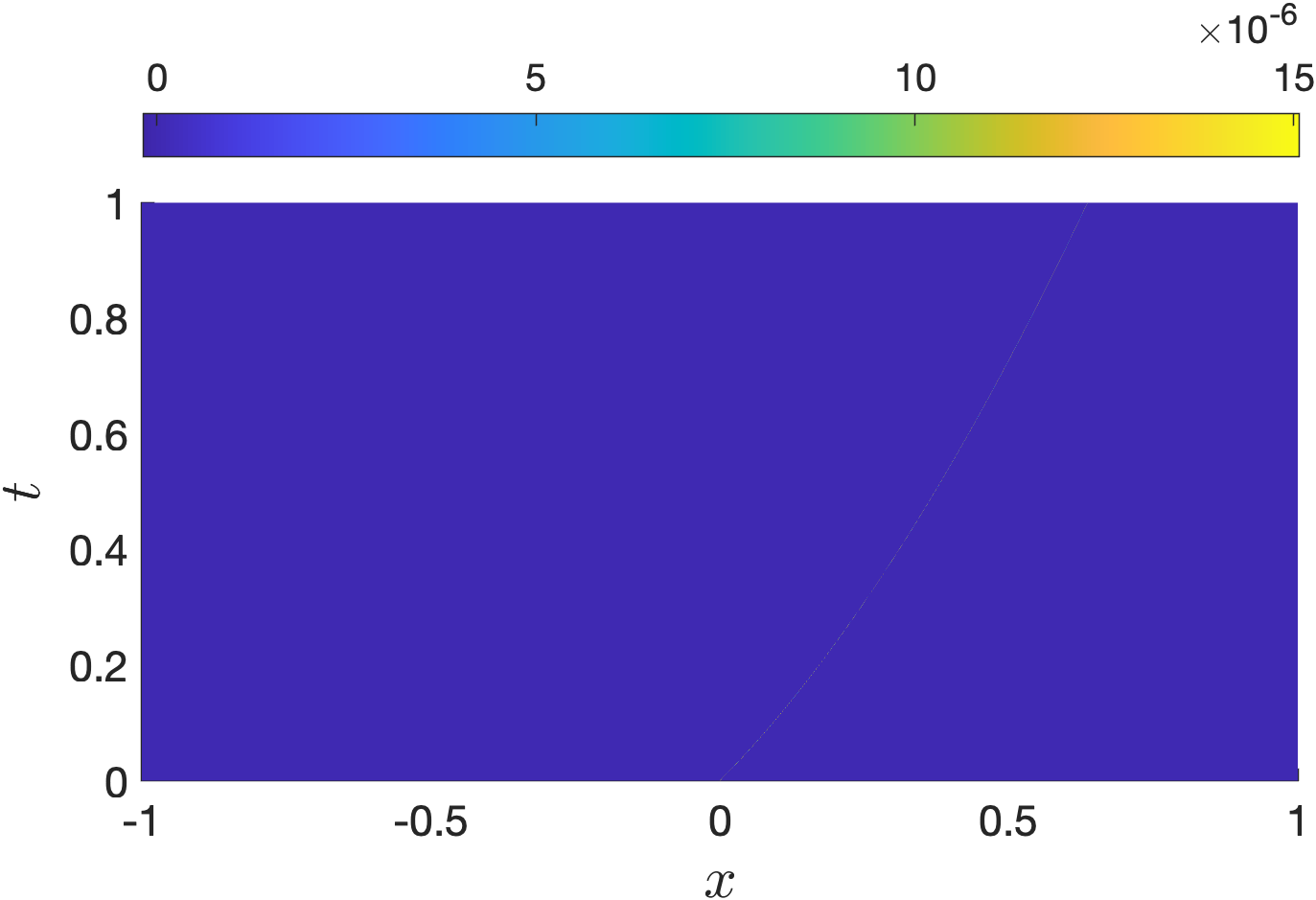}
		\caption{Artificial viscosity}
	\end{subfigure}
	\hfill
	\begin{subfigure}[b]{0.48\textwidth}
		\centering
		\includegraphics[width=\textwidth]{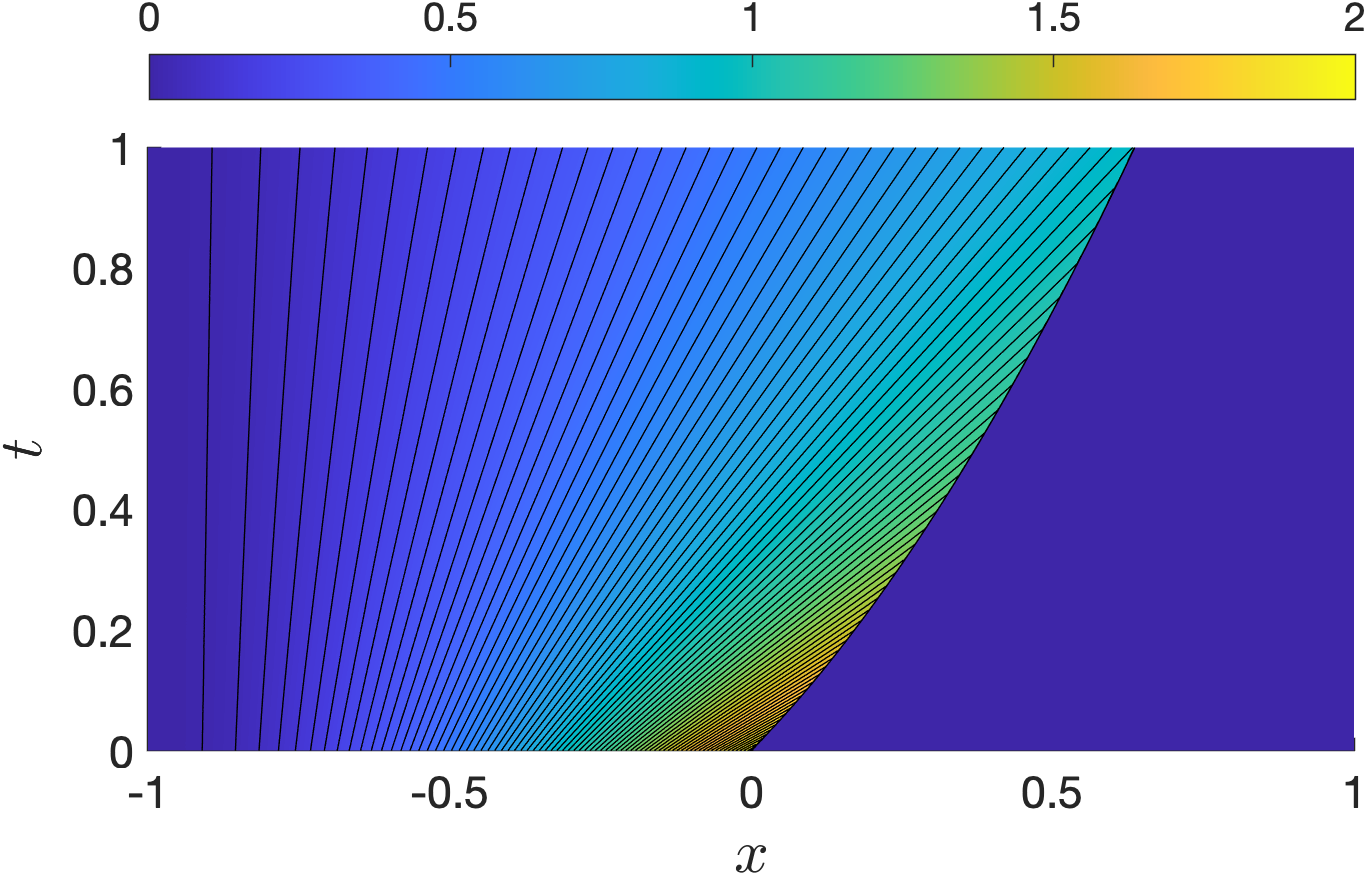}
		\caption{Approximate solution}
	\end{subfigure} 
\caption {\label{figbg3w} Numerical solution computed on the second shock-aligned mesh.}
\end{figure}

\revise{Next, we demonstrate that the artificial viscosity can be further reduced by using   the numerical solution on the first shock-aligned mesh to generate a second shock-aligned mesh. We show in Figure \ref{figbg3w} the numerical solution computed on the second shock-aligned mesh. We see that the artificial viscosity on the second shock aligned mesh is about 10 times smaller than that on the first shock-aligned mesh. Hence, the approximate solution on the second shock aligned mesh should be more accurate than that on the first shock-aligned mesh.}

\begin{figure}[h]
\centering
	\centering
	\begin{subfigure}[b]{0.49\textwidth}
		\centering		\includegraphics[width=\textwidth]{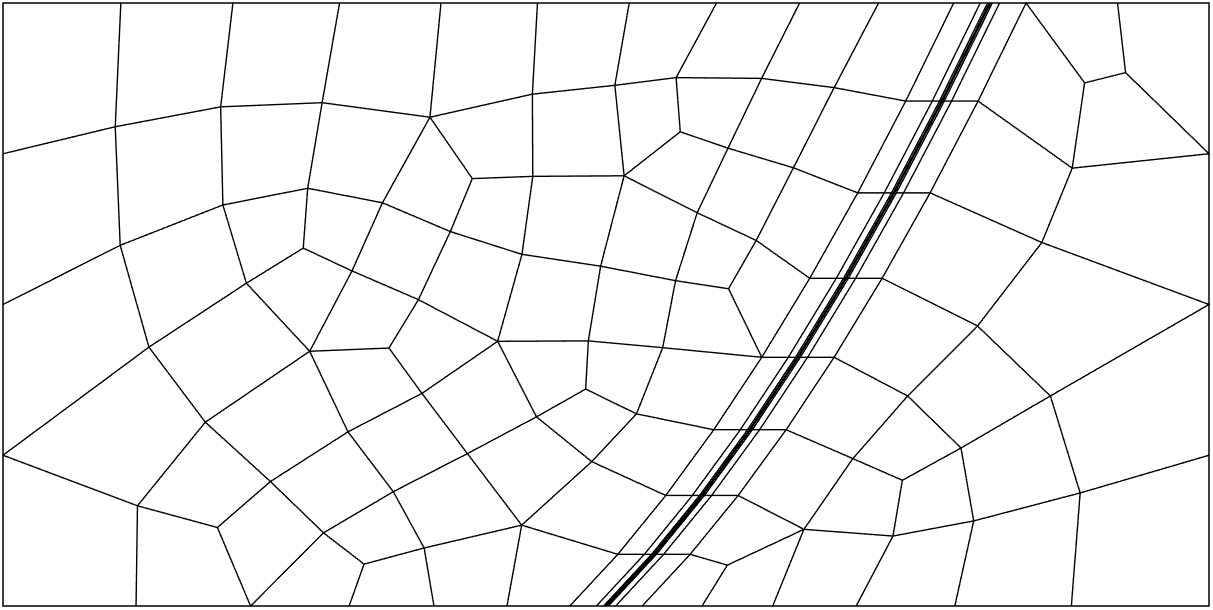}
		\caption{Initial mesh}
	\end{subfigure}
	\hfill
	\begin{subfigure}[b]{0.49\textwidth}
		\centering
		\includegraphics[width=\textwidth]{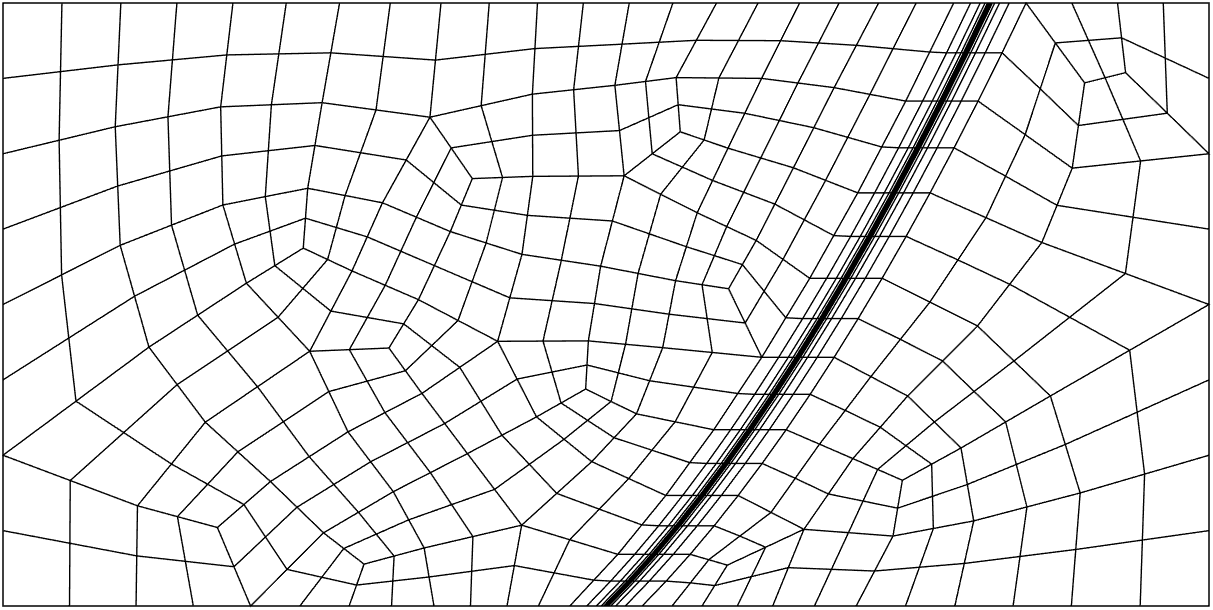}
		\caption{First refined mesh}
	\end{subfigure} 
\caption {\label{figbg1w} Meshes used to perform the convergence study of the numerical solution for the inviscid Burgers' equation.}
\end{figure}

\revise{It is interesting to study the convergence rate of the numerical solution in the smooth region. To this end, we consider an intial shock-aligned mesh shown in Figure \ref{figbg1w} and successively refine this mesh by subdividing each element into 4 smaller elements. We define the error as 
\begin{equation}
\mbox{error}^2 =  \int_{\Omega_{\rm smooth}} (u - u_n)^2 d \bm x    
\end{equation}
where $\Omega_{\rm smooth} =  \{K \in \mathcal{T}_h \ : \ \int_K  \bar{\eta}(\bm x) d \bm x  < \bar{\eta}_{\rm T}  |K| \}$ is the smooth region associated with the numerical solution on the initial mesh and polynomial degree $k=1$. Table \ref{tabEx1a} shows the errors for $k=1, 2, 3$ at different refinement levels. We see that the numerical solution in the smooth region converges optimally with order $k+1$.
}

\begin{table}[thbp]
  \begin{center}
\scalebox{0.9}{%
    $\begin{array}{|c||c c  | c c|  c c|  c c|}
    \hline
   \mbox{Refinement} & \multicolumn{2}{|c}{k=1} &  \multicolumn{2}{|c}{k=2} & \multicolumn{2}{|c|}{k=3} \\
     \mbox{level} & \mbox{error}  & \mbox{order} & \mbox{error}  & \mbox{order} & \mbox{error}  & \mbox{order}  \\
    \hline
  0  &  6.04\mbox{e-}3  &  --  &  4.93\mbox{e-}4  &  --  &  4.74\mbox{e-}5  &  --  \\  
  1  &  1.68\mbox{e-}3  &  1.85  &  7.41\mbox{e-}5  &  2.73  &  4.30\mbox{e-}6  &  3.46  \\  
  2  &  4.65\mbox{e-}4  &  1.85  &  1.06\mbox{e-}5  &  2.80  &  3.50\mbox{e-}7  &  3.62  \\  
  3  &  1.26\mbox{e-}4  &  1.89  &  1.45\mbox{e-}6  &  2.87  &  2.35\mbox{e-}8  &  3.90  \\  
  4  &  3.32\mbox{e-}5  &  1.93  &  1.90\mbox{e-}7  &  2.94  &  1.53\mbox{e-}9  &  3.95  \\  
\hline
     \end{array} $
}
\end{center}{$\phantom{|}$}
     \caption{\revise{Convergence rates of the numerical solution in the smooth region for the inviscid Burgers' equation.}}
   \label{tabEx1a}
\end{table}

\subsection{Inviscid transonic flow past NACA 0012 airfoil}

\begin{figure}[htbp]
\centering
	\centering
	\begin{subfigure}[b]{0.49\textwidth}
		\centering		\includegraphics[width=\textwidth]{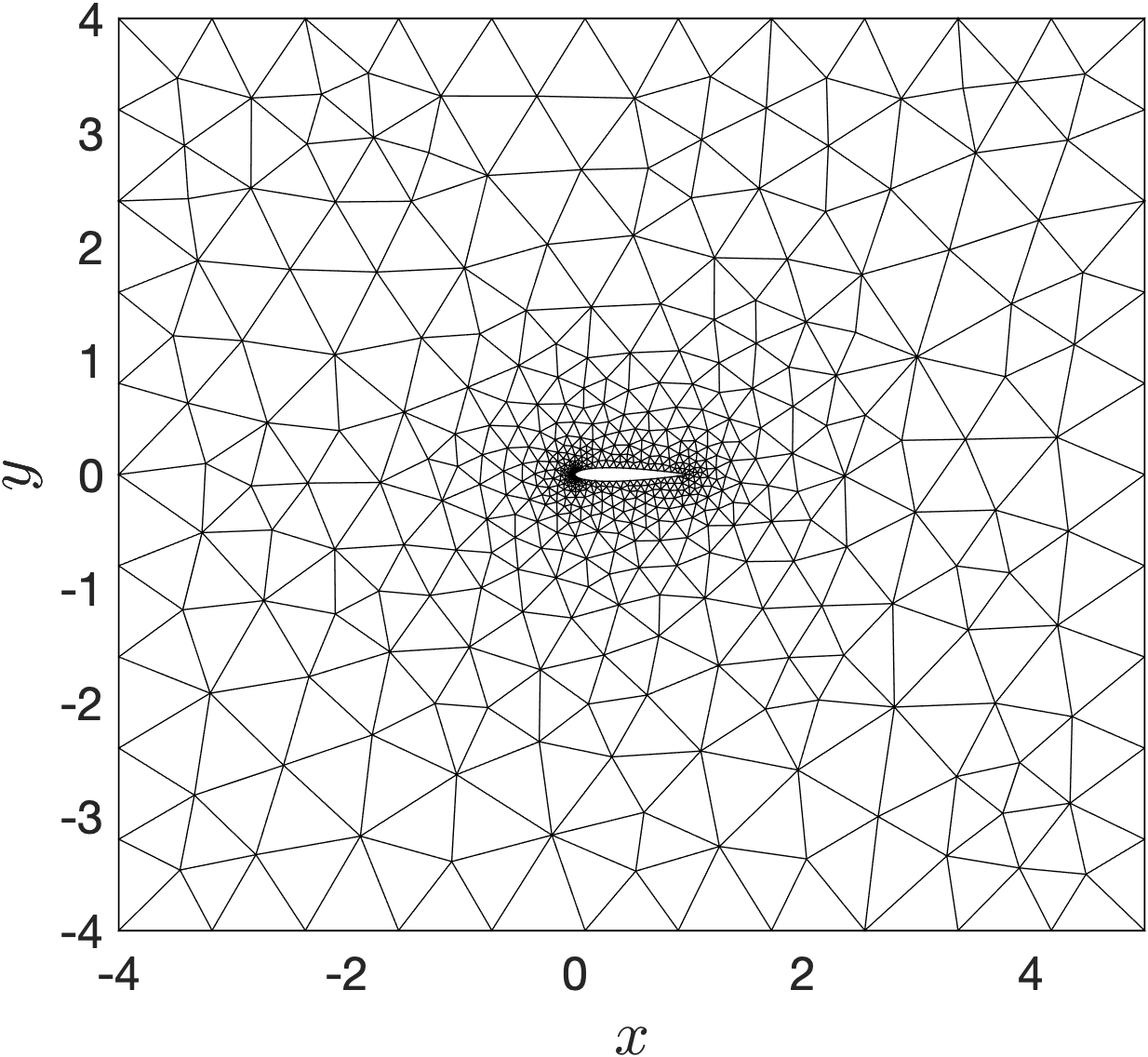}
		\caption{Regular mesh of 1082 $k=4$ elements}
	\end{subfigure}
	\hfill
	\begin{subfigure}[b]{0.49\textwidth}
		\centering
		\includegraphics[width=\textwidth]{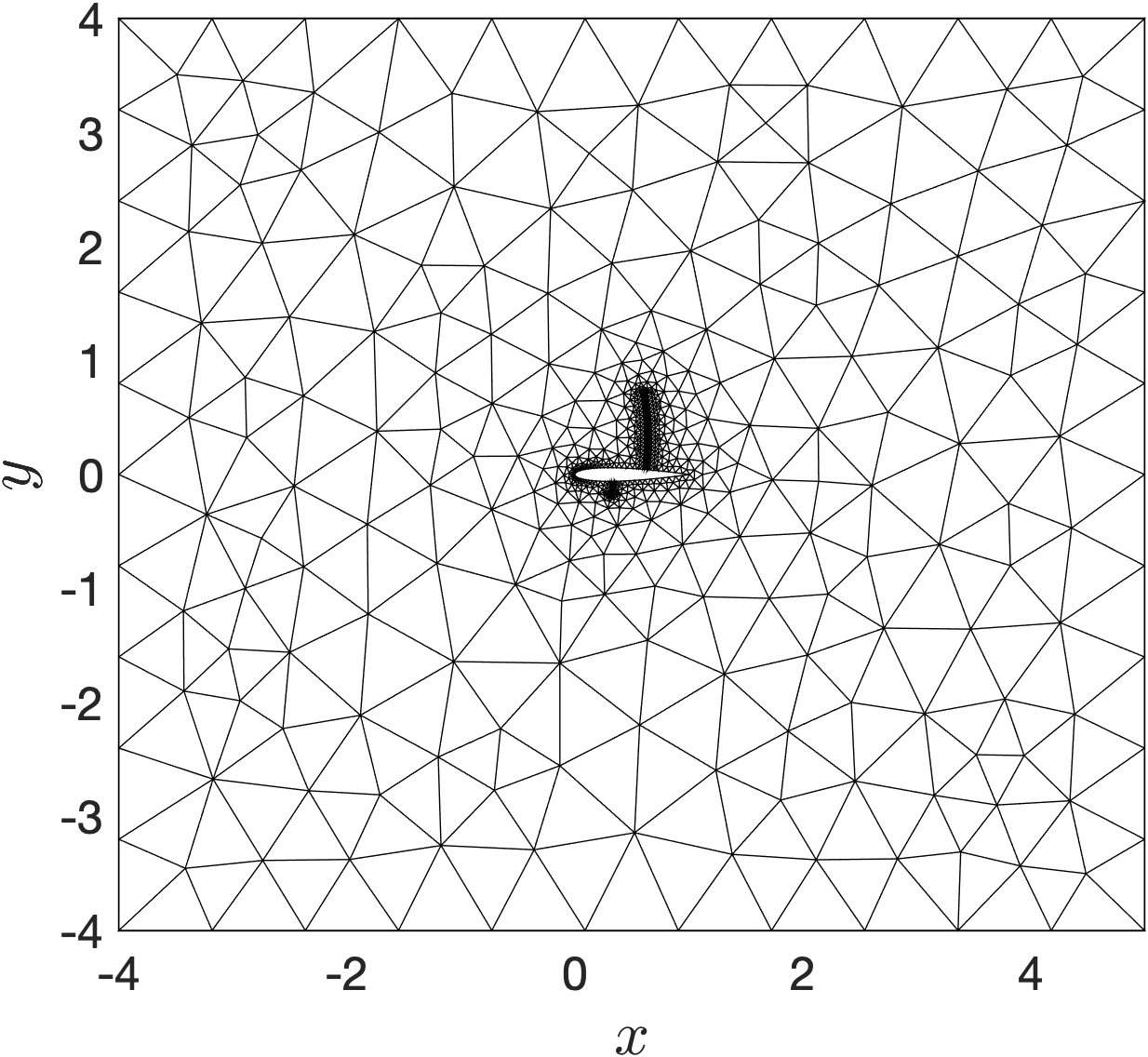}
		\caption{Shock-aligned mesh of 1769 $k=4$ elements}
	\end{subfigure} 
\caption {\label{fignaca1} Computational meshes for inviscid flow past NACA 0012 airfoil at $M_\infty = 0.8$ and $1.5^{\rm o}$ angle of attack.}
\end{figure}

The second example involves a case of transonic flow past a NACA 0012 airfoil at angle of attack $\alpha = 1.5^{\rm o}$ and freestream Mach number $M_\infty = 0.8$ \cite{Nguyen2011a}. A  shock is formed on the upper surface, while another  weaker shock is formed under the lower surface. Figure~\ref{fignaca1} depicts the initial unstructured grid of 1082 elements and a shock-aligned mesh of 1769 elements. 

\begin{figure}[htbp]
\centering
	\centering
	\begin{subfigure}[b]{0.49\textwidth}
		\centering		\includegraphics[width=\textwidth]{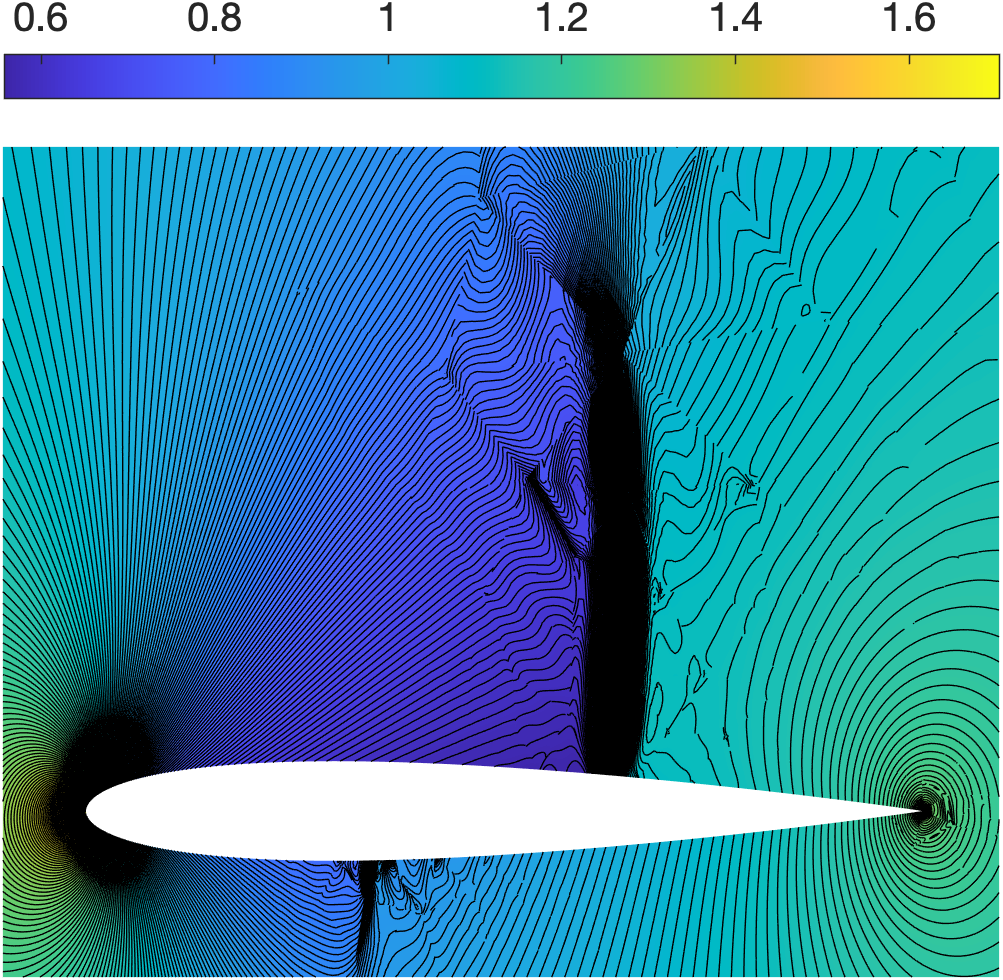}
		\caption{Regular mesh}
	\end{subfigure}
	\hfill
	\begin{subfigure}[b]{0.49\textwidth}
		\centering
		\includegraphics[width=\textwidth]{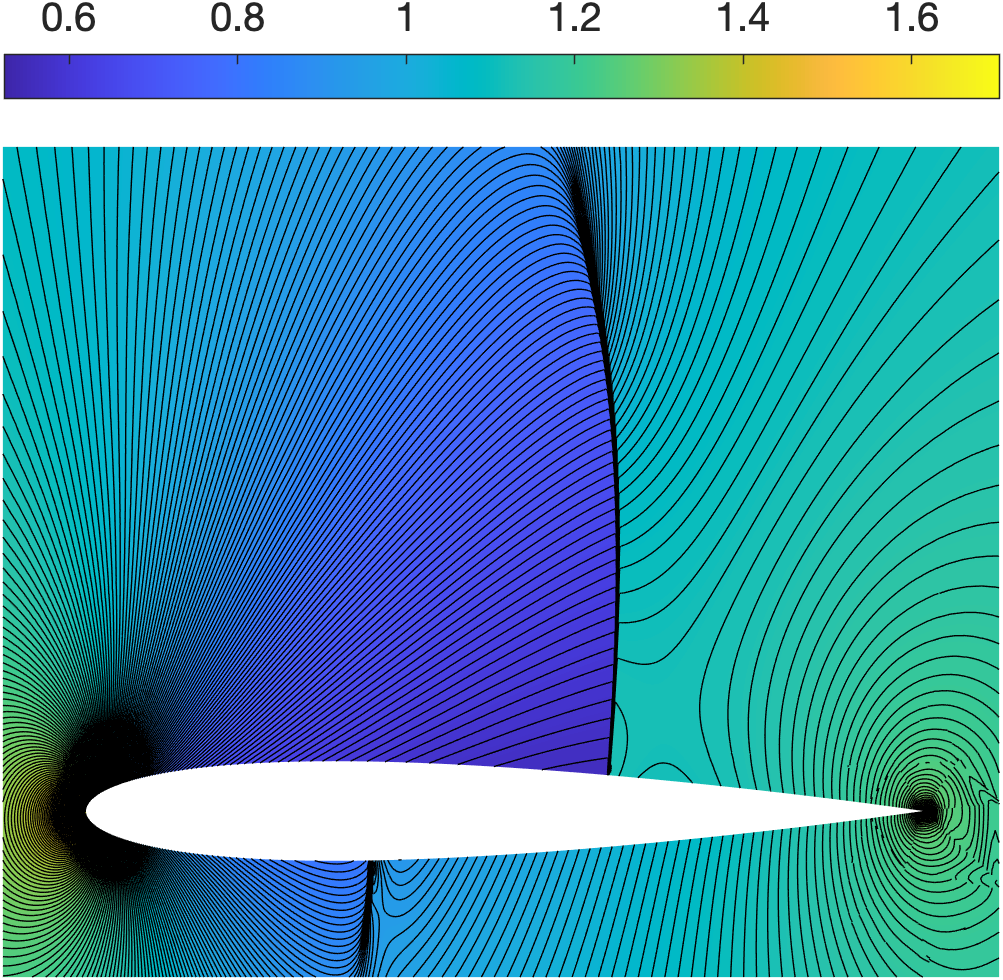}
		\caption{Shock-aligned mesh}
	\end{subfigure} 
\caption {\label{fignaca2} Computed pressure for inviscid flow past NACA 0012 airfoil at $M_\infty = 0.8$ and $1.5^{\rm o}$ angle of attack.}
\end{figure}

\begin{figure}[htbp]
\centering
	\centering
	\begin{subfigure}[b]{0.32\textwidth}
		\centering		\includegraphics[width=\textwidth]{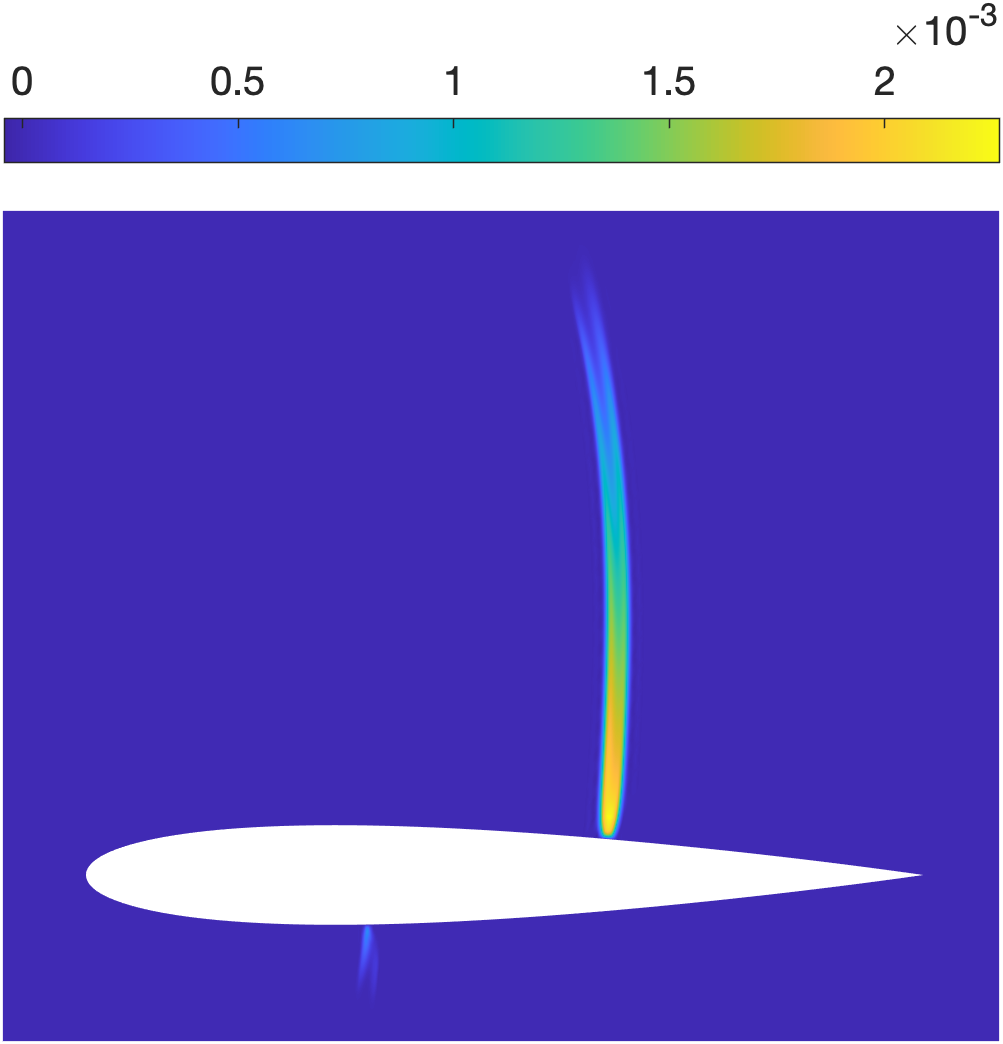}
	\end{subfigure}
	\hfill
	\begin{subfigure}[b]{0.32\textwidth}
		\centering
		\includegraphics[width=\textwidth]{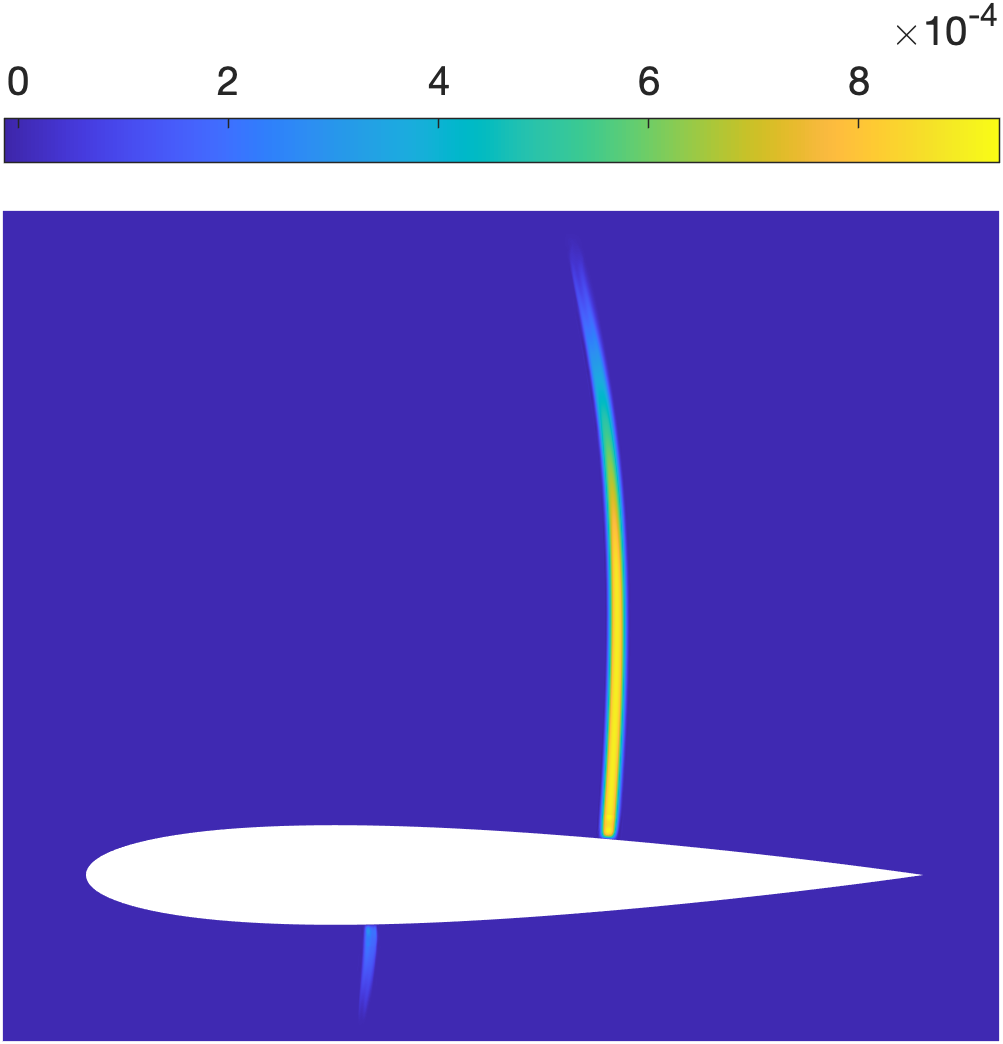}
	\end{subfigure} 
        \hfill
	\begin{subfigure}[b]{0.32\textwidth}
		\centering
		\includegraphics[width=\textwidth]{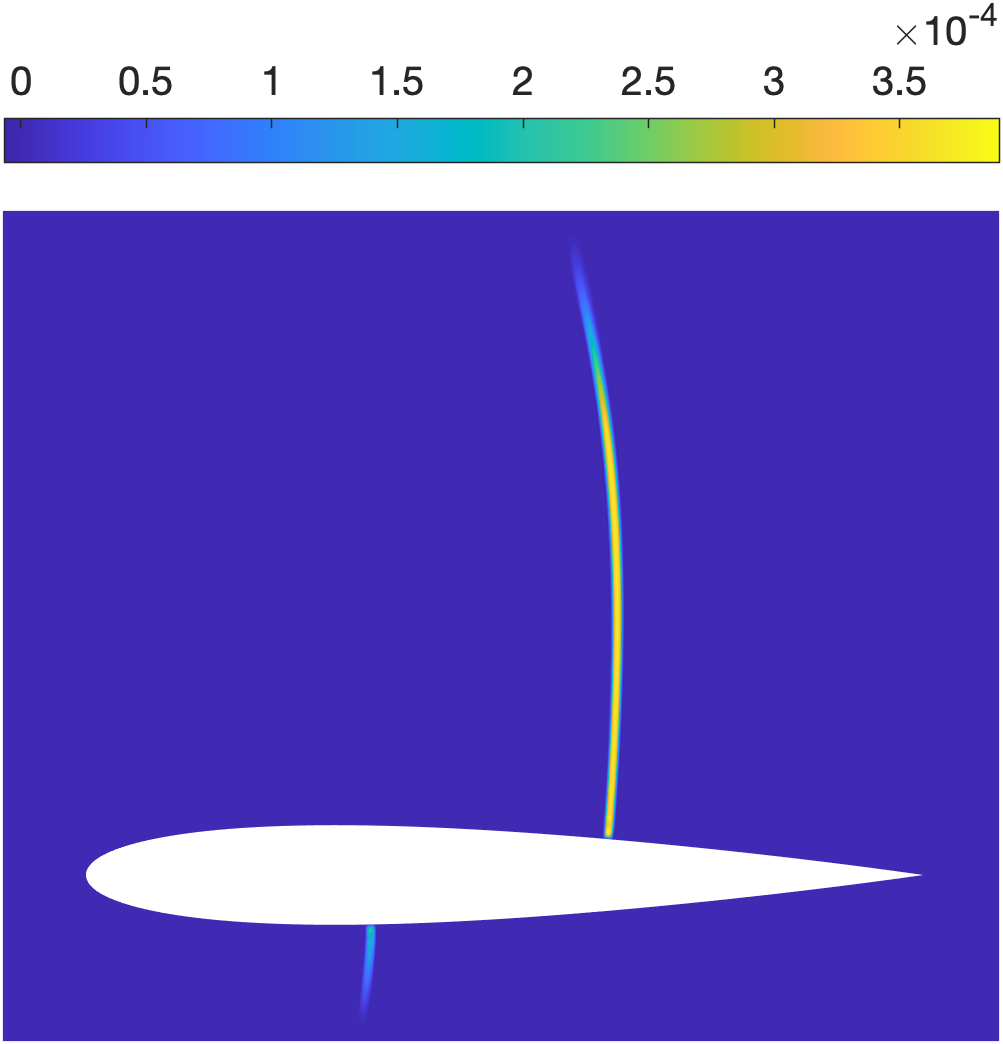}
	\end{subfigure} \\[2ex]
 	\begin{subfigure}[b]{0.32\textwidth}
		\centering		\includegraphics[width=\textwidth]{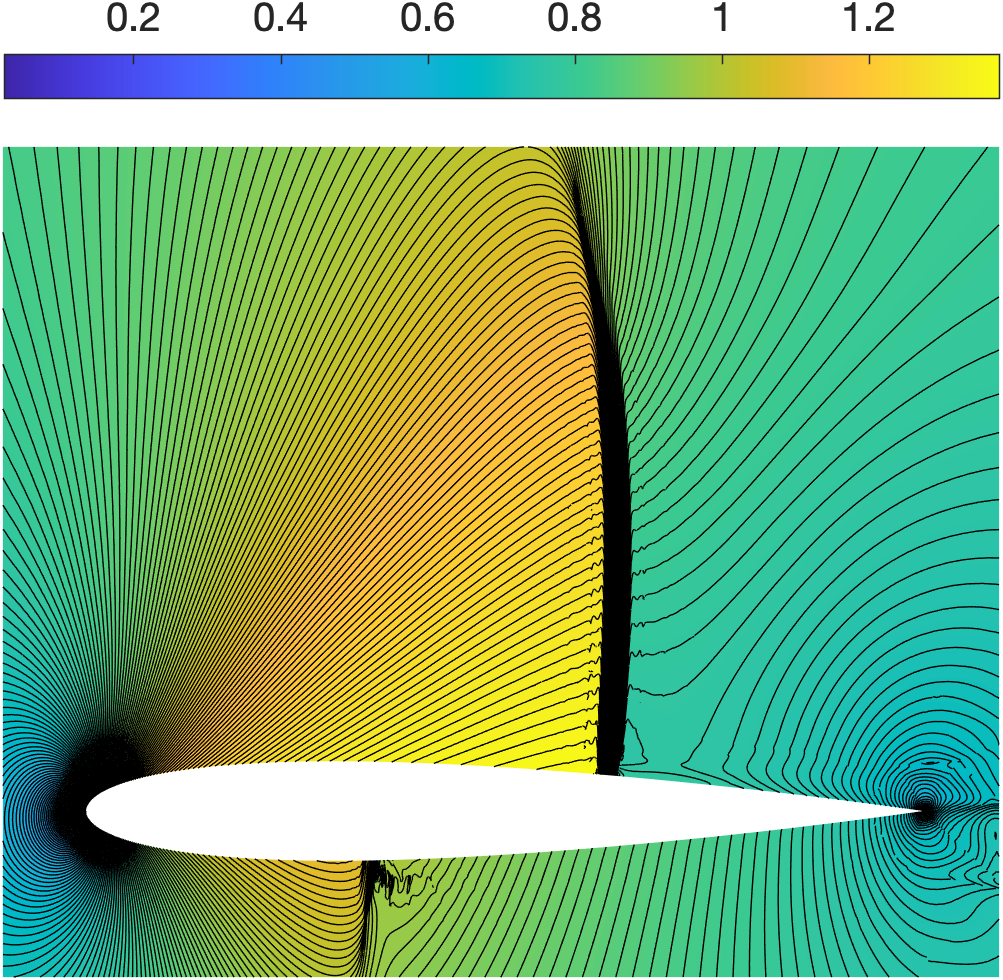}
  \caption{$n=1$}    
	\end{subfigure}
	\hfill
	\begin{subfigure}[b]{0.32\textwidth}
		\centering
		\includegraphics[width=\textwidth]{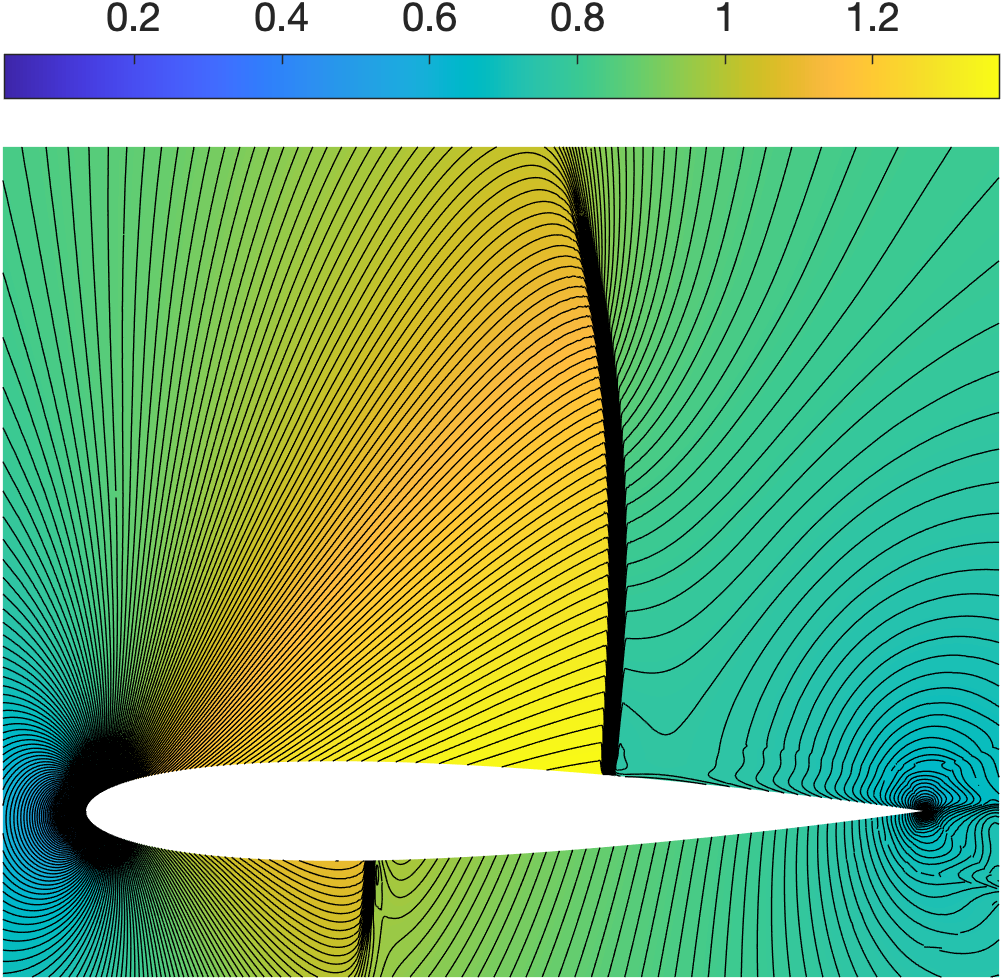}
        \caption{$n=5$}    
	\end{subfigure} 
        \hfill
	\begin{subfigure}[b]{0.32\textwidth}
		\centering
		\includegraphics[width=\textwidth]{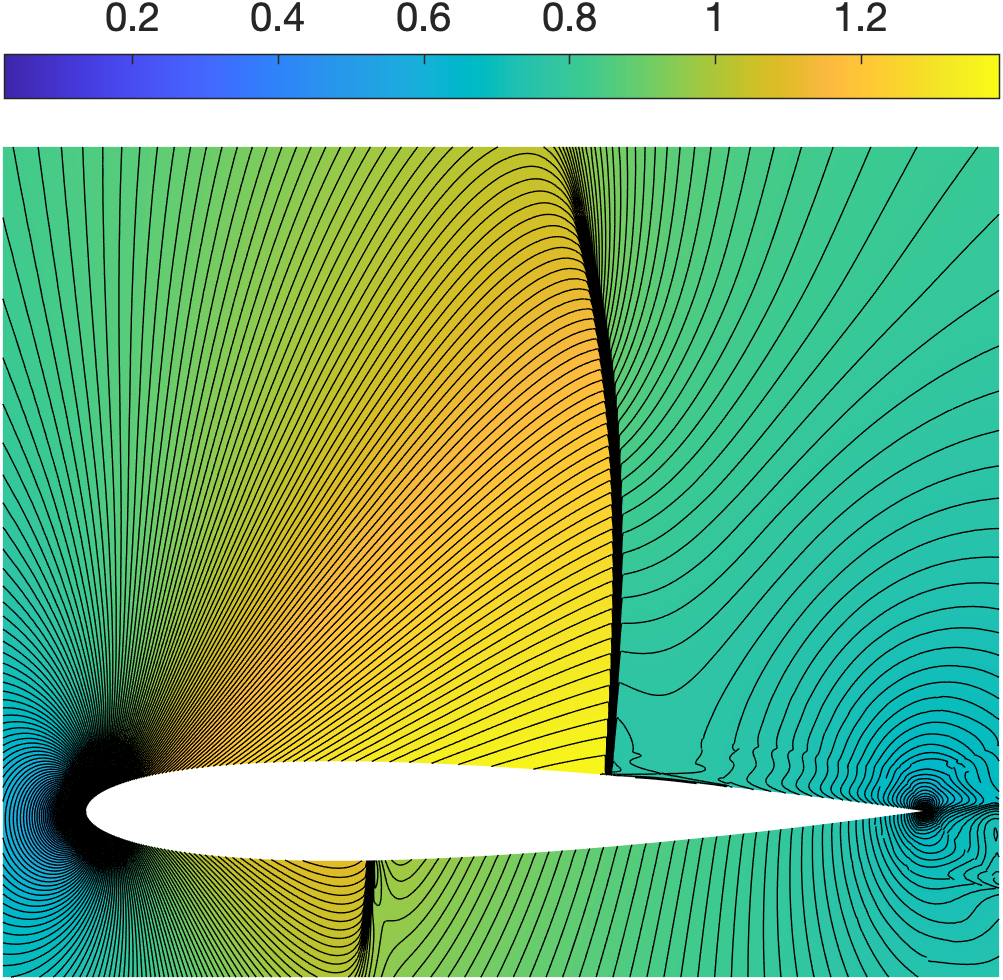}
        \caption{$n=9$}    
	\end{subfigure} 
\caption {\label{fignaca3} Artificial viscosity (top row) and Mach number (bottom row) at different homotopy iterations on the shock-aligned mesh for inviscid flow past NACA 0012 airfoil at $M_\infty = 0.8$ and $1.5^{\rm o}$ angle of attack.}
\end{figure}

\begin{figure}[htbp]
\centering
	\centering
	\begin{subfigure}[b]{0.49\textwidth}
		\centering		\includegraphics[width=\textwidth]{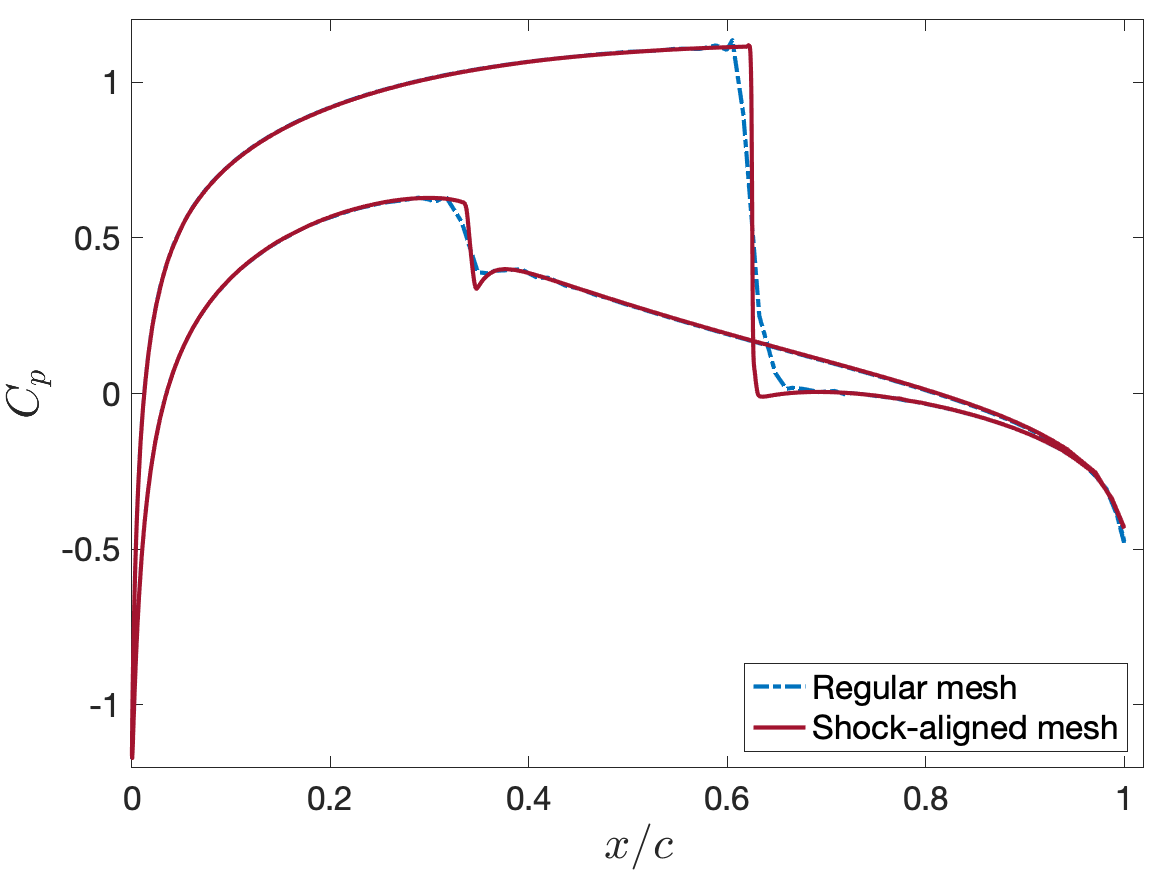}		
	\end{subfigure}
	\hfill
	\begin{subfigure}[b]{0.49\textwidth}
		\centering
		\includegraphics[width=\textwidth]{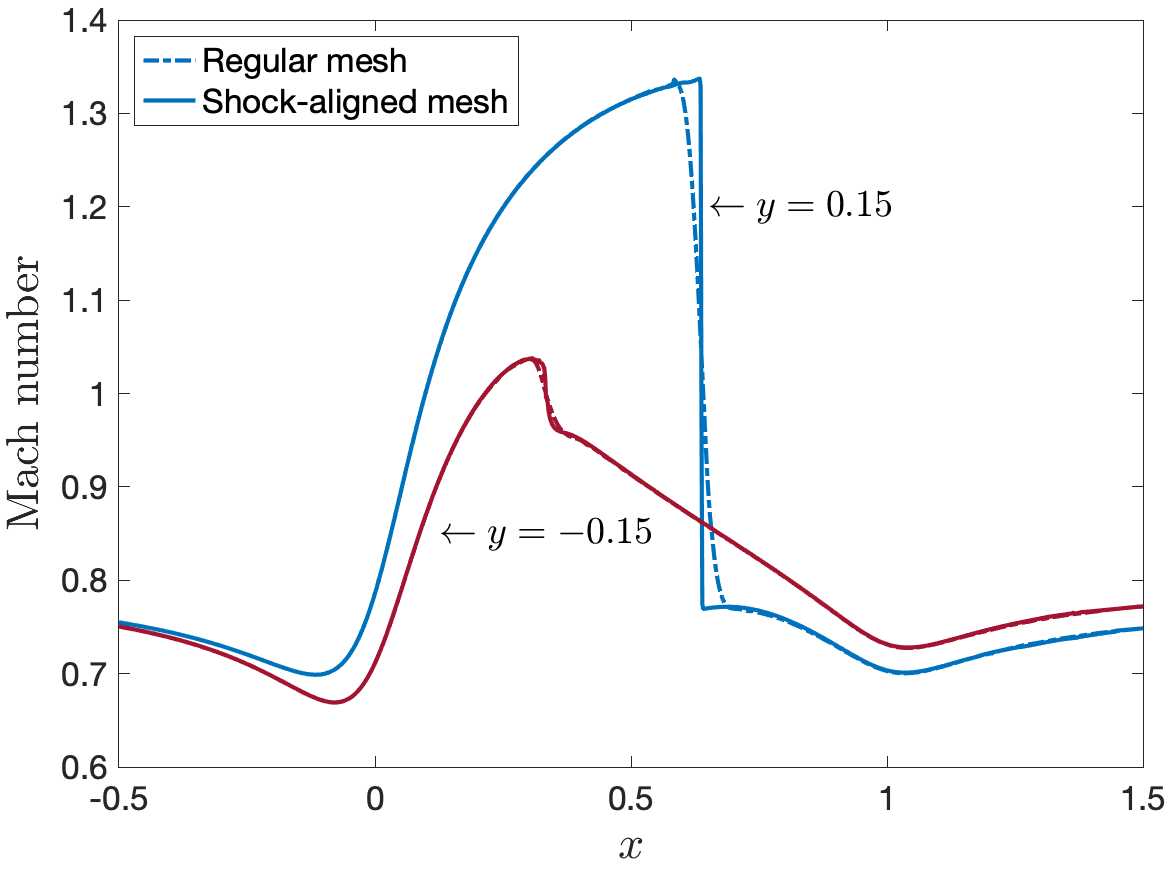}		
	\end{subfigure} 
\caption {\label{fignaca4} Profiles of computed pressure on the upper and lower surfaces of the airfoil (left) and Mach number along horizontal lines $y=\pm 0.15$ (right) for inviscid flow past NACA 0012 airfoil at $M_\infty = 0.8$ and $1.5^{\rm o}$ angle of attack.}
\end{figure}

Figure \ref{fignaca2} shows the pressure computed on the regular mesh and the shock-aligned mesh. As expected, the shock-aligned mesh yields a sharper, smoother, and more accurate solution than the regular mesh. This can be clearly seen from the profiles of the computed pressure and Mach number in Figure \ref{fignaca4}. We see that the shocks are captured very well by using our method to compute the solution on the shock-aligned mesh. Away from the shock region, the computed solutions match with each other. Figure \ref{fignaca3} depicts artificial viscosity and Mach number computed on the shock-aligned mesh at different homotopy iterations. We note that the amount of artificial viscosity is reduced as $n$ increases, resulting in sharper shock profiles and more accurate solutions.

\subsection{Ringleb flow}

\revise{We next consider the Ringleb flow to demonstrate the optimal accuracy of the numerical solution for smooth problems. The Ringleb flow is an exact smooth solution of the Euler equations. For any given $(x,y)$, we  obtain the radial velocity $V$ by solving the following nonlinear equation
\begin{equation*}
(x - 0.5 L^2) + y^2 = \frac{1}{4 \rho^2 V^4},
\end{equation*}
where
\begin{equation*}
\displaystyle c  =  \sqrt{1 - \frac{V^2}{5}} \ , \quad \rho  =  c^5 \ ,\quad L =   \frac{1}{c} + \frac{1}{3c^3} + \frac{1}{5c^5} - \frac{1}{2} \ln \frac{1+c}{1-c} .
\end{equation*}
We then compute the exact solution as
\begin{equation*}
\rho  =  c^5 \ , \quad p  =  c^7/\gamma \ ,  \quad v_1  =  V \cos(\theta) \ ,  \quad v_2  =  V \sin(\theta) ,
\end{equation*}
where 
\begin{equation*}
\displaystyle \psi  = \sqrt{\frac{1}{2V^2} - (x-0.5L) \rho} \ , \qquad \theta  =  \arcsin (\psi V) .
\end{equation*}
Since the exact solution can be determined for any spatial point, we take the domain $\Omega$ to be $(-2,-1)\times(1,2)$. The boundary condition is prescribed by setting the freestream value $\bm{u}_\infty$ to the exact solution on the boundary of the domain. Triangular meshes are used by splitting a regular $m \times m$ Cartesian grid into $2 m^2$ triangles.} 

\revise{The initial values for the regularization parameters are set to $\bm \lambda_0 = (10^{-3}, 1)$ and the homotopy parameter is set to $\zeta = 0.1$. We stop the homotopy continuation after $n=12$ iterations so that $\bm \lambda_n = (10^{-15}, 1)$. Although we can let the homotopy continuation run further, it does not make any difference to the accuracy of the numerical solution. We present in Table~\ref{Eulertab1} the $L^2$ error and convergence rate of the numerical solution as a function of $h$ and $k$. The $L^2$ error is defined as $\|\bm u_{e} - \bm u_n\|_{\Omega}$, where $\bm u_e$ is the exact solution and $\bm u_n$ is the numerical solution. We observe that the numerical solution converges with the optimal order $k+1$. The adaptive viscosity method is capable of yielding solutions with optimal convergence rates because the method can drive the artificial viscosity to zero for smooth problems.
}


\begin{table}[th]
  \begin{center}
\scalebox{0.95}{%
    $\begin{array}{|c||c c  | c c|  c c|  c c| c c| c c|}
    \hline
 \mbox{mesh} & \multicolumn{2}{|c|}{k=1} &  \multicolumn{2}{|c}{k=2} &  \multicolumn{2}{|c}{k=3}  & \multicolumn{2}{|c|}{k=4}   \\
     1/h & \mbox{error}  & \mbox{order} & \mbox{error}  & \mbox{order} & \mbox{error}  & \mbox{order} & \mbox{error}  & \mbox{order}  \\
     \hline     
  2  &  4.35\mbox{e-}3  &  --  &  3.24\mbox{e-}4  &  --  &  2.35\mbox{e-}5  &  --  &  2.08\mbox{e-}6  &  --  \\  
  4  &  1.10\mbox{e-}3  &  1.98  &  4.85\mbox{e-}5  &  2.74  &  1.43\mbox{e-}6  &  4.04  &  7.90\mbox{e-}8  &  4.72  \\  
  8  &  2.80\mbox{e-}4  &  1.98  &  6.92\mbox{e-}6  &  2.81  &  8.63\mbox{e-}8  &  4.05  &  2.80\mbox{e-}9  &  4.82  \\  
  16  &  7.06\mbox{e-}5  &  1.99  &  9.37\mbox{e-}7  &  2.88  &  5.18\mbox{e-}9  &  4.06  &  9.36\mbox{e-}11  &  4.90  \\  
  32  &  1.77\mbox{e-}5  &  2.00  &  1.22\mbox{e-}7  &  2.94  &  3.16\mbox{e-}10  &  4.03  &  2.99\mbox{e-}12  &  4.97  \\  
\hline
 \end{array} $
}
\end{center}{$\phantom{|}$}
     \caption{\revise{Convergence rates of the numerical solution for the Ringleb flow.}}
   \label{Eulertab1}
\end{table}

\subsection{Inviscid supersonic flow past unit circular cylinder}

The third test case is the supersonic flow past a unit circular cylinder at $M_\infty = 3$. A strong bow shock forms in front of the cylinder. Part of the flow region behind the shock is subsonic. The cylinder wall is modeled with inviscid wall boundary condition. Supersonic outflow boundary conditions are used at the outflow boundaries, while the rest of the boundary features supersonic inflow conditions. This test case serves to demonstrate  the effectiveness of our approach for supersonic shocks. 

\begin{figure}[htbp]
	\centering
	\begin{subfigure}[b]{0.49\textwidth}
		\centering		\includegraphics[width=\textwidth]{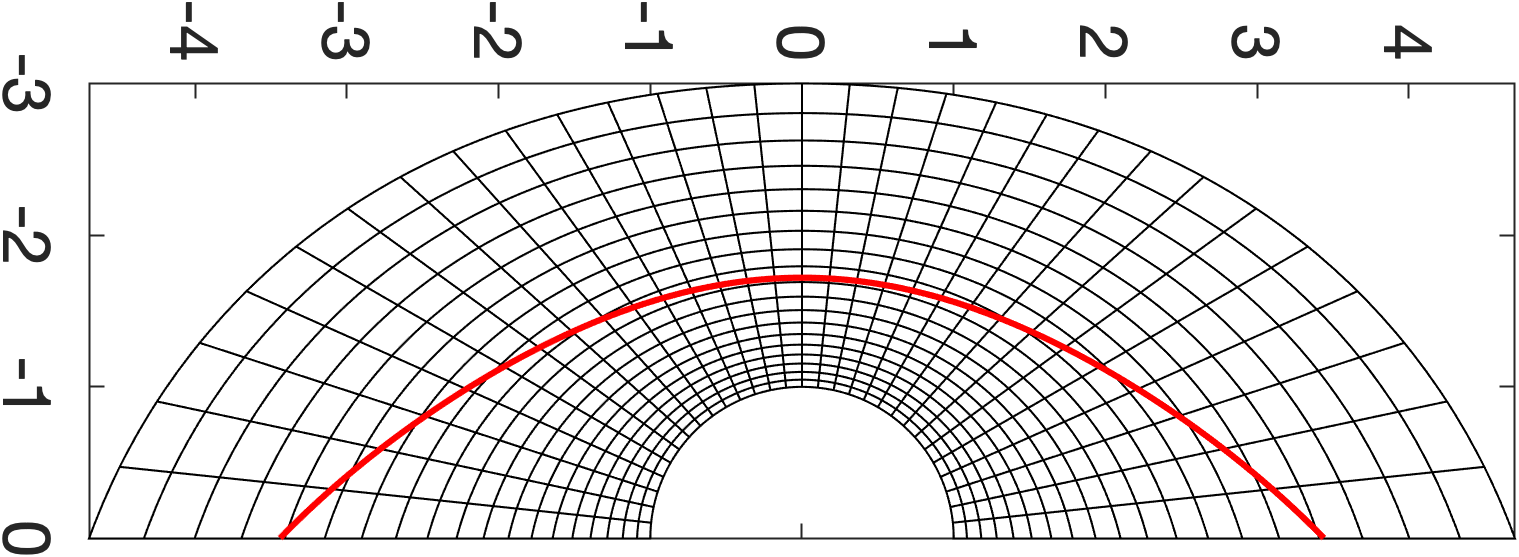}
		\caption{Regular mesh of 570 $k=4$ elements}
	\end{subfigure}
	\hfill
	\begin{subfigure}[b]{0.49\textwidth}
		\centering
		\includegraphics[width=\textwidth]{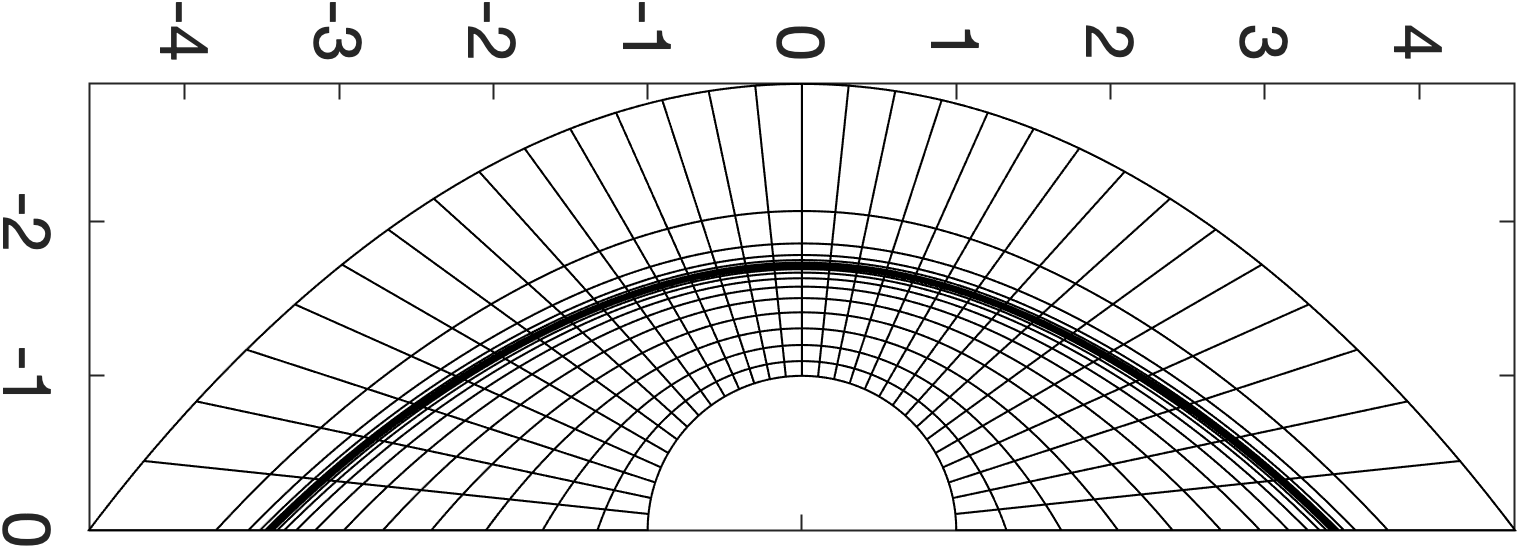}
		\caption{Shock-aligned mesh of 600 $k=4$ elements}
	\end{subfigure} 
	\caption{\label{figcyl3a} Computational meshes for inviscid flow past the circular cylinder at $M_\infty = 3$. The red curve indicates the location of the bow shock identified using the approximate solution on the regular mesh. The meshes are rotated 90 degrees to save spaces.}
\end{figure}

\begin{figure}[h]
	\centering
	\begin{subfigure}[b]{0.49\textwidth}
		\centering		\includegraphics[width=\textwidth]{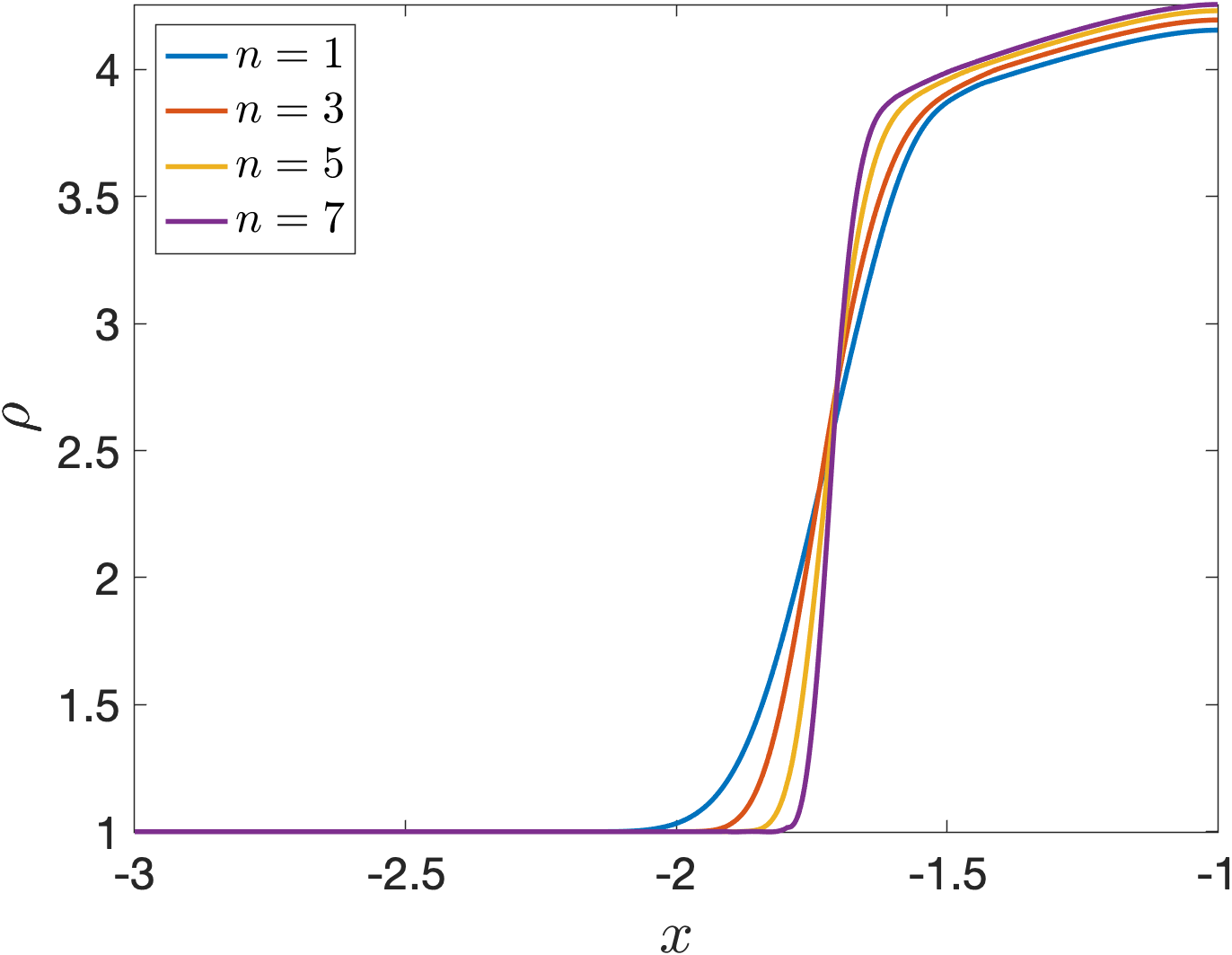}
	\end{subfigure}
	\hfill
	\begin{subfigure}[b]{0.49\textwidth}
		\centering
		\includegraphics[width=\textwidth]{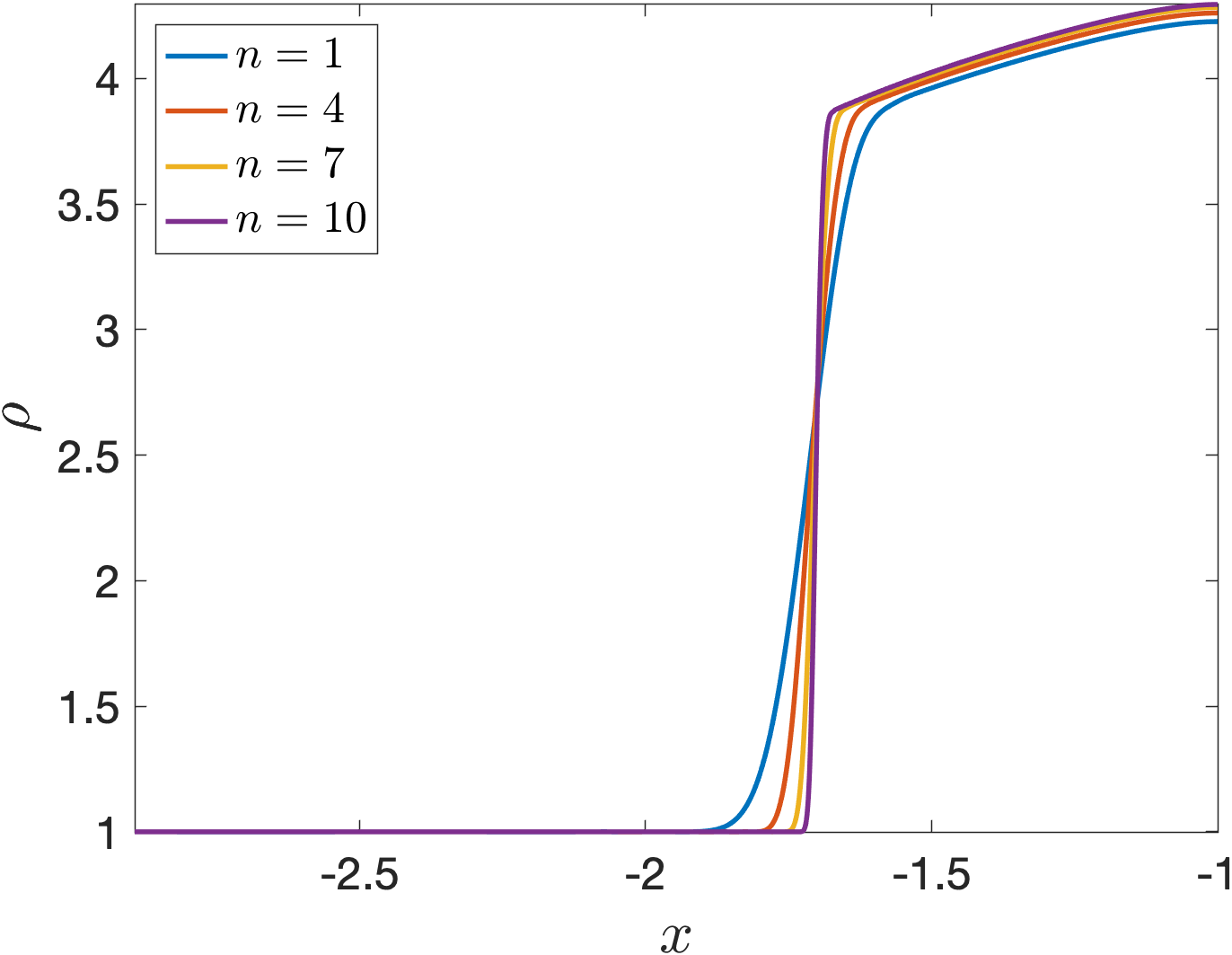}
	\end{subfigure} 
 	\begin{subfigure}[b]{0.49\textwidth}
		\centering		\includegraphics[width=\textwidth]{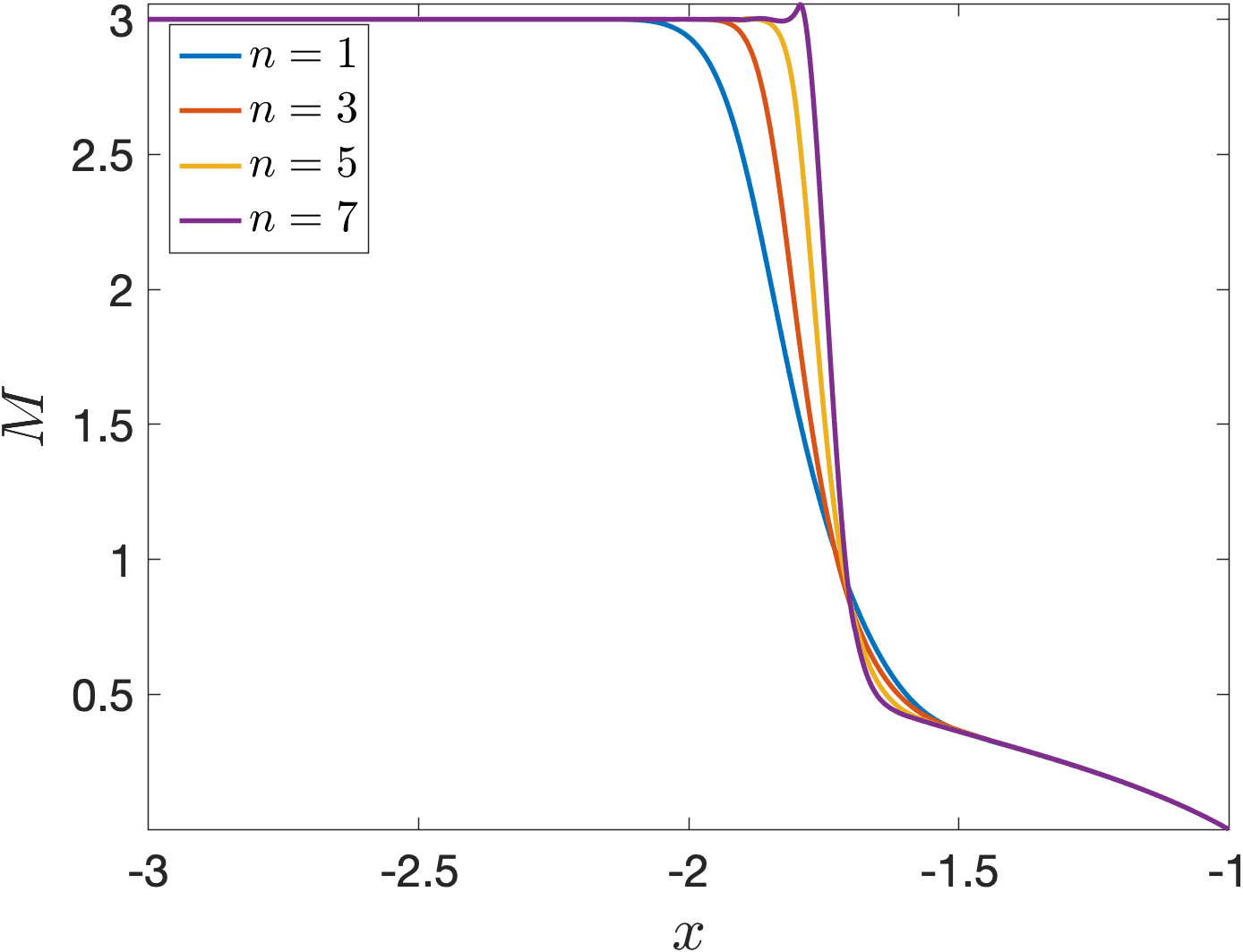}
		\caption{Regular mesh}
	\end{subfigure}
	\hfill
	\begin{subfigure}[b]{0.49\textwidth}
		\centering
		\includegraphics[width=\textwidth]{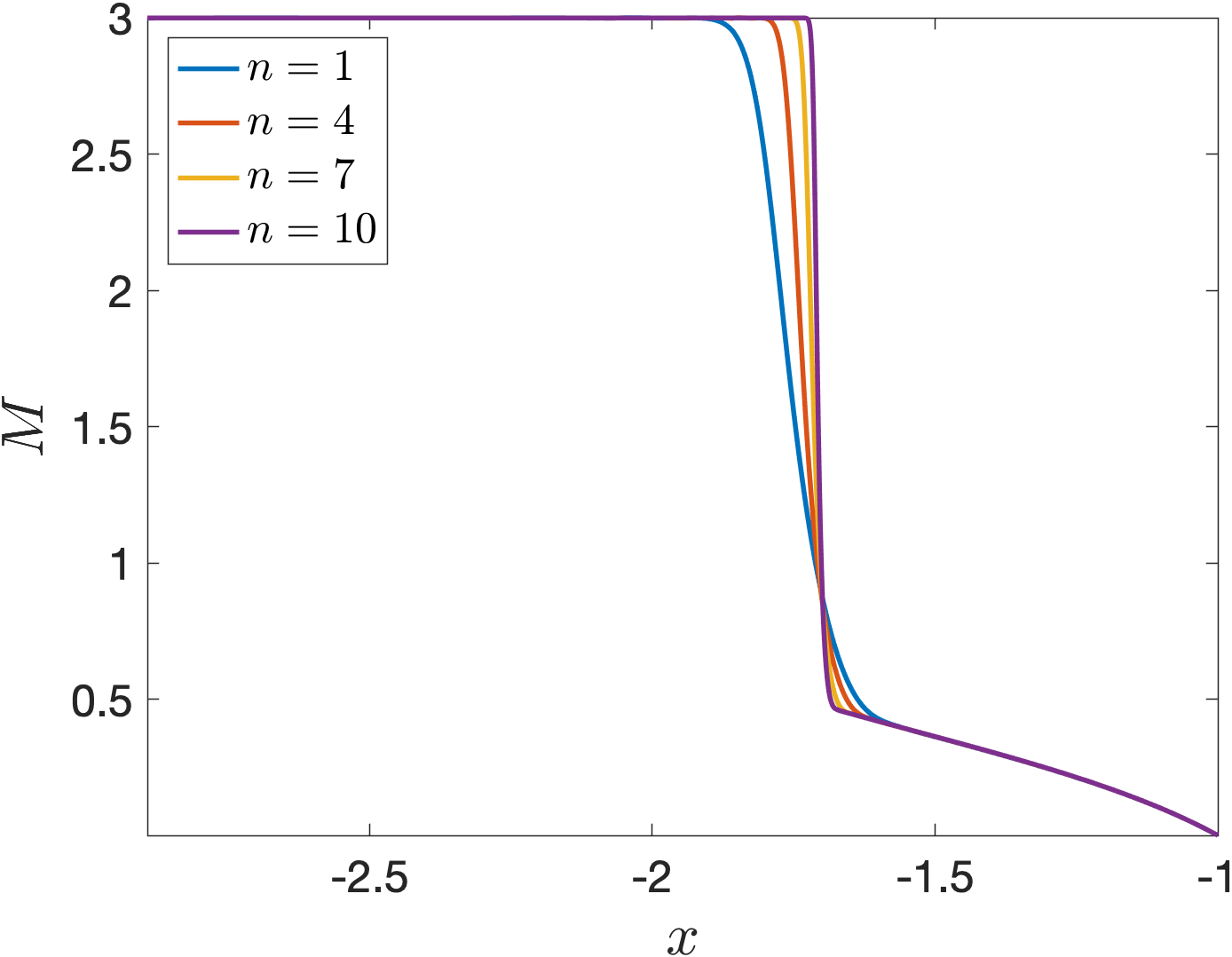}
		\caption{Shock-aligned mesh}
	\end{subfigure} 
	\caption{\label{figcyl3b} Profiles of density (top) and Mach number (bottom) along $y=0$ at different homptopy iterations for inviscid flow past the cylinder at $M_\infty=3$.}
\end{figure}

\begin{figure}[htbp]
\centering
	\begin{subfigure}[b]{0.18\textwidth}
		\centering		\includegraphics[width=\textwidth]{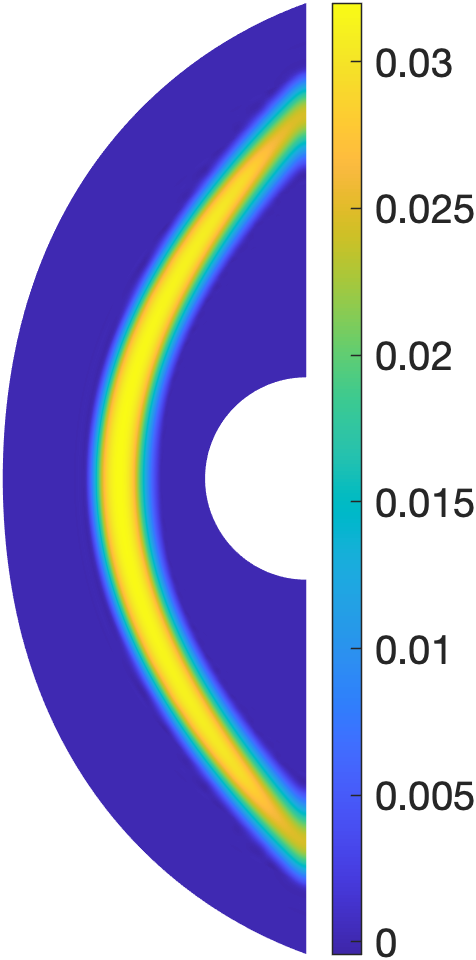}
	\end{subfigure}
	\hfill
	\begin{subfigure}[b]{0.18\textwidth}
		\centering		\includegraphics[width=\textwidth]{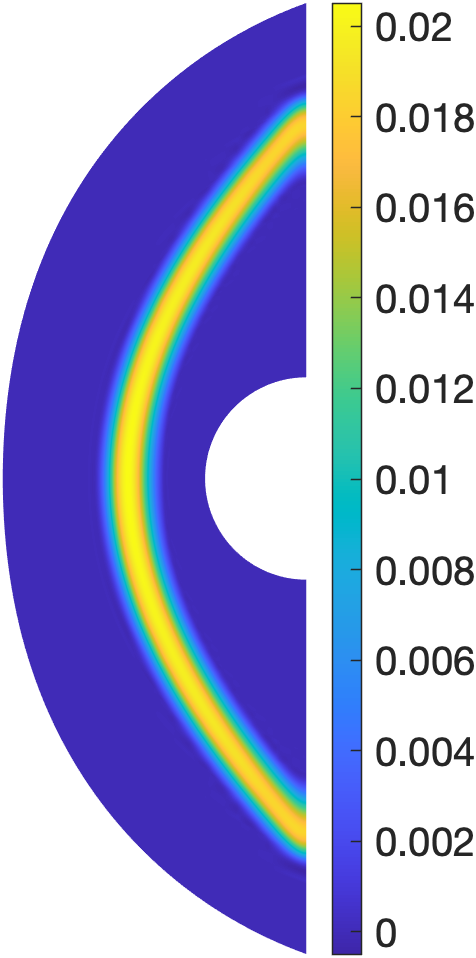}
	\end{subfigure}
        \hfill
        \begin{subfigure}[b]{0.18\textwidth}
		\centering		\includegraphics[width=\textwidth]{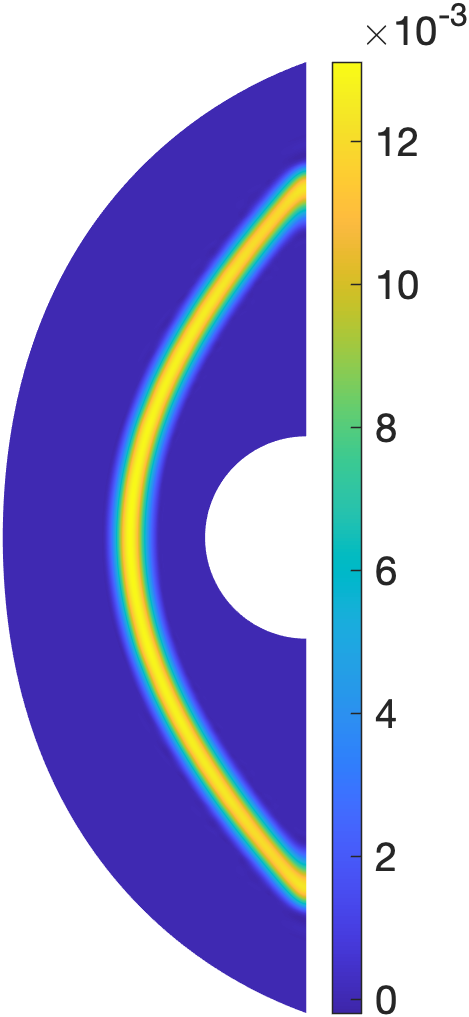}
	\end{subfigure}
        \hfill
        \begin{subfigure}[b]{0.18\textwidth}
		\centering		\includegraphics[width=\textwidth]{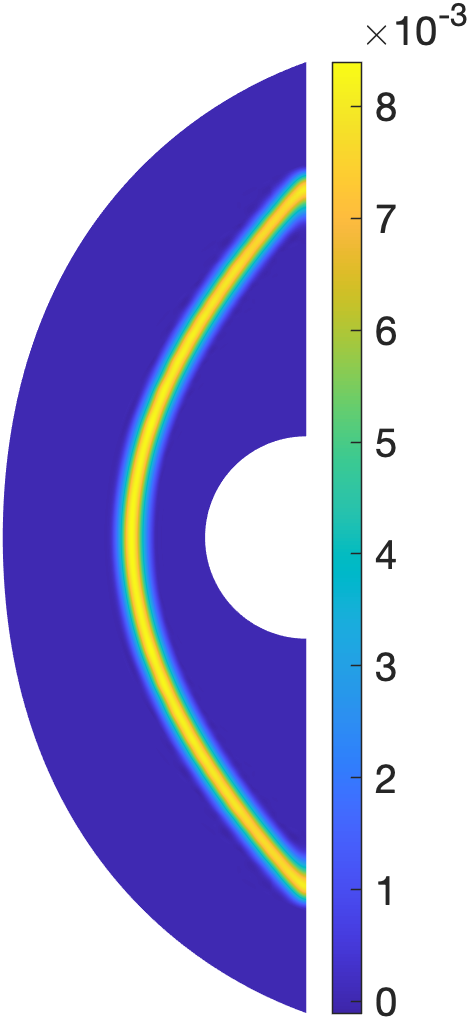}
	\end{subfigure} \\[2ex]
        \begin{subfigure}[b]{0.17\textwidth}
		\centering		\includegraphics[width=\textwidth]{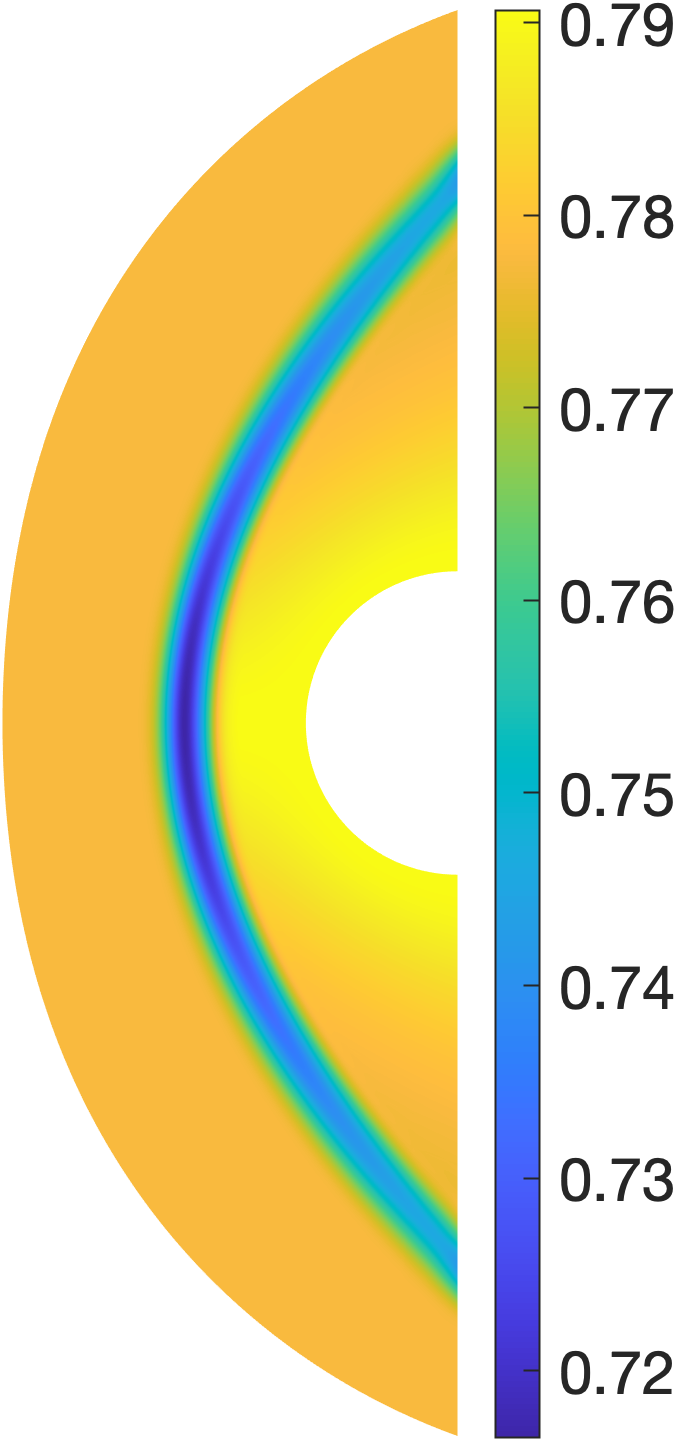}
	\end{subfigure}
	\hfill
	\begin{subfigure}[b]{0.17\textwidth}
		\centering		\includegraphics[width=\textwidth]{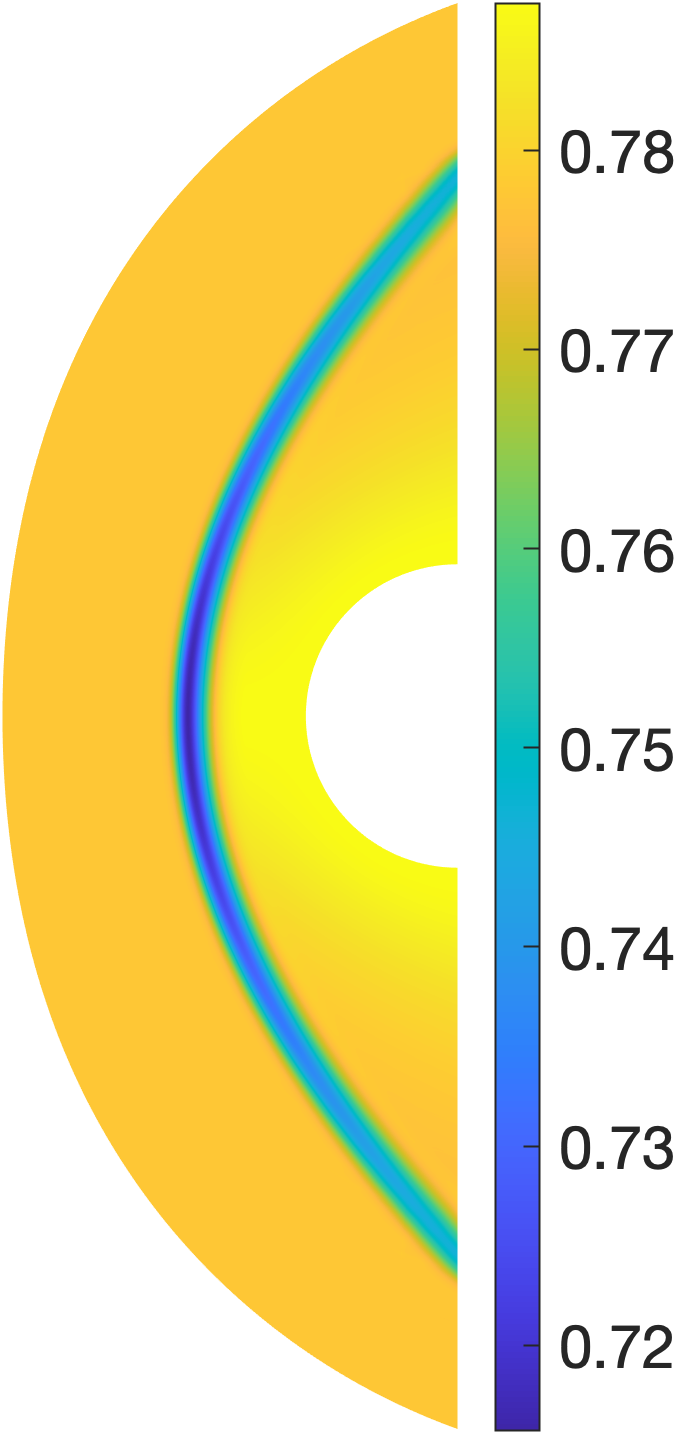}
	\end{subfigure}
        \hfill
        \begin{subfigure}[b]{0.17\textwidth}
		\centering		\includegraphics[width=\textwidth]{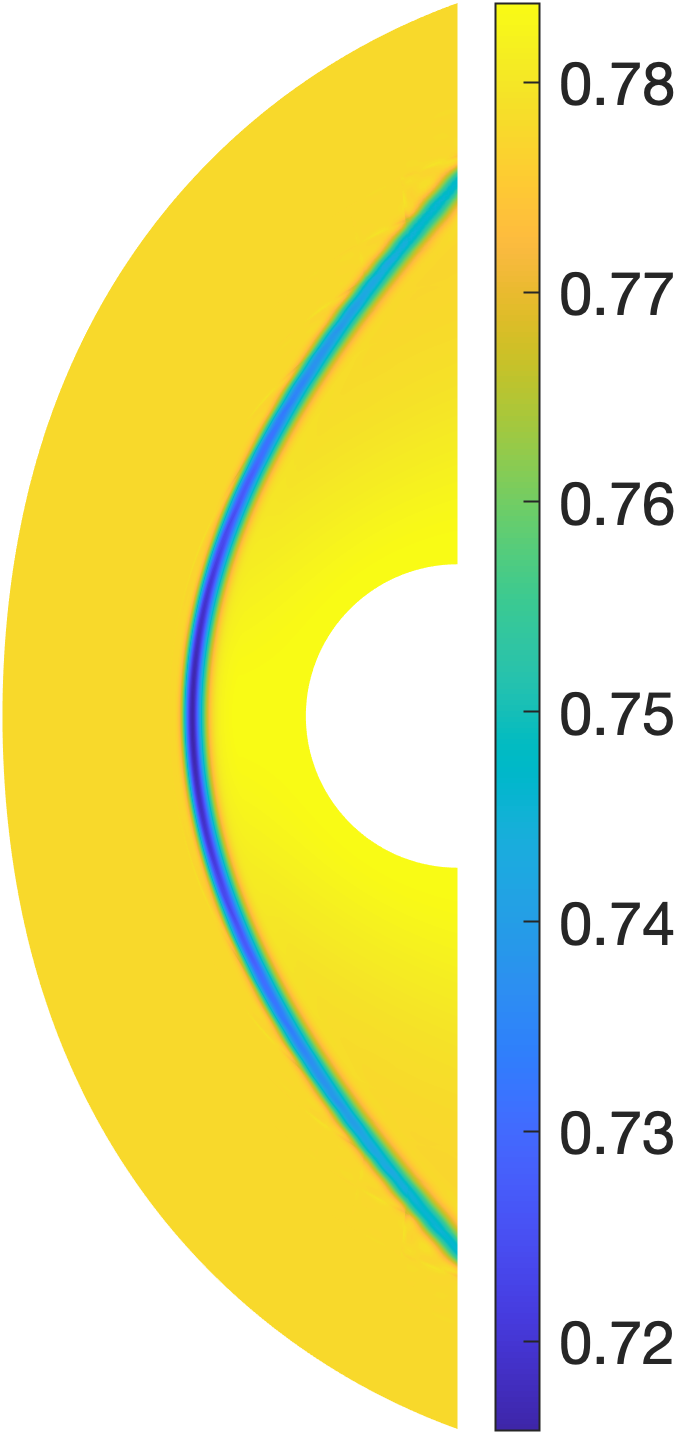}
	\end{subfigure}
        \hfill
        \begin{subfigure}[b]{0.17\textwidth}
		\centering		\includegraphics[width=\textwidth]{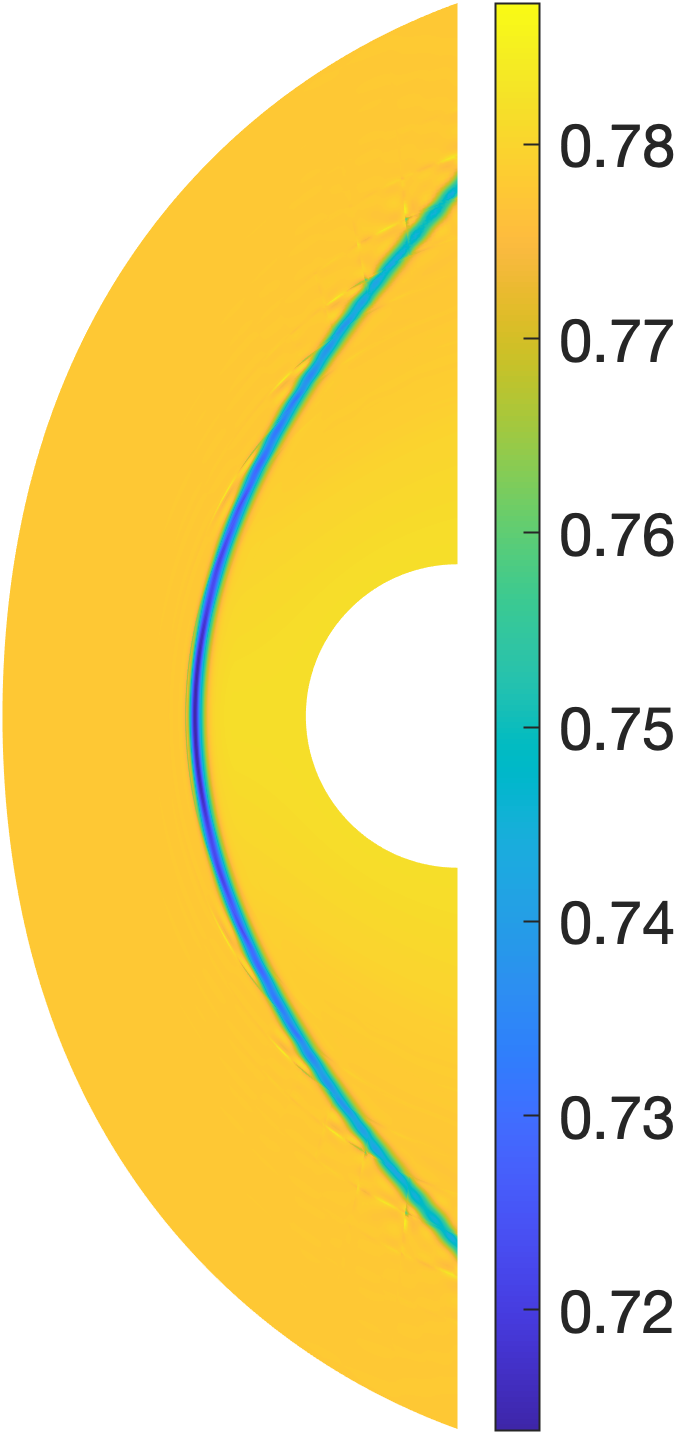}
	\end{subfigure} \\[2ex]
         \begin{subfigure}[b]{0.16\textwidth}
		\centering		\includegraphics[width=\textwidth]{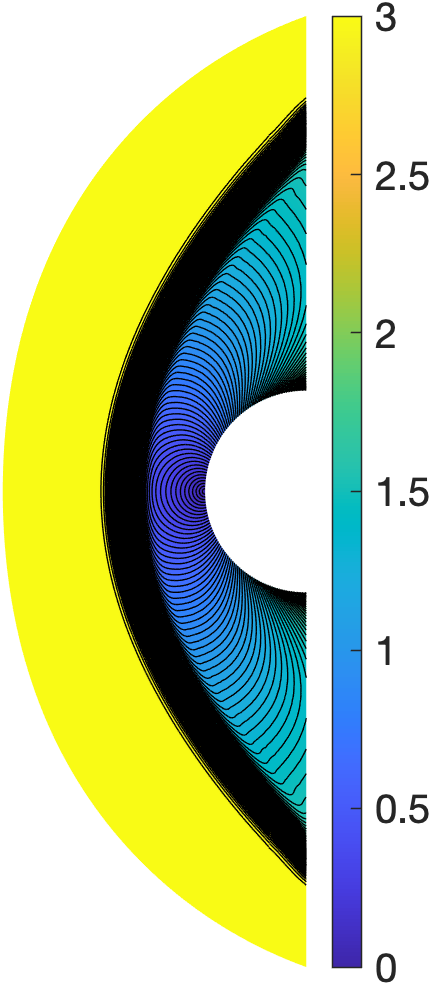}
    \caption{n=1}
	\end{subfigure}
	\hfill
	\begin{subfigure}[b]{0.16\textwidth}
		\centering		\includegraphics[width=\textwidth]{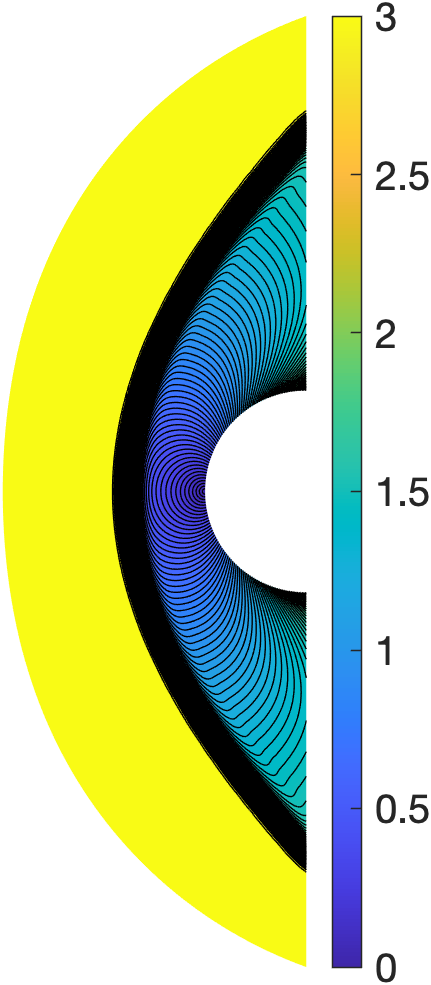}
  \caption{n=3}
	\end{subfigure}
        \hfill
        \begin{subfigure}[b]{0.16\textwidth}
		\centering		\includegraphics[width=\textwidth]{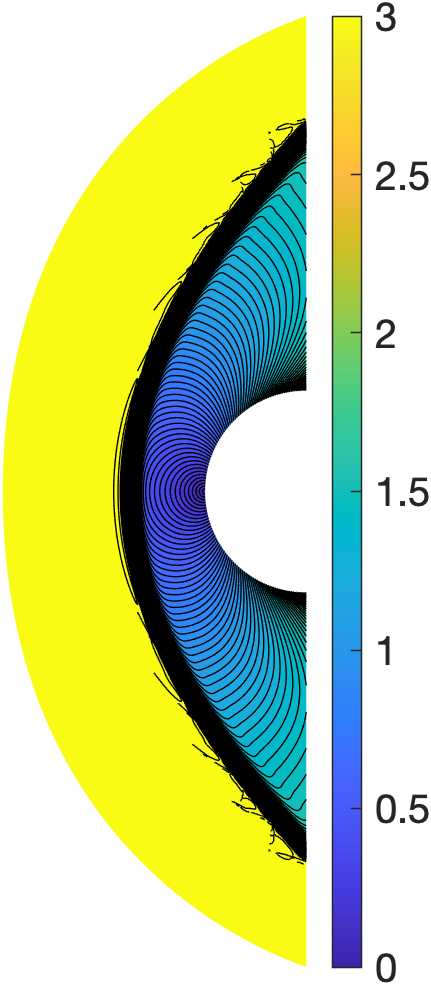}
  \caption{n=5}
	\end{subfigure}
        \hfill
        \begin{subfigure}[b]{0.16\textwidth}
		\centering		\includegraphics[width=\textwidth]{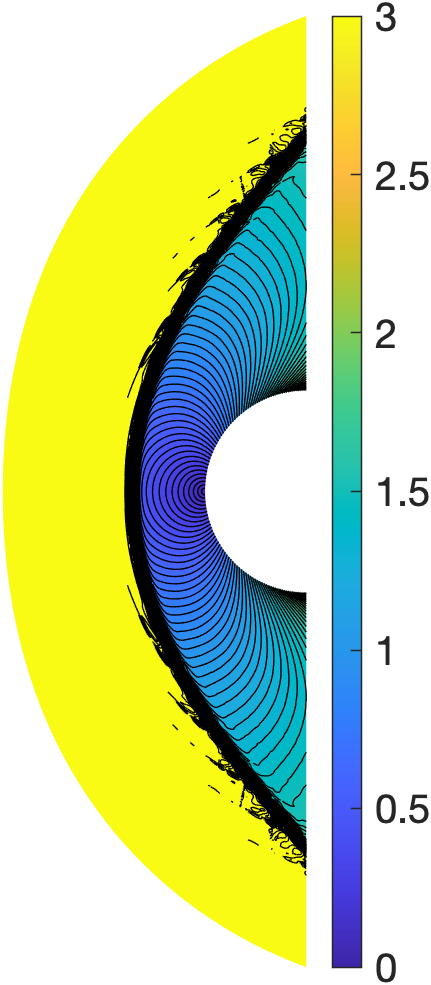}
          \caption{n=7}
	\end{subfigure} 
\caption{Computed artificial viscosity (top row), enthalpy (middle row), and Mach number (bottom row)   at different homotopy iterations on the regular mesh for inviscid supersonic flow past the cylinder at $M_\infty=3$.}
\label{cyl3d}	
\vspace{-0.5cm}	
\end{figure}

\begin{figure}[htbp]
\centering
	\begin{subfigure}[b]{0.18\textwidth}
		\centering		\includegraphics[width=\textwidth]{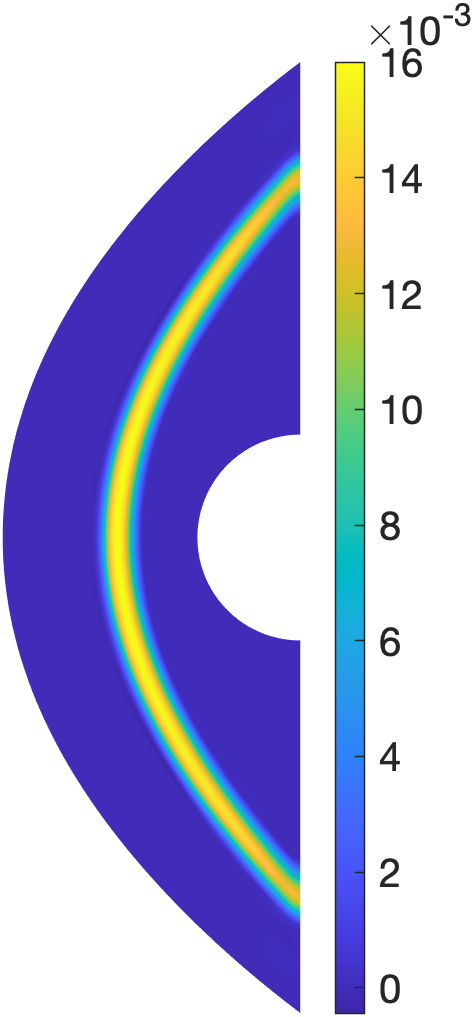}
	\end{subfigure}
	\hfill
	\begin{subfigure}[b]{0.18\textwidth}
		\centering		\includegraphics[width=\textwidth]{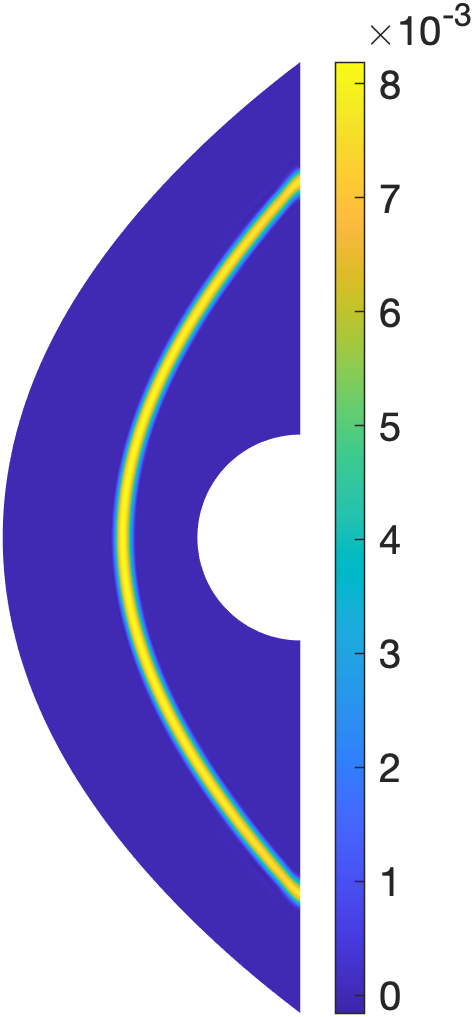}
	\end{subfigure}
        \hfill
        \begin{subfigure}[b]{0.18\textwidth}
		\centering		\includegraphics[width=\textwidth]{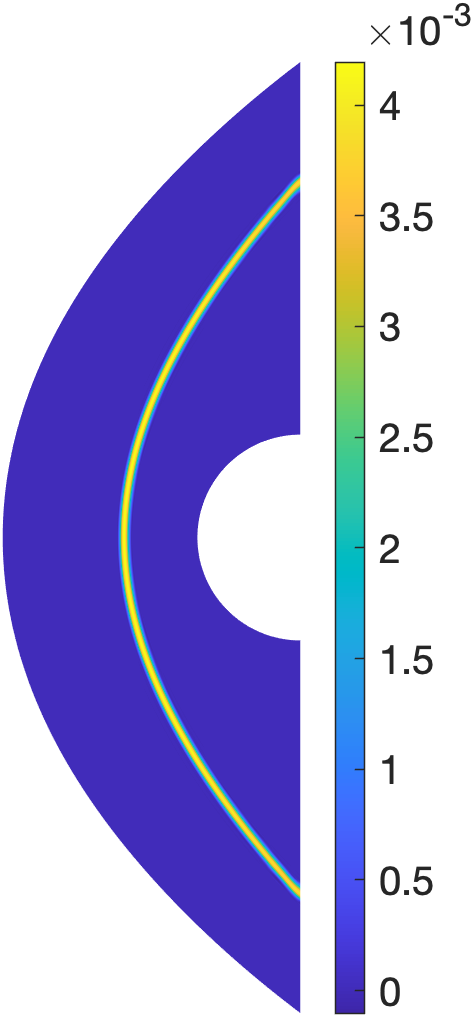}
	\end{subfigure}
        \hfill
        \begin{subfigure}[b]{0.18\textwidth}
		\centering		\includegraphics[width=\textwidth]{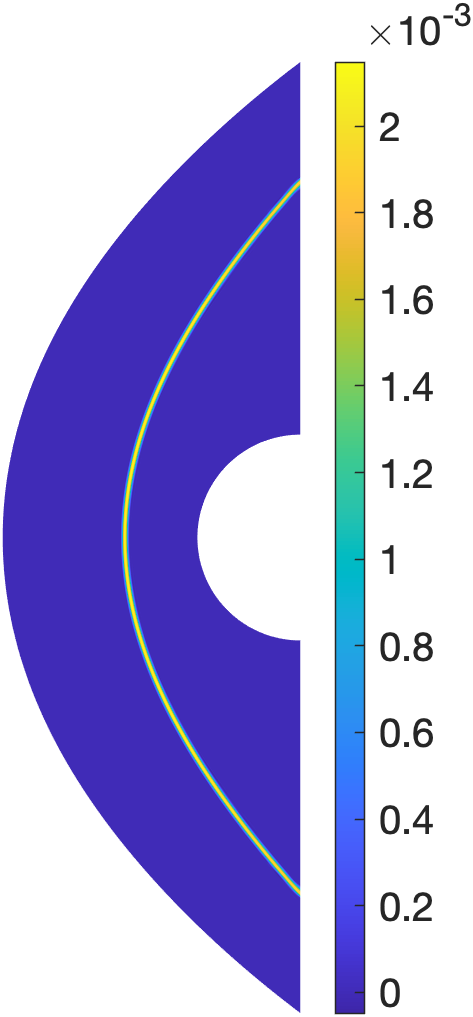}
	\end{subfigure} \\[2ex]
        \begin{subfigure}[b]{0.17\textwidth}
		\centering		\includegraphics[width=\textwidth]{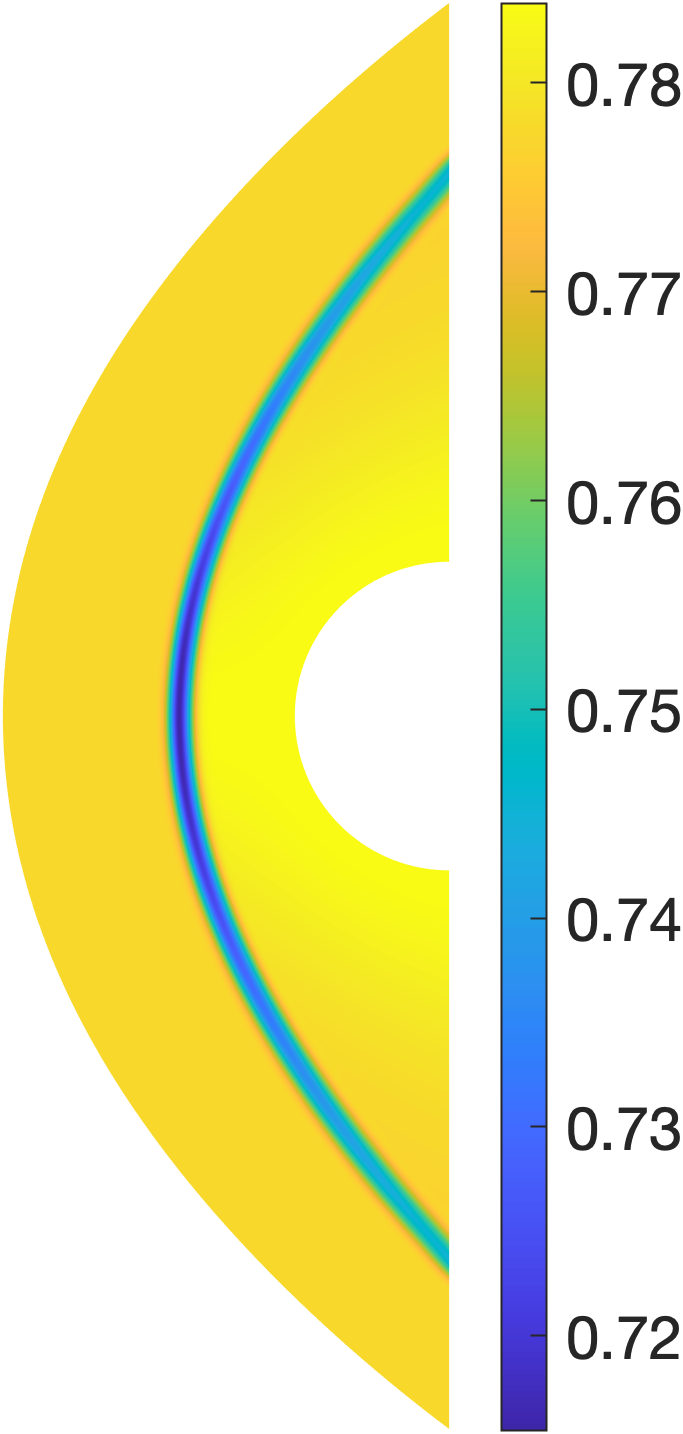}
	\end{subfigure}
	\hfill
	\begin{subfigure}[b]{0.17\textwidth}
		\centering		\includegraphics[width=\textwidth]{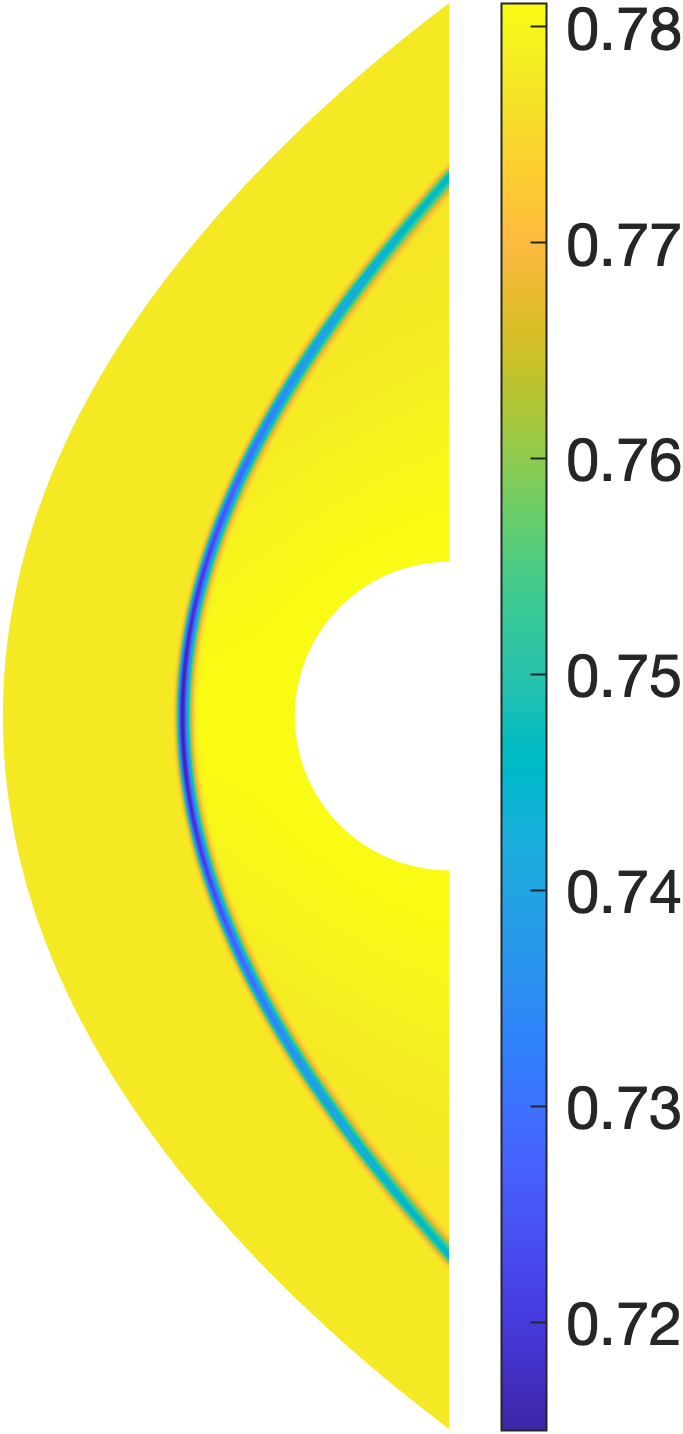}
	\end{subfigure}
        \hfill
        \begin{subfigure}[b]{0.17\textwidth}
		\centering		\includegraphics[width=\textwidth]{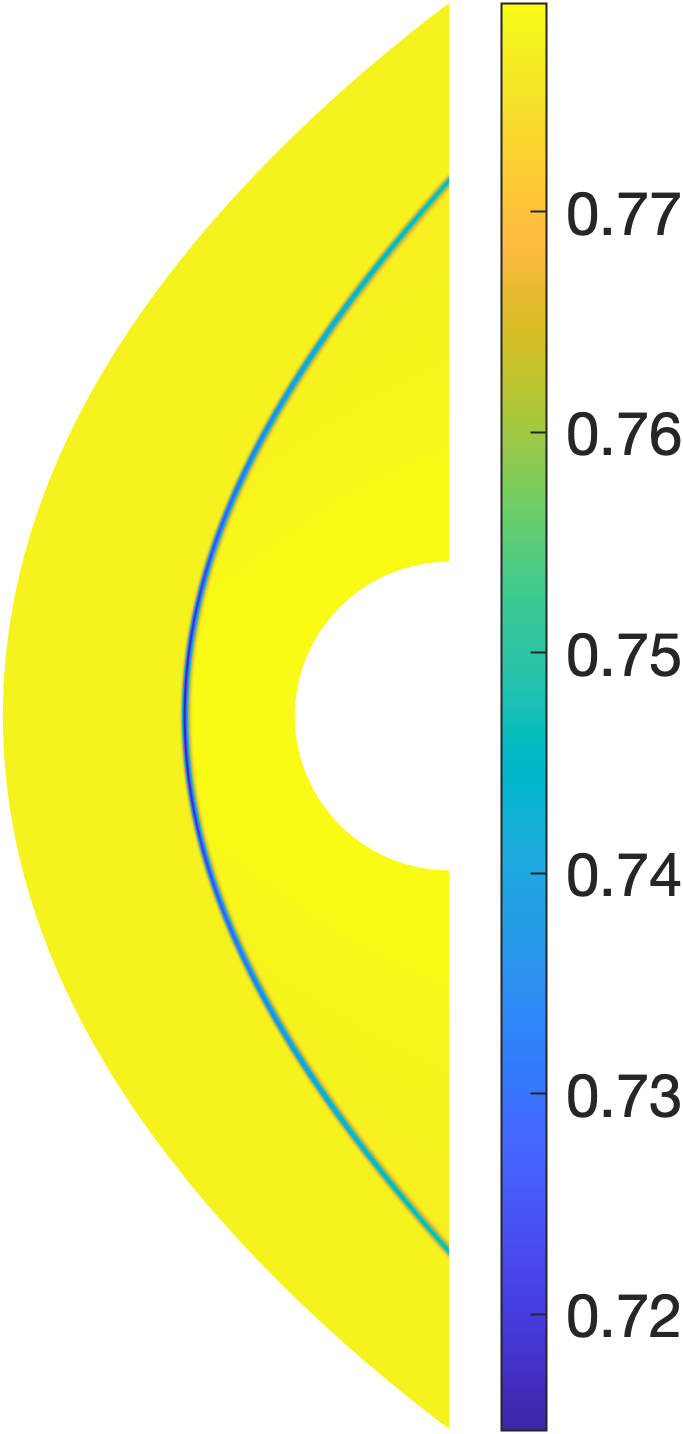}
	\end{subfigure}
        \hfill
        \begin{subfigure}[b]{0.17\textwidth}
		\centering		\includegraphics[width=\textwidth]{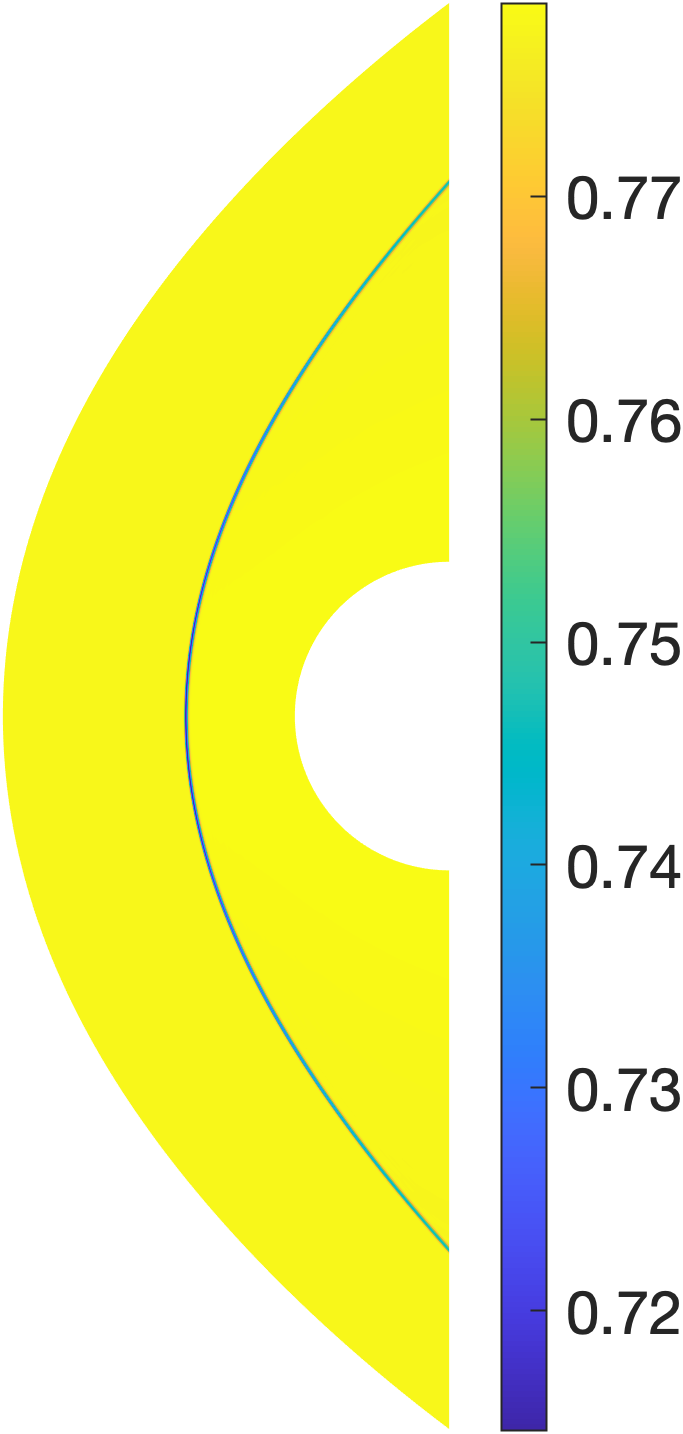}
	\end{subfigure} \\[2ex]
         \begin{subfigure}[b]{0.16\textwidth}
		\centering		\includegraphics[width=\textwidth]{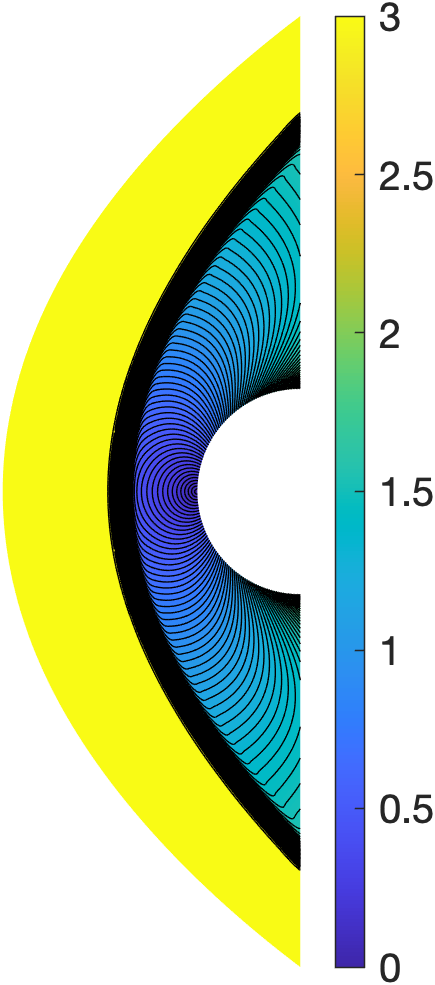}
    \caption{n=1}
	\end{subfigure}
	\hfill
	\begin{subfigure}[b]{0.16\textwidth}
		\centering		\includegraphics[width=\textwidth]{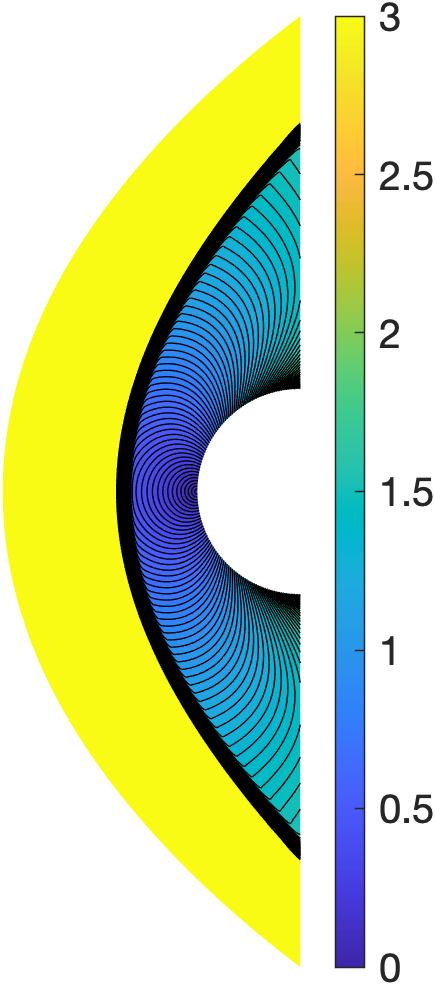}
    \caption{n=4}
	\end{subfigure}
        \hfill
        \begin{subfigure}[b]{0.16\textwidth}
		\centering		\includegraphics[width=\textwidth]{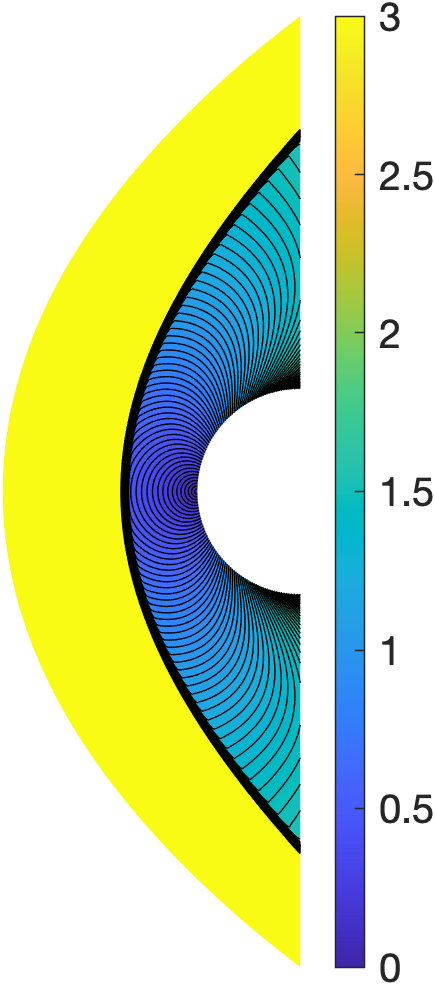}
    \caption{n=7}
	\end{subfigure}
        \hfill
        \begin{subfigure}[b]{0.16\textwidth}
		\centering		\includegraphics[width=\textwidth]{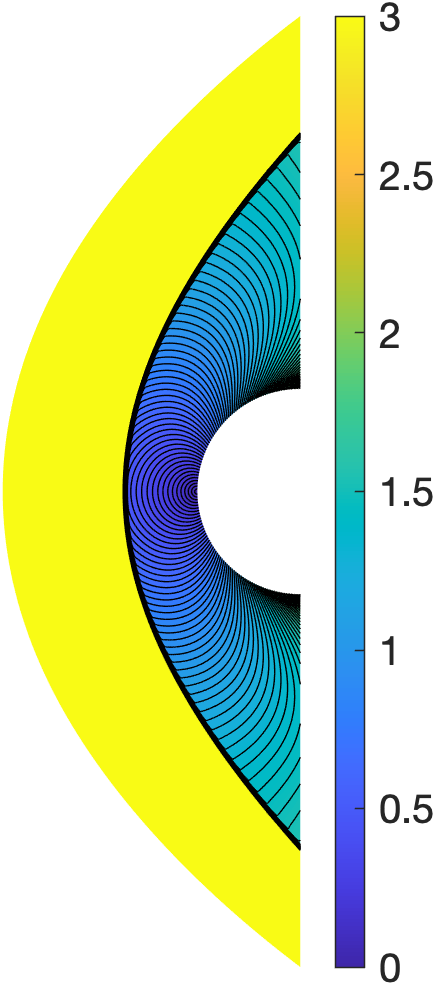}
  \caption{n=10}
	\end{subfigure} 
\caption{Computed artificial viscosity (top row), enthalpy (middle row), and Mach number (bottom row)   at different homotopy iterations on the shock-aligned mesh for inviscid supersonic flow past the cylinder at $M_\infty=3$.}
\label{cyl3e}	
\vspace{-0.5cm}	
\end{figure}

We show the regular mesh and the shock-align mesh in Figure \ref{figcyl3a}, and profiles of density and Mach number along $y=0$ in Figure  \ref{figcyl3b} at different homotopy iterations. We see that the shock profiles get sharper as $n$ increases. Furthermore, the shock profiles on the shock-aligned mesh are considerably sharper than those on the regular mesh. Figure \ref{cyl3d} and Figure \ref{cyl3e} depict the  solutions computed on the regular mesh and the shock-aligned mesh, respectively. The magnitude and width of artificial viscosity are considerably reduced as $n$ increases, indicating a significant reduction of the amount of artificial viscosity. Furthermore, the shock-aligned mesh yields much smaller and narrower artificial viscosity than the regular mesh. As a result, the solutions computed on the shock-aligned mesh are more accurate than those on the regular mesh, as it can be clearly seen from the computed enthalpy and Mach number. 


Table \ref{tabcyl32} tabulates relevant quantities of interest as a function of the homotopy iteration $n$ on the shock-aligned mesh.
Here $\theta_n = \sigma(\bm \lambda_n)/\sigma(\bm \lambda_1)$, $\|\varepsilon_n\|_{\Omega}$ is the $L_2$ norm of the artificial viscosity field on the physical domain, $\|M_n\|_{\infty}$ is the $L_\infty$ norm of the Mach number, $e_n^H = \|H_n - H_{\rm exact}\|_{\Omega \backslash {\Omega}^{\rm shock}}$ is the $L_2$ norm of the enthalpy error on the smooth domain,  $\rho_n(\bm x^*)$ and $p_n(\bm x^*)$ are the values of density and pressure at the stagnation point $\bm x^* = (-1, 0)$. The homotopy continuation ends at $n=10$ where $\theta_{10} = 5.4388$ is greater than the specified threshold. We see that the amount of artificial viscosity (namely, $\|\varepsilon_n\|_{\Omega}$) and the enthapy error (namely, $e_n^H$) decrease as $n$ increases. The results clearly show that the amount of artificial viscosity affects the accuracy of the approximate solution not only in the shock region but also away from the shock region. In particular, both the density and pressure at the stagnation point increase as the amount of artificial viscosity decreases. Moreover, the bow shock is very well captured since the overshoot in the Mach number is quite small as $\|M_n\|_{\infty}$ is close to $M_\infty = 3$.


\begin{table}[htbp]
\centering
\begin{small}
	\begin{tabular}{|c|cccccccc|}
		\hline
		$n$ & $\lambda_{n,1}$ & $\lambda_{n,2}$ & $\theta_n$ & $\|\varepsilon_n\|_{\Omega}$ &  $e^H_n$  & $\|M_n\|_{\infty}$ &  $\rho_n(\bm x^*)$ & $p_n(\bm x^*)$  \\
		\hline
 1  &  0.0200  &  5.0000  &  1.0000  &  0.0186  &  0.0096  &  3.0057  &  4.2263  &  0.9465  \\  
 2  &  0.0160  &  4.2000  &  1.9725  &  0.0137  &  0.0084  &  3.0169  &  4.2364  &  0.9478  \\  
 3  &  0.0128  &  3.5600  &  1.8190  &  0.0099  &  0.0066  &  3.0017  &  4.2498  &  0.9495  \\  
 4  &  0.0102  &  3.0480  &  2.6473  &  0.0072  &  0.0055  &  3.0002  &  4.2607  &  0.9509  \\  
 5  &  0.0082  &  2.6384  &  2.8580  &  0.0052  &  0.0043  &  3.0025  &  4.2697  &  0.9520  \\  
 6  &  0.0066  &  2.3107  &  3.2740  &  0.0038  &  0.0038  &  3.0043  &  4.2770  &  0.9530  \\  
 7  &  0.0052  &  2.0486  &  3.5356  &  0.0027  &  0.0030  &  3.0026  &  4.2828  &  0.9538  \\  
 8  &  0.0042  &  1.8389  &  3.7983  &  0.0020  &  0.0025  &  3.0019  &  4.2876  &  0.9544  \\  
 9  &  0.0034  &  1.6711  &  4.1949  &  0.0015  &  0.0020  &  3.0014  &  4.2914  &  0.9550  \\  
 10  &  0.0027  &  1.5369  &  5.4388  &  0.0011  &  0.0017  &  3.0007  &  4.2945  &  0.9554  \\  
		\hline
	\end{tabular}
 \end{small}
	\caption{Relevant quantities of interest as a function of homotopy iteration $n$ for the approximate solutions computed on the shock-aligned mesh for inviscid hypersonic flow past the cylinder at $M_\infty=3$.} 
	\label{tabcyl32}
\end{table}

\subsection{Inviscid hypersonic flow past unit circular cylinder}

\begin{figure}[htbp]
	\centering
	\begin{subfigure}[b]{0.52\textwidth}
		\centering		\includegraphics[width=\textwidth]{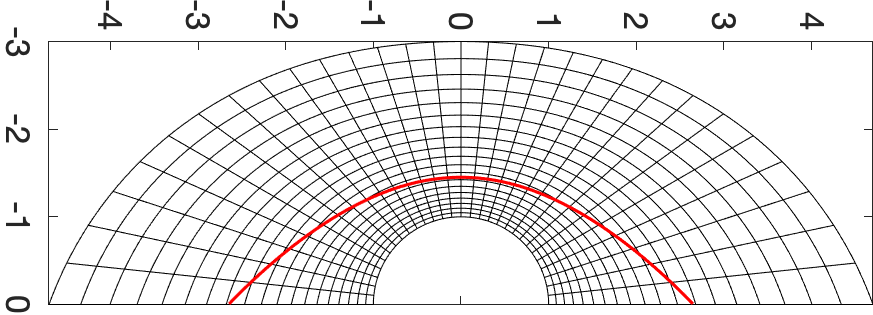}
		\caption{Regular mesh of 570 $k=4$ elements}
	\end{subfigure}
	\hfill
	\begin{subfigure}[b]{0.47\textwidth}
		\centering
		\includegraphics[width=\textwidth]{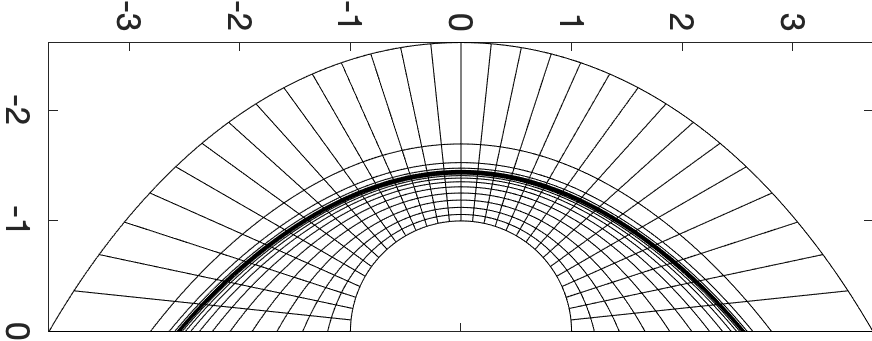}
		\caption{Shock-aligned mesh of 600 $k=4$ elements}
	\end{subfigure} 
	\caption{Meshes for inviscid hypersonic flow past the circular cylinder at $M_\infty = 7$.}
\label{figcyl7q} 
\end{figure}

\begin{figure}[htbp]
	\centering
	\begin{subfigure}[b]{0.49\textwidth}
		\centering		\includegraphics[width=\textwidth]{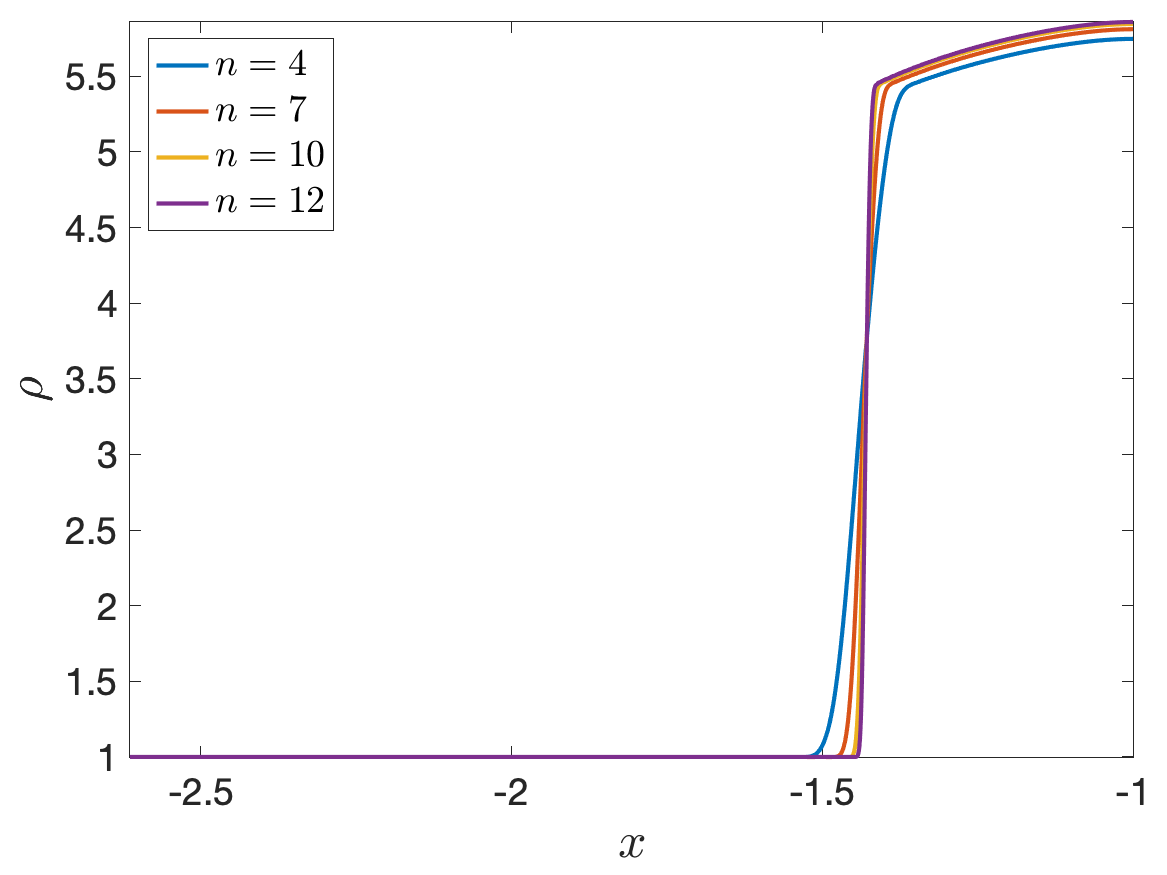}
		\caption{Along the line $y=0$}
	\end{subfigure}
	\hfill
	\begin{subfigure}[b]{0.49\textwidth}
		\centering
		\includegraphics[width=\textwidth]{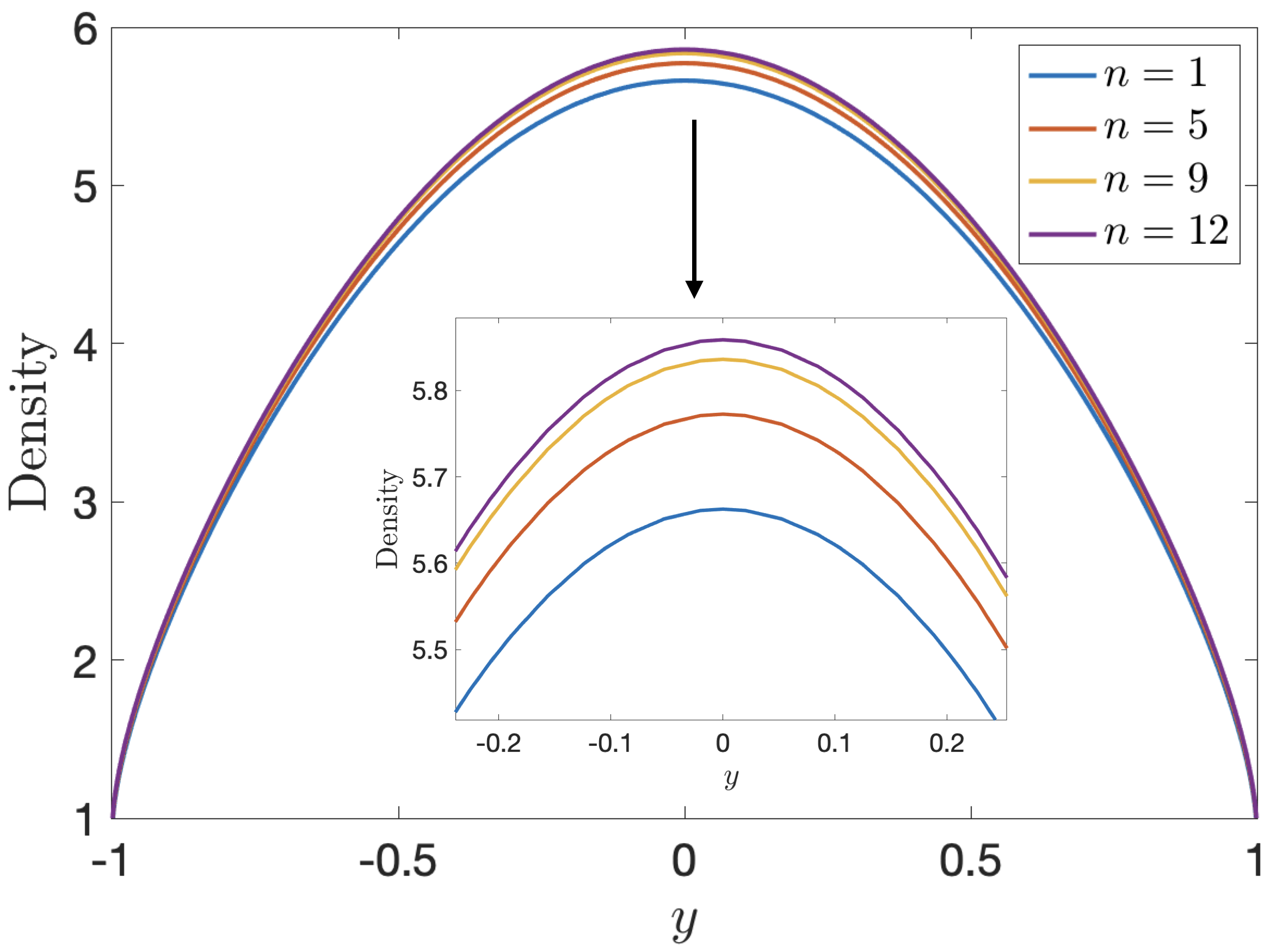}
		\caption{Along the cylinder $\sqrt{x^2+y^2} = 1$}
	\end{subfigure} 
	\caption{Profiles of density computed on the shock-aligned mesh for hypersonic flow past the cylinder at $M_\infty=7$.}
\label{figcyl7a} 
\end{figure}

\begin{figure}[htbp]
\centering
	\begin{subfigure}[b]{0.2\textwidth}
		\centering		\includegraphics[width=\textwidth]{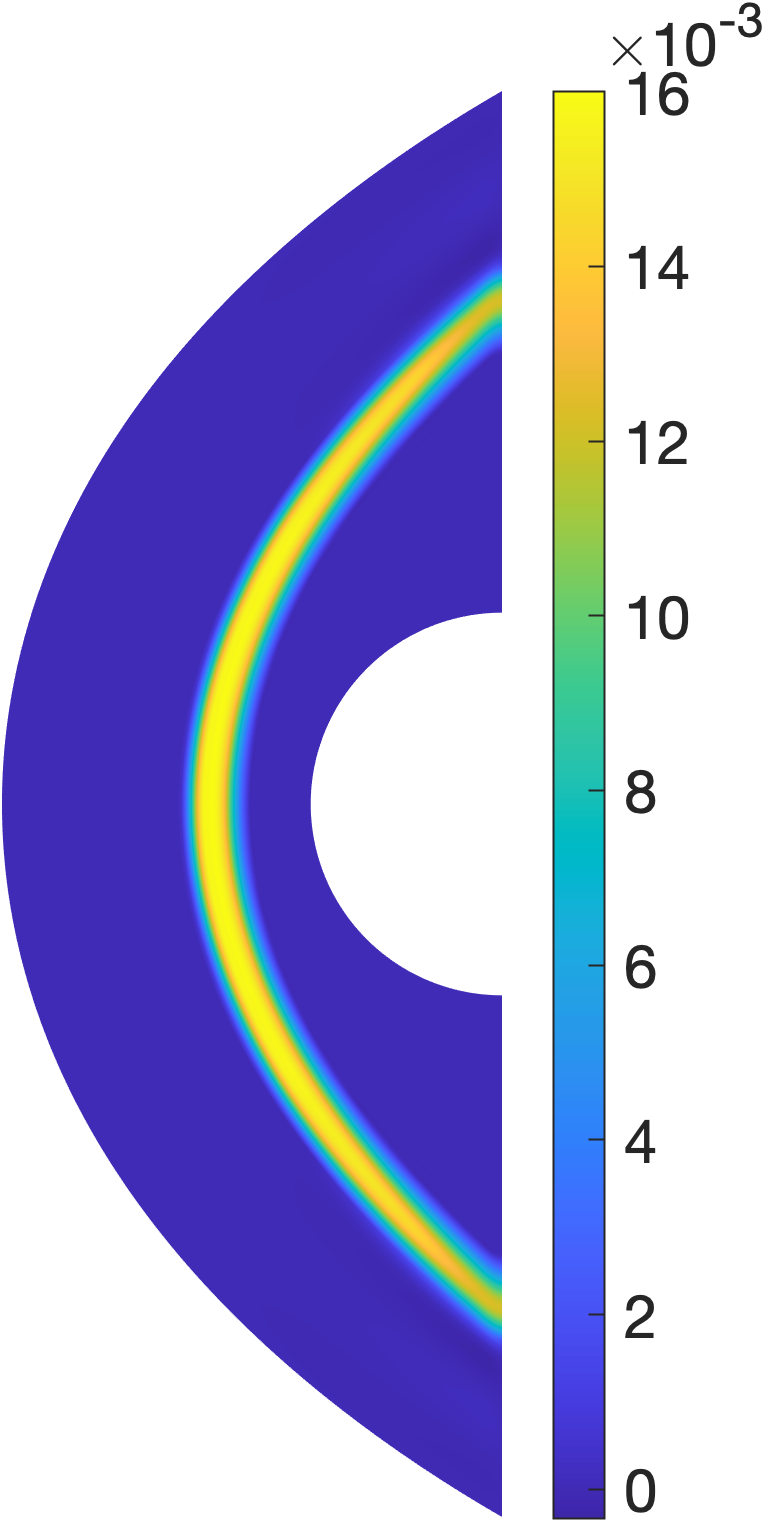}
	\end{subfigure}
	\hfill
	\begin{subfigure}[b]{0.2\textwidth}
		\centering		\includegraphics[width=\textwidth]{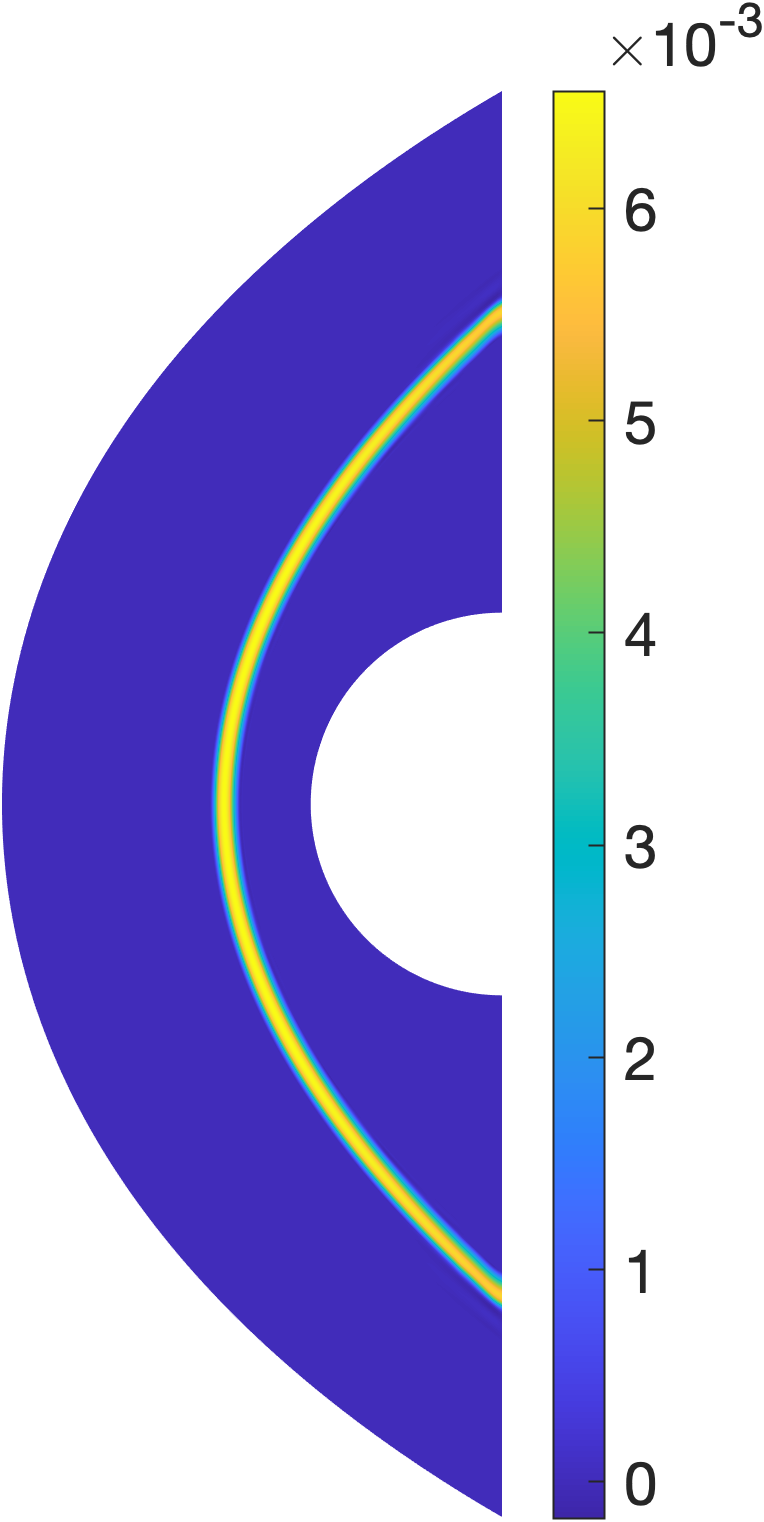}
	\end{subfigure}
        \hfill
        \begin{subfigure}[b]{0.2\textwidth}
		\centering		\includegraphics[width=\textwidth]{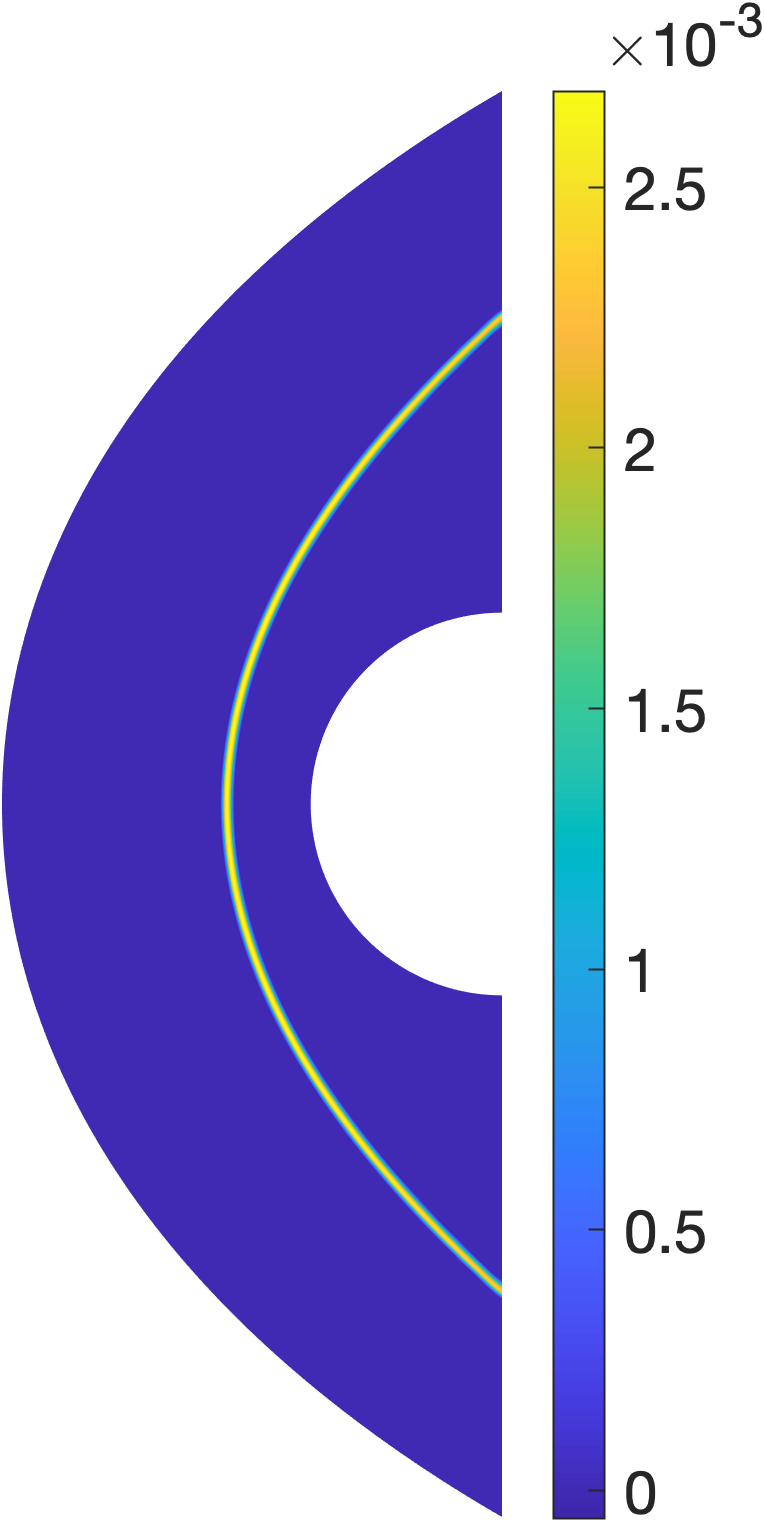}
	\end{subfigure}
        \hfill
        \begin{subfigure}[b]{0.2\textwidth}
		\centering		\includegraphics[width=\textwidth]{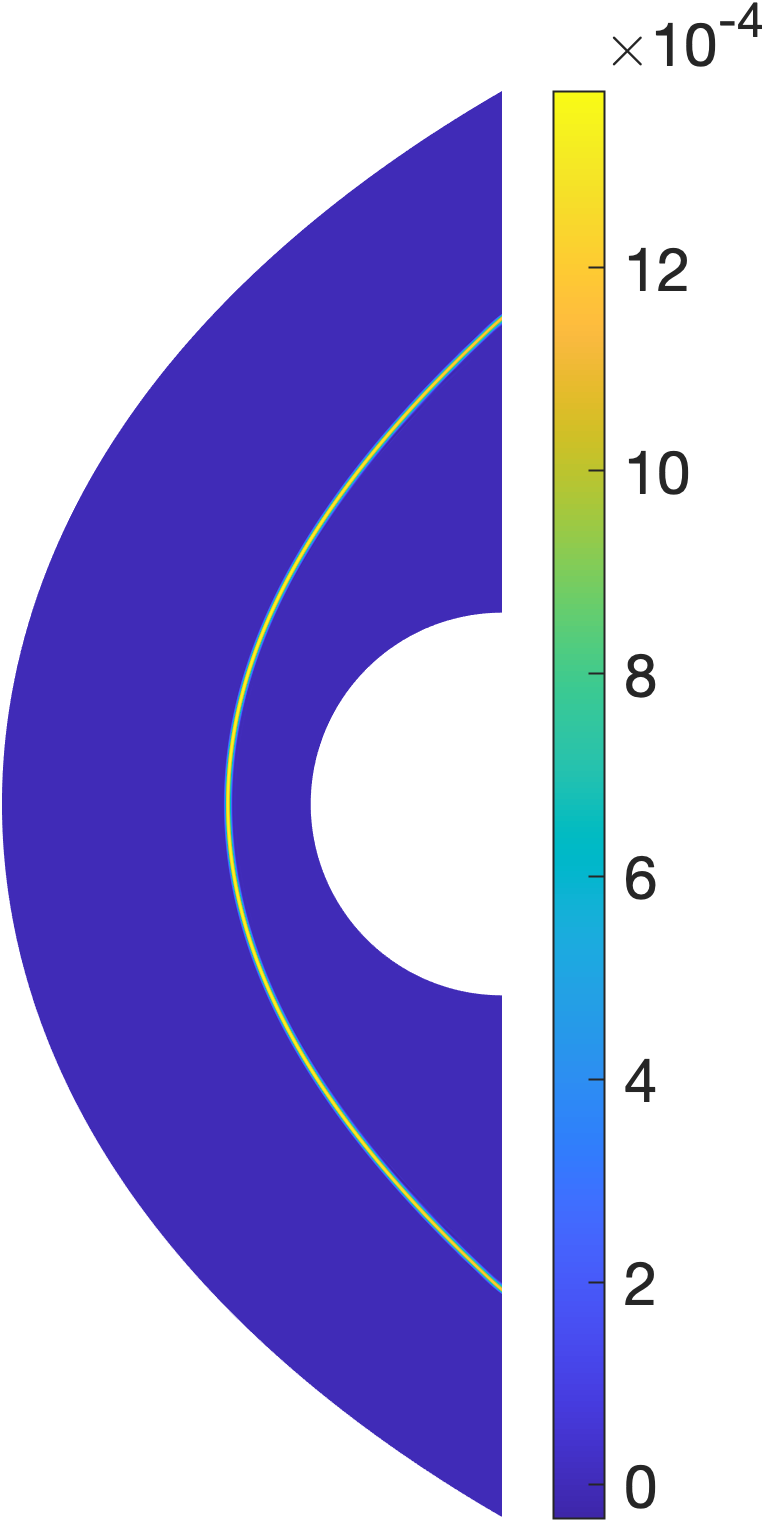}
	\end{subfigure} \\[2ex]
        \begin{subfigure}[b]{0.19\textwidth}
		\centering		\includegraphics[width=\textwidth]{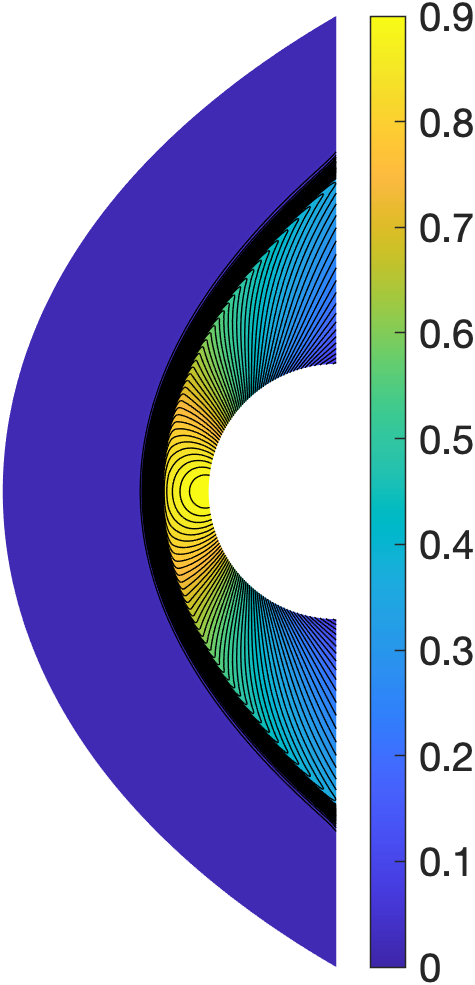}
	\end{subfigure}
	\hfill
	\begin{subfigure}[b]{0.19\textwidth}
		\centering		\includegraphics[width=\textwidth]{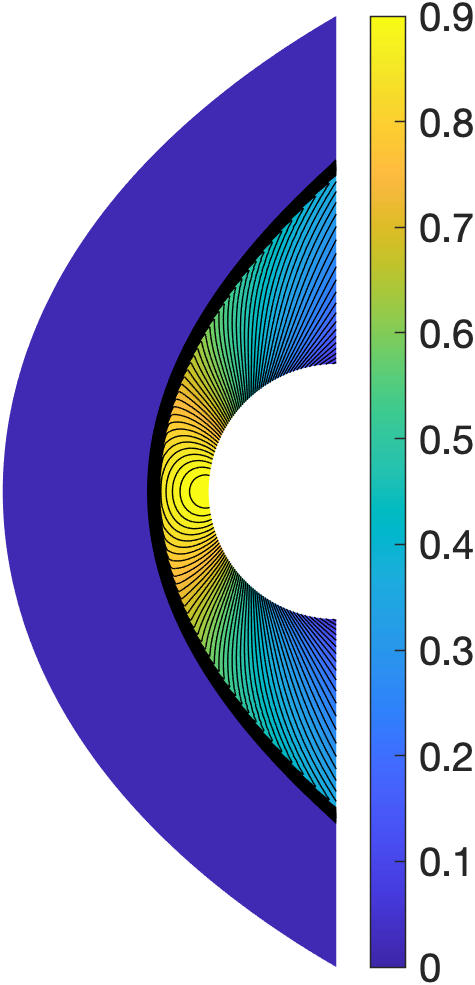}
	\end{subfigure}
        \hfill
        \begin{subfigure}[b]{0.19\textwidth}
		\centering		\includegraphics[width=\textwidth]{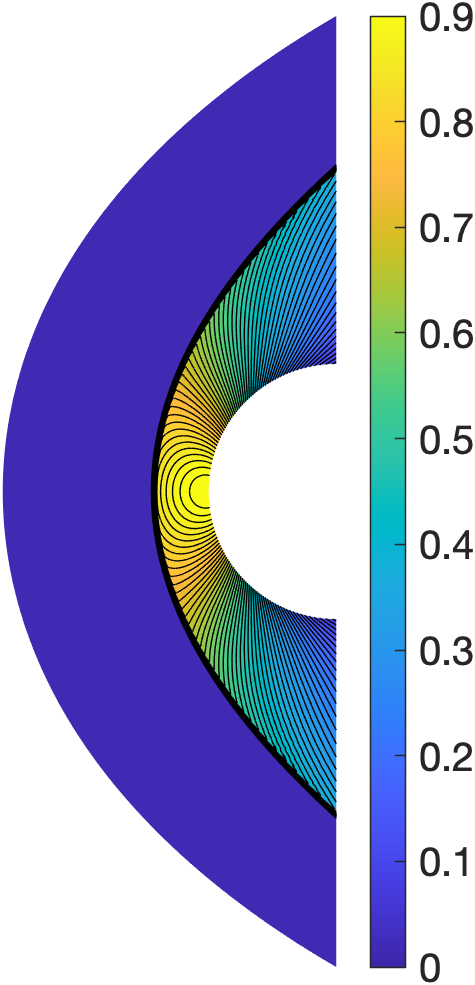}
	\end{subfigure}
        \hfill
        \begin{subfigure}[b]{0.19\textwidth}
		\centering		\includegraphics[width=\textwidth]{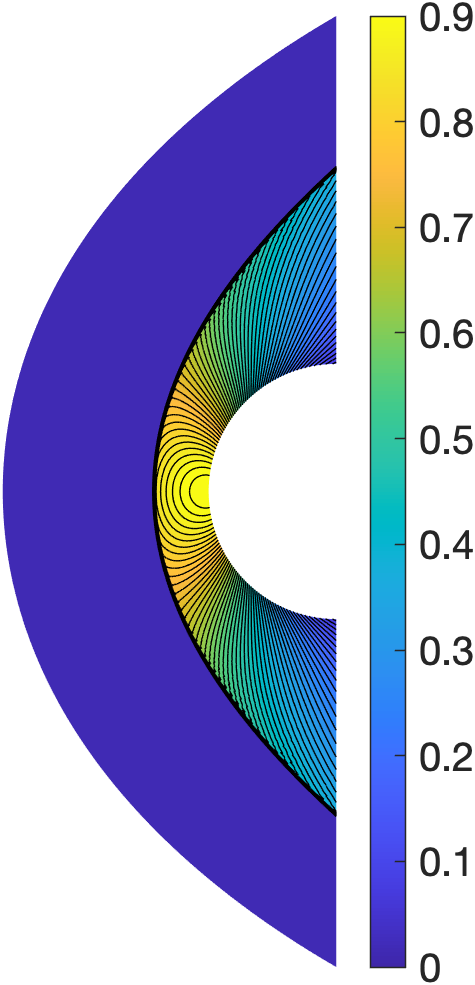}        
	\end{subfigure} \\[2ex]
        \begin{subfigure}[b]{0.18\textwidth}
		\centering		\includegraphics[width=\textwidth]{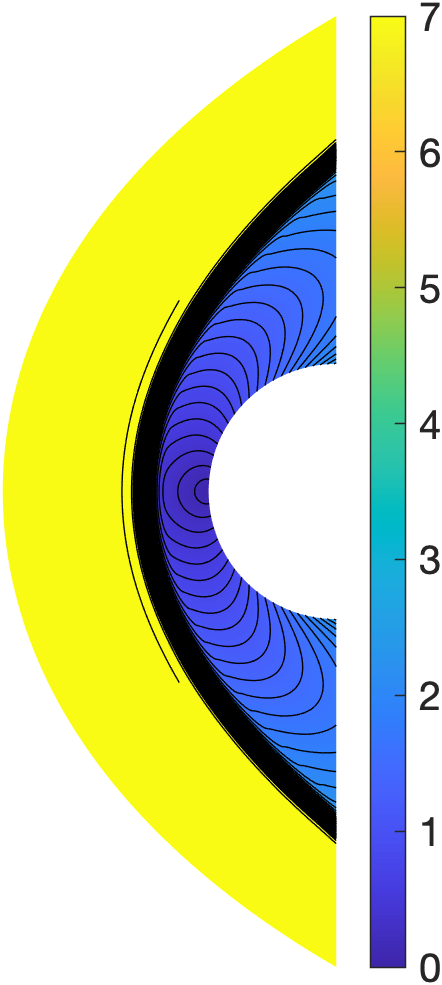}
  \caption{n=1}
	\end{subfigure}
	\hfill
	\begin{subfigure}[b]{0.18\textwidth}
		\centering		\includegraphics[width=\textwidth]{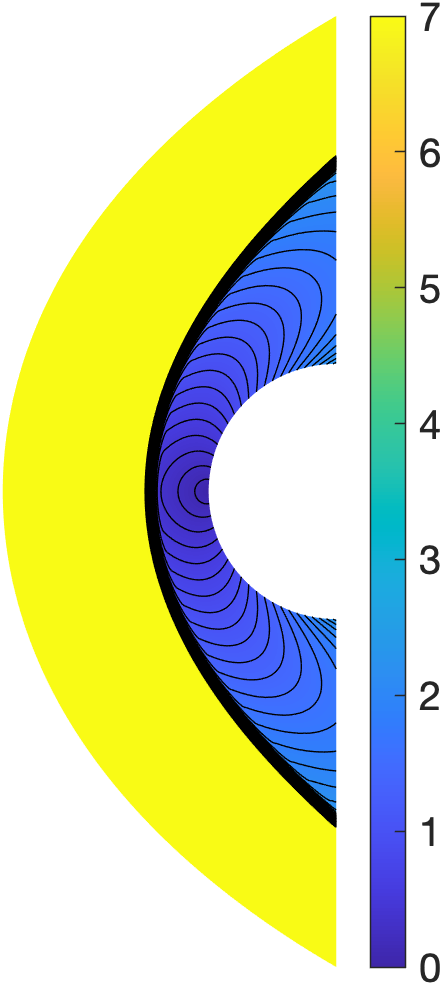}
  \caption{n=5}
	\end{subfigure}
        \hfill
        \begin{subfigure}[b]{0.18\textwidth}
		\centering		\includegraphics[width=\textwidth]{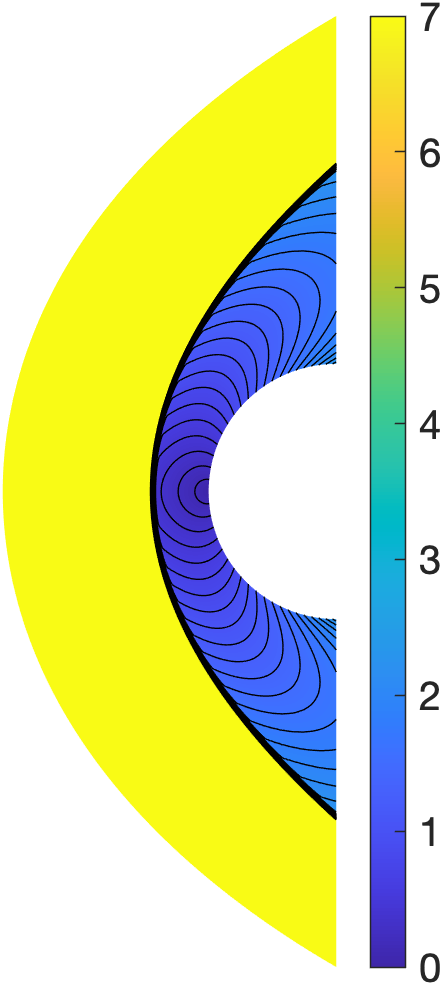}
  \caption{n=9}
	\end{subfigure}
        \hfill
        \begin{subfigure}[b]{0.18\textwidth}
		\centering		\includegraphics[width=\textwidth]{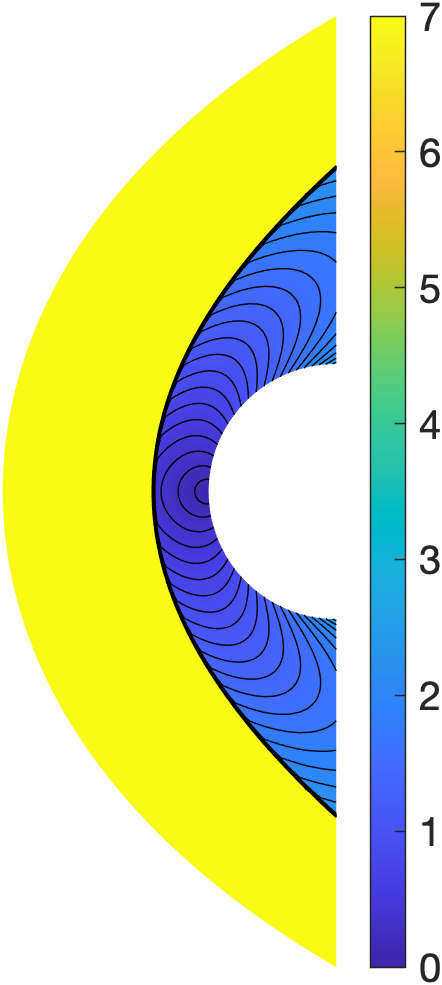}
  \caption{n=12}
	\end{subfigure} 
\caption{Computed artificial viscosity (top row), pressure (middle row), and Mach number (bottom row) at different homotopy iterations on the shock-aligned mesh for inviscid hypersonic flow past the cylinder at $M_\infty=7$.}
\label{cyl71}	
\vspace{-0.5cm}	
\end{figure}

The last test case involving hypersonic flow past a unit circular cylinder at $M_\infty = 7$ demonstrates the effectiveness of our approach for very strong shocks in the hypersonic regime. The boundary conditions are the same as those in the previous test case. The regular and the shock-aligned meshes are shown in Figure \ref{figcyl7q}. Profiles of density computed on the shock-aligned mesh are shown in Figure \ref{figcyl7a} for different homotopy iterations.  We see that the density profiles converge and get sharper as $n$ increases. Figure \ref{cyl71} depicts the  solution computed on the shock-aligned mesh. The artificial viscosity is reduced significantly as $n$ increases. Furthermore, the solution converged at $n=12$ is smooth and sharp.  Table \ref{tabcyl72} tabulates relevant quantities of interest as a function of the homotopy iteration $n$ on the shock-aligned mesh. The homotopy continuation ends at $n=13$ where $\theta_{13}$ exceeds the specified threshold. We see that the amount of artificial viscosity (namely, $\|\varepsilon_n\|_{\Omega}$) and the enthapy error (namely, $e_n^H$) decrease as $n$ increases. The results clearly show that the amount of artificial viscosity affects the accuracy of the approximate solution not only in the shock region but also away from the shock region. In particular, both the density and pressure at the stagnation point increase as the amount of artificial viscosity decreases. Moreover, the bow shock is very well captured since the overshoot in the Mach number is quite small as $\|M_n\|_{\infty}$ is close to $M_\infty = 7$. These observations are similar to those on the supersonic test case at $M_\infty = 3$.

\begin{figure}[htbp]
\centering
	\centering
	\begin{subfigure}[b]{0.32\textwidth}
		\centering		\includegraphics[width=\textwidth]{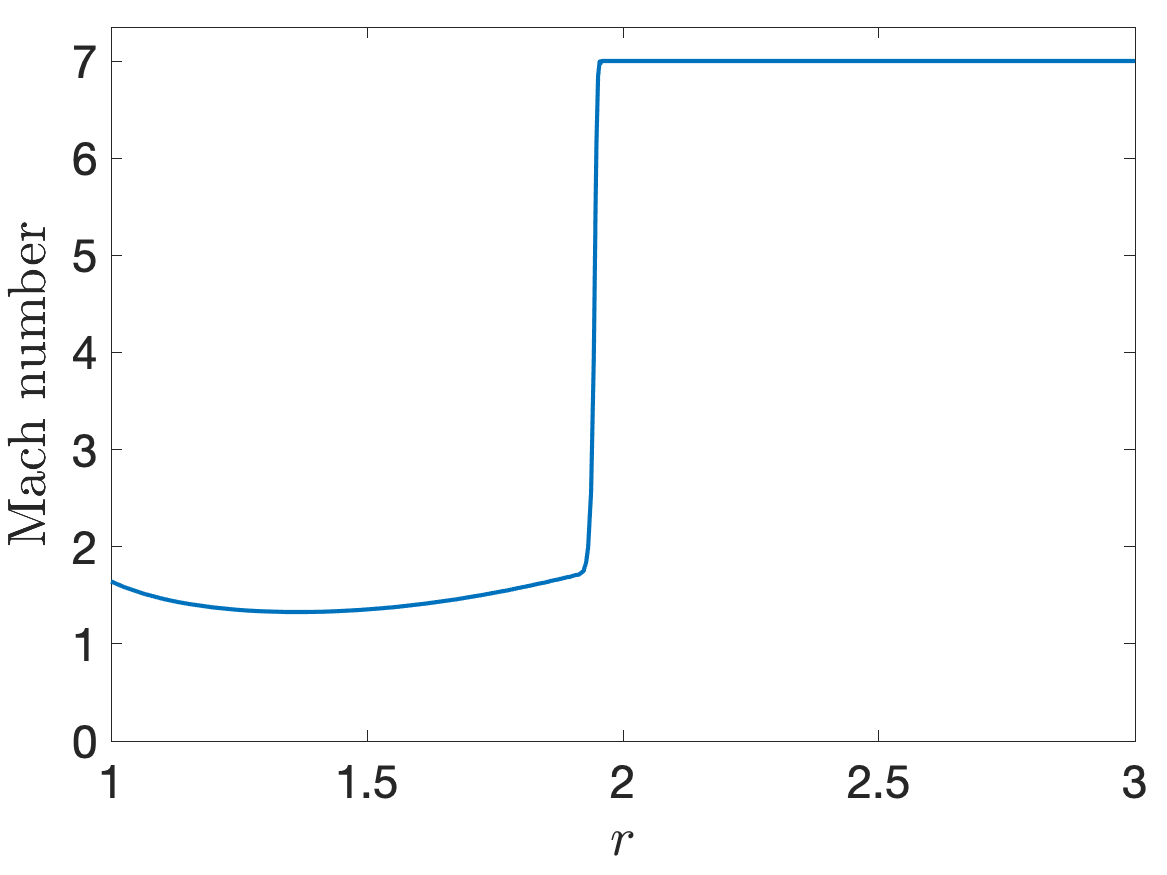}
		\caption{$n=12$}
	\end{subfigure}
	\hfill
	\begin{subfigure}[b]{0.32\textwidth}
		\centering
		\includegraphics[width=\textwidth]{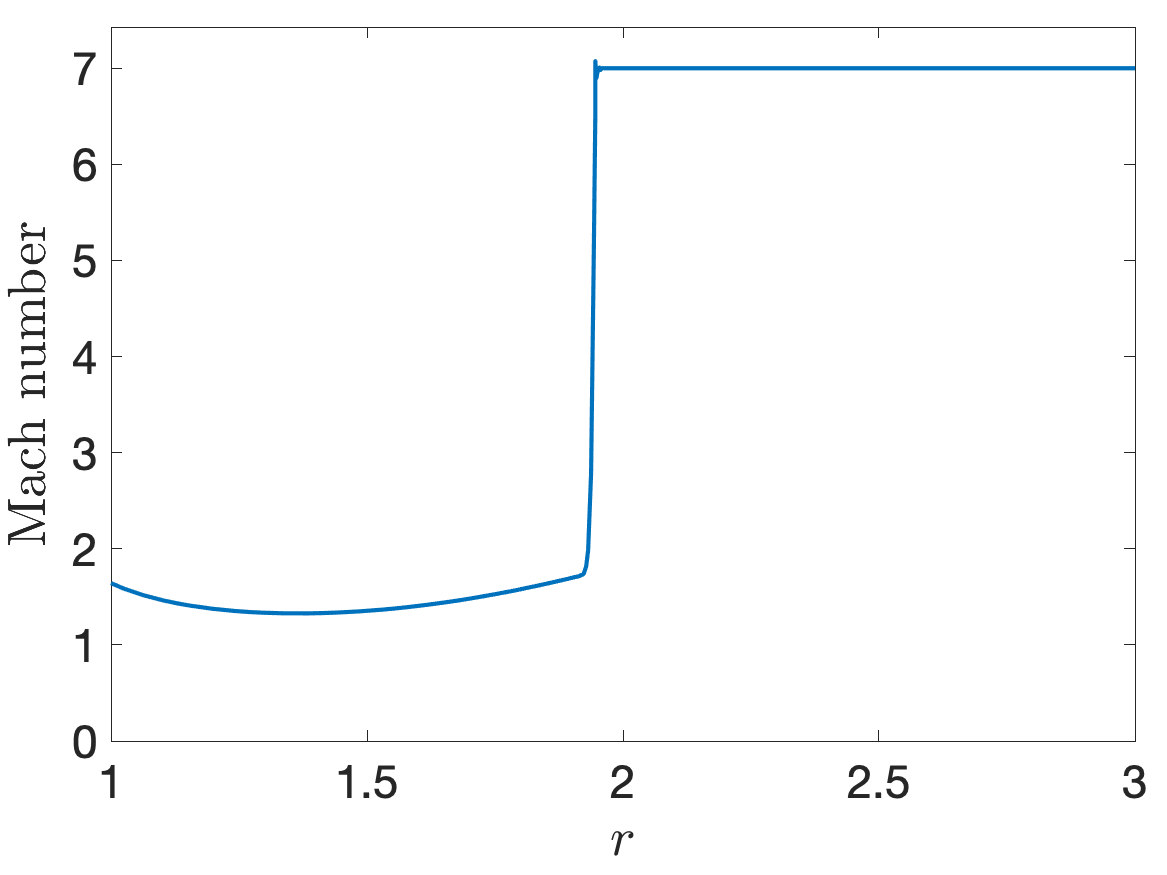}
		\caption{$n=13$}
	\end{subfigure} 
 \hfill
	\begin{subfigure}[b]{0.32\textwidth}
		\centering
		\includegraphics[width=\textwidth]{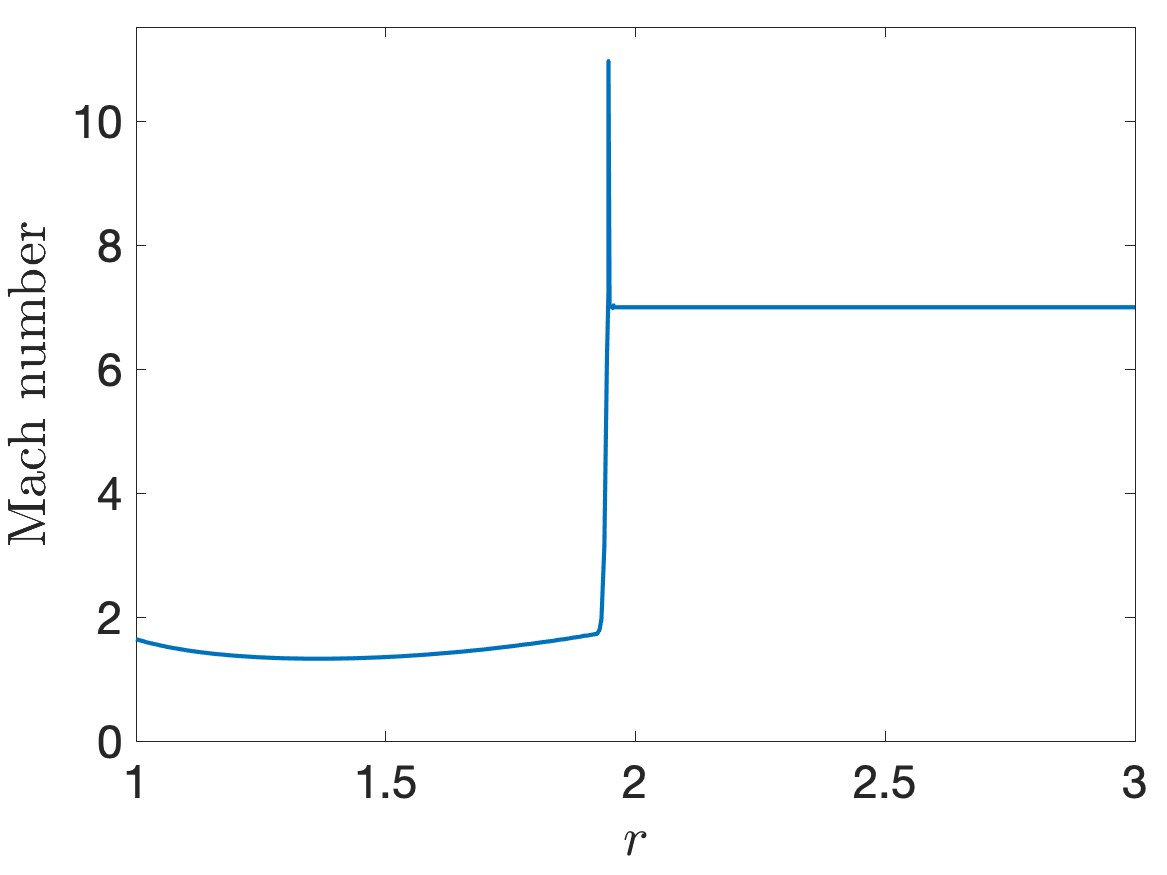}
		\caption{$n=14$}
	\end{subfigure} 
\caption {\label{figbg3a} Profiles of the Mach number along the line $y = 2.6051 x$ as a function of $r = \sqrt{x^2 + y^2}$ for $n=12, 13,$ and 14.}
\end{figure}

\revise{The overshoot in the Mach number  is noticeable for $n \ge 12$. We expect the overshoot to increase as the artificial viscosity decreases during the homotopy continuation. To clearly see the effect of reducing the artificial viscosity on the numerical solution, we continue the homotopy iteration up to $n=14$. We see that the overshoot in the Mach number increases rapidly after $n=12$. This can be attributed to high polynomial degrees ($k=4$) and high Mach number flows in the hypersonic regime ($M_\infty = 7$), which render the numerical solution sensitive to the amount of artificial viscosity. Figure \ref{figbg3a} shows profiles of the  Mach number along the line $y = 2.6051 x$ for the last three homotopy iterations. We observe that the Mach number profile is smooth for $n=12$, slightly oscillatory at the shock location for $n=13$, and largely overshoots at the shock location for $n=14$. For a fixed grid resolution, if the amount of artificial viscosity drops below an optimal value, then the quality of the numerical solution can deteriorate quickly when polynomials of high degree are used to represent the numerical solution. Therefore, it is important to stop the homotopy continuation when the numerical solution is still smooth and sharp. In this regard, we note that $\theta_n = 6.9563$ exceeds the threshold $C_{\sigma} = 5.0$ at $n = 13$. As a result, we end the homotopy continuation at $n=13$ to satisfy the constraints (18) and  accept $\bm u_{12}$ as the numerical solution of the problem.    
}

\begin{table}[htbp]
\centering
\begin{small}
	\begin{tabular}{|c|cccccccc|}
		\hline
		$n$ & $\lambda_{n,1}$ & $\lambda_{n,2}$ & $\theta_n$ & $\|\varepsilon_n\|_{\Omega}$ &  $e^H_n$  & $\|M_n\|_{\infty}$ &  $\rho_n(\bm x^*)$ & $p_n(\bm x^*)$  \\
		\hline
 1  &  0.0200  &  5.0000  &  1.0026  &  0.0153  &  0.0111  &  7.0533  &  5.6627  &  0.9069  \\  
 2  &  0.0160  &  4.2000  &  1.0000  &  0.0109  &  0.0108  &  7.5869  &  5.6755  &  0.9083  \\  
 3  &  0.0128  &  3.5600  &  1.1959  &  0.0080  &  0.0087  &  7.1191  &  5.7146  &  0.9114  \\  
 4  &  0.0102  &  3.0480  &  1.4298  &  0.0059  &  0.0069  &  7.0422  &  5.7472  &  0.9141  \\  
 5  &  0.0082  &  2.6384  &  1.6255  &  0.0042  &  0.0056  &  7.0113  &  5.7727  &  0.9162  \\  
 6  &  0.0066  &  2.3107  &  1.8670  &  0.0030  &  0.0046  &  7.0233  &  5.7937  &  0.9181  \\  
 7  &  0.0052  &  2.0486  &  2.0450  &  0.0022  &  0.0036  &  7.0038  &  5.8111  &  0.9196  \\  
 8  &  0.0042  &  1.8389  &  2.1374  &  0.0016  &  0.0030  &  7.0056  &  5.8249  &  0.9209  \\  
 9  &  0.0034  &  1.6711  &  2.2507  &  0.0011  &  0.0024  &  7.0023  &  5.8363  &  0.9219  \\  
 10  &  0.0027  &  1.5369  &  2.1059  &  0.0008  &  0.0020  &  7.0005  &  5.8454  &  0.9227  \\  
 11  &  0.0021  &  1.4295  &  2.8960  &  0.0006  &  0.0016  &  7.0006  &  5.8528  &  0.9234  \\  
 12  &  0.0017  &  1.3436  &  4.1467  &  0.0004  &  0.0013  &  7.0133  &  5.8588  &  0.9240  \\  
 13 &  0.0014  &  1.2749    & 6.9563   & 0.0003  &  0.0014 &    7.2261 &  5.8637  &  0.9245 \\
   14 &    0.0011  &  1.2199   & 12.402 &    0.0002   & 0.0010 &   10.974  & 5.8676  &    0.9248 \\
		\hline
	\end{tabular}
 \end{small}
	\caption{Relevant quantities of interest as a function of homotopy iteration $n$ on the shock-aligned mesh for inviscid hypersonic flow past the cylinder at $M_\infty=7$.} 
	\label{tabcyl72}
\end{table}

\section{Concluding remarks} \label{sec:conclusions}
We have presented an adaptive artificial viscosity regularization approach for the numerical approximation of shock waves.
The method couples a set of nonlinear conservation laws with an additional Helmholtz equation that defines a PDE-based artificial viscosity field.
The procedure features a homotopy continuation of the regularization parameters that minimizes the amount of artificial viscosity needed to stabilize the shock waves, while enforcing positivity-preserving and smoothness constraints on the numerical solution.
The approach is general for any kind of mesh but it is also combined with a mesh adaptation procedure that aligns the grid with shocks in order to further reduce the amount of viscosity and provide sharper and more accurate approximations.

The numerical methodology is solved using finite element methods. In particular, the HDG method used to discretize the set of governing equations is combined with the CG method, employed to solve the Helmholtz equation and provide an inherently continuous viscosity field.
The approach is found to be robust and efficient enough so that no time or polynomial continuation processes are required for steady-state problems, converging to the optimal shock approximation in around 10 homotopy iterations.

We have presented a set of numerical examples featuring the inviscid Burgers' equation and inviscid transonic, supersonic and supersonic flows in two dimensions in order to demonstrate the performance of the method.
The approach is able to produce a sequence of approximate solutions that converge to the exact solution, producing sharp representations of the shock waves, and smooth and non-oscillatory fields everywhere else.

The procedure here presented can be extended to the solution of compressible viscous flows or chemically reacting flows without loss of generality.
To this end, different variants of the regularized viscosity can be devised, including physics-based artificial viscosity terms that augment the molecular viscous components.
On the other hand, the proposed methodology can be also coupled to shock-alignment of shock-fitting strategies, providing increased robustness in the iterative grid adaptation process.
In this sense, an iterative procedure stemming from the proposed shock-aligned mesh generation algorithm could be automated.

\section*{Acknowledgements} \label{}

We gratefully acknowledge the United States  Department of Energy under contract DE-NA0003965, the National Science Foundation for supporting this work (under grant number NSF-PHY-2028125), and the Air Force Office of Scientific Research under Grant No. FA9550-22-1-0356 for supporting this work.


 \bibliographystyle{elsarticle-num} 
\bibliography{library}

\begin{thebibliography}{10}
\expandafter\ifx\csname url\endcsname\relax
  \def\url#1{\texttt{#1}}\fi
\expandafter\ifx\csname urlprefix\endcsname\relax\def\urlprefix{URL }\fi
\expandafter\ifx\csname href\endcsname\relax
  \def\href#1#2{#2} \def\path#1{#1}\fi

\bibitem{Cook2004}
A.~W. Cook, W.~H. Cabot,
  \href{http://linkinghub.elsevier.com/retrieve/pii/S0021999103005746}{{A
  high-wavenumber viscosity for high-resolution numerical methods}}, Journal of
  Computational Physics 195~(2) (2004) 594--601.
\newblock \href {https://doi.org/10.1016/j.jcp.2003.10.012}
  {\path{doi:10.1016/j.jcp.2003.10.012}}.
\newline\urlprefix\url{http://linkinghub.elsevier.com/retrieve/pii/S0021999103005746}

\bibitem{Cook2005}
A.~W. Cook, W.~H. Cabot,
  \href{http://linkinghub.elsevier.com/retrieve/pii/S0021999104004000}{{Hyperviscosity
  for shock-turbulence interactions}}, Journal of Computational Physics 203~(2)
  (2005) 379--385.
\newblock \href {https://doi.org/10.1016/j.jcp.2004.09.011}
  {\path{doi:10.1016/j.jcp.2004.09.011}}.
\newline\urlprefix\url{http://linkinghub.elsevier.com/retrieve/pii/S0021999104004000}

\bibitem{Fiorina2007}
B.~Fiorina, S.~K. Lele, {An artificial nonlinear diffusivity method for
  supersonic reacting flows with shocks}, Journal of Computational Physics
  222~(1) (2007) 246--264.
\newblock \href {https://doi.org/10.1016/j.jcp.2006.07.020}
  {\path{doi:10.1016/j.jcp.2006.07.020}}.

\bibitem{Kawai2008}
S.~Kawai, S.~Lele,
  \href{http://linkinghub.elsevier.com/retrieve/pii/S0021999108003641}{{Localized
  artificial diffusivity scheme for discontinuity capturing on curvilinear
  meshes}}, Journal of Computational Physics 227~(22) (2008) 9498--9526.
\newblock \href {https://doi.org/10.1016/j.jcp.2008.06.034}
  {\path{doi:10.1016/j.jcp.2008.06.034}}.
\newline\urlprefix\url{http://linkinghub.elsevier.com/retrieve/pii/S0021999108003641}

\bibitem{Kawai2010}
S.~Kawai, S.~K. Shankar, S.~K. Lele,
  \href{http://linkinghub.elsevier.com/retrieve/pii/S0021999109006160}{{Assessment
  of localized artificial diffusivity scheme for large-eddy simulation of
  compressible turbulent flows}}, Journal of Computational Physics 229~(5)
  (2010) 1739--1762.
\newblock \href {https://doi.org/10.1016/j.jcp.2009.11.005}
  {\path{doi:10.1016/j.jcp.2009.11.005}}.
\newline\urlprefix\url{http://linkinghub.elsevier.com/retrieve/pii/S0021999109006160}

\bibitem{Klockner2011}
A.~Kl{\"{o}}ckner, T.~Warburton, J.~S. Hesthaven, {Viscous shock capturing in a
  time-explicit discontinuous Galerkin method}, Mathematical Modelling of
  Natural Phenomena 6~(3) (2011) 57--83.
\newblock \href {http://arxiv.org/abs/1102.3190} {\path{arXiv:1102.3190}},
  \href {https://doi.org/10.1051/mmnp/20116303}
  {\path{doi:10.1051/mmnp/20116303}}.

\bibitem{MR2056921}
L.~Krivodonova, J.~Xin, J.-F. Remacle, N.~Chevaugeon, J.~E. Flaherty,
  \href{http://dx.doi.org/10.1016/j.apnum.2003.11.002}{{Shock detection and
  limiting with discontinuous Galerkin methods for hyperbolic conservation
  laws}}, Appl. Numer. Math. 48~(3-4) (2004) 323--338.
\newblock \href {https://doi.org/10.1016/j.apnum.2003.11.002}
  {\path{doi:10.1016/j.apnum.2003.11.002}}.
\newline\urlprefix\url{http://dx.doi.org/10.1016/j.apnum.2003.11.002}

\bibitem{Mani2009}
A.~Mani, J.~Larsson, P.~Moin,
  \href{http://linkinghub.elsevier.com/retrieve/pii/S0021999109003623}{{Suitability
  of artificial bulk viscosity for large-eddy simulation of turbulent flows
  with shocks}}, Journal of Computational Physics 228~(19) (2009) 7368--7374.
\newblock \href {https://doi.org/10.1016/j.jcp.2009.06.040}
  {\path{doi:10.1016/j.jcp.2009.06.040}}.
\newline\urlprefix\url{http://linkinghub.elsevier.com/retrieve/pii/S0021999109003623}

\bibitem{Olson2013}
B.~J. Olson, S.~K. Lele, {Directional artificial fluid properties for
  compressible large-eddy simulation}, Journal of Computational Physics 246
  (2013) 207--220.
\newblock \href {https://doi.org/10.1016/j.jcp.2013.03.026}
  {\path{doi:10.1016/j.jcp.2013.03.026}}.

\bibitem{persson06:_shock_capturing}
P.~O. Persson, J.~Peraire, {Sub-cell shock capturing for discontinuous Galerkin
  methods}, in: Collection of Technical Papers - 44th AIAA Aerospace Sciences
  Meeting, Vol.~2, Reno, Neveda, 2006, pp. 1408--1420.
\newblock \href {https://doi.org/10.2514/6.2006-112}
  {\path{doi:10.2514/6.2006-112}}.

\bibitem{Persson2013}
P.~O. Persson, {Shock capturing for high-order discontinuous Galerkin
  simulation of transient flow problems}, in: 21st AIAA Computational Fluid
  Dynamics Conference, San Diego, CA, 2013, p. 3061.
\newblock \href {https://doi.org/10.2514/6.2013-3061}
  {\path{doi:10.2514/6.2013-3061}}.

\bibitem{Premasuthan2013}
S.~Premasuthan, C.~Liang, A.~Jameson, {Computation of Flows With Shocks using
  the Spectral Difference method with Artificial Viscosity: Part I}, Computers
  {\&} Fluids 98 (2013) 111--121.

\bibitem{Premasuthan2014}
S.~Premasuthan, C.~Liang, A.~Jameson, {Computation of flows with shocks using
  the Spectral Difference method with artificial viscosity, II: Modified
  formulation with local mesh refinement}, Computers {\&} Fluids 98 (2014)
  122--133.

\bibitem{Fernandez2018}
P.~Fernandez, N.~C. Nguyen, J.~Peraire, {A physics-based shock capturing method
  for unsteady laminar and turbulent flows}, in: 56th AIAA Aerospace Sciences
  Meeting, Orlando, Florida, 2018, pp. AIAA--2018--0062.

\bibitem{Moro2016}
D.~Moro, N.~C. Nguyen, J.~Peraire, {Dilation-based shock capturing for
  high-order methods}, International Journal for Numerical Methods in Fluids
  82~(7) (2016) 398--416.
\newblock \href {https://doi.org/10.1002/fld.4223}
  {\path{doi:10.1002/fld.4223}}.

\bibitem{Nguyen2011a}
N.~C. Nguyen, J.~Peraire,
  \href{http://arc.aiaa.org/doi/abs/10.2514/6.2011-3060}{{An adaptive
  shock-capturing HDG method for compressible flows}}, in: 20th AIAA
  Computational Fluid Dynamics Conference 2011, American Institute of
  Aeronautics and Astronautics, Reston, Virigina, 2011, pp. AIAA 2011--3060.
\newblock \href {https://doi.org/10.2514/6.2011-3060}
  {\path{doi:10.2514/6.2011-3060}}.
\newline\urlprefix\url{http://arc.aiaa.org/doi/abs/10.2514/6.2011-3060}

\bibitem{Burbeau2001}
A.~Burbeau, P.~Sagaut, C.~H. Bruneau, {A Problem-Independent Limiter for
  High-Order Runge-Kutta Discontinuous Galerkin Methods}, Journal of
  Computational Physics 169~(1) (2001) 111--150.
\newblock \href {https://doi.org/10.1006/jcph.2001.6718}
  {\path{doi:10.1006/jcph.2001.6718}}.

\bibitem{Cockburn1989}
B.~Cockburn, C.-W. Shu, {TVB Runge-Kutta Local Projection Discontinuous
  Galerkin Finite Element Method for Conservation Laws II: General Framework},
  Mathematics of Computation 52~(186) (1989) 411.
\newblock \href {https://doi.org/10.2307/2008474} {\path{doi:10.2307/2008474}}.

\bibitem{Krivodonova2007}
L.~Krivodonova, {Limiters for high-order discontinuous Galerkin methods},
  Journal of Computational Physics 226~(1) (2007) 879--896.

\bibitem{Cockburn1998a}
B.~Cockburn, C.~W. Shu, {The Runge-Kutta Discontinuous Galerkin Method for
  Conservation Laws V: Multidimensional Systems}, Journal of Computational
  Physics 141~(2) (1998) 199--224.
\newblock \href {https://doi.org/10.1006/jcph.1998.5892}
  {\path{doi:10.1006/jcph.1998.5892}}.

\bibitem{Lv2015}
Y.~Lv, M.~Ihme, {Entropy-bounded discontinuous Galerkin scheme for Euler
  equations}, Journal of Computational Physics 295 (2015) 715--739.
\newblock \href {http://arxiv.org/abs/1411.5044} {\path{arXiv:1411.5044}},
  \href {https://doi.org/10.1016/j.jcp.2015.04.026}
  {\path{doi:10.1016/j.jcp.2015.04.026}}.

\bibitem{Sonntag2017}
M.~Sonntag, C.~D. Munz, {Efficient Parallelization of a Shock Capturing for
  Discontinuous Galerkin Methods using Finite Volume Sub-cells}, Journal of
  Scientific Computing 70~(3) (2017) 1262--1289.
\newblock \href {https://doi.org/10.1007/s10915-016-0287-5}
  {\path{doi:10.1007/s10915-016-0287-5}}.

\bibitem{Luo2007}
H.~Luo, J.~D. Baum, R.~L{\"{o}}hner, {A Hermite WENO-based limiter for
  discontinuous Galerkin method on unstructured grids}, Journal of
  Computational Physics 225~(1) (2007) 686--713.
\newblock \href {https://doi.org/10.1016/j.jcp.2006.12.017}
  {\path{doi:10.1016/j.jcp.2006.12.017}}.

\bibitem{Qiu2005}
J.~Qiu, C.~W. Shu, {Runge-Kutta discontinuous Galerkin method using WENO
  limiters}, SIAM Journal on Scientific Computing 26~(3) (2005) 907--929.
\newblock \href {https://doi.org/10.1137/S1064827503425298}
  {\path{doi:10.1137/S1064827503425298}}.

\bibitem{Zhu2008}
J.~Zhu, J.~Qiu, C.~W. Shu, M.~Dumbser, {Runge-Kutta discontinuous Galerkin
  method using WENO limiters II: Unstructured meshes}, Journal of Computational
  Physics 227~(9) (2008) 4330--4353.
\newblock \href {https://doi.org/10.1016/j.jcp.2007.12.024}
  {\path{doi:10.1016/j.jcp.2007.12.024}}.

\bibitem{Zhu2013}
J.~Zhu, X.~Zhong, C.~W. Shu, J.~Qiu, {Runge-Kutta discontinuous Galerkin method
  using a new type of WENO limiters on unstructured meshes}, Journal of
  Computational Physics 248 (2013) 200--220.
\newblock \href {https://doi.org/10.1016/j.jcp.2013.04.012}
  {\path{doi:10.1016/j.jcp.2013.04.012}}.

\bibitem{Barter2010}
G.~E. Barter, D.~L. Darmofal,
  \href{http://linkinghub.elsevier.com/retrieve/pii/S0021999109006299}{{Shock
  capturing with PDE-based artificial viscosity for DGFEM: Part I.
  Formulation}}, Journal of Computational Physics 229~(5) (2010) 1810--1827.
\newblock \href {https://doi.org/10.1016/j.jcp.2009.11.010}
  {\path{doi:10.1016/j.jcp.2009.11.010}}.
\newline\urlprefix\url{http://linkinghub.elsevier.com/retrieve/pii/S0021999109006299}

\bibitem{Hartmann2013}
R.~Hartmann, {Higher-order and adaptive discontinuous Galerkin methods with
  shock-capturing applied to transonic turbulent delta wing flow},
  International Journal for Numerical Methods in Fluids 72 (2013) 883--894.

\bibitem{Lv2016}
Y.~Lv, Y.~C. See, M.~Ihme, {An entropy-residual shock detector for solving
  conservation laws using high-order discontinuous Galerkin methods}, Journal
  of Computational Physics 322 (2016) 448--472.
\newblock \href {https://doi.org/10.1016/j.jcp.2016.06.052}
  {\path{doi:10.1016/j.jcp.2016.06.052}}.

\bibitem{Abbassi2014}
H.~Abbassi, F.~Mashayek, G.~B. Jacobs, {Shock capturing with entropy-based
  artificial viscosity for staggered grid discontinuous spectral element
  method}, Computers {\&} Fluids 98 (2014) 152--163.

\bibitem{Chaudhuri2017}
A.~Chaudhuri, G.~B. Jacobs, W.~S. Don, H.~Abbassi, F.~Mashayek, {Explicit
  discontinuous spectral element method with entropy generation based
  artificial viscosity for shocked viscous flows}, Journal of Computational
  Physics 332 (2017) 99--117.
\newblock \href {https://doi.org/10.1016/j.jcp.2016.11.042}
  {\path{doi:10.1016/j.jcp.2016.11.042}}.

\bibitem{Johnsen2010}
E.~Johnsen, J.~Larsson, A.~V. Bhagatwala, W.~H. Cabot, P.~Moin, B.~J. Olson,
  P.~S. Rawat, S.~K. Shankar, B.~Sj{\"{o}}green, H.~Yee, X.~Zhong, S.~K. Lele,
  \href{http://linkinghub.elsevier.com/retrieve/pii/S0021999109005804}{{Assessment
  of high-resolution methods for numerical simulations of compressible
  turbulence with shock waves}}, Journal of Computational Physics 229~(4)
  (2010) 1213--1237.
\newblock \href {https://doi.org/10.1016/j.jcp.2009.10.028}
  {\path{doi:10.1016/j.jcp.2009.10.028}}.
\newline\urlprefix\url{http://linkinghub.elsevier.com/retrieve/pii/S0021999109005804}

\bibitem{VonNeumann1950}
J.~{Von Neumann}, R.~D. Richtmyer, {A method for the numerical calculation of
  hydrodynamic shocks}, Journal of Applied Physics 21 (1950) 232--237.

\bibitem{Jameson1995}
A.~Jameson, {Analysis and Design of Numerical Schemes for Gas Dynamics, 2:
  Artificial Diffusion and Discrete Shock Structure}, International Journal of
  Computational Fluid Dynamics 5 (1995) 1--38.

\bibitem{HughesMalletMisukamiII86}
T.~J. Hughes, M.~Mallet, M.~Akira, {A new finite element formulation for
  computational fluid dynamics: II. Beyond SUPG}, Computer Methods in Applied
  Mechanics and Engineering 54~(3) (1986) 341--355.
\newblock \href {https://doi.org/10.1016/0045-7825(86)90110-6}
  {\path{doi:10.1016/0045-7825(86)90110-6}}.

\bibitem{madtad93}
Y.~Maday, S.~O. Kaber, E.~Tadmor, {Legendre pseudospectral viscosity method for
  nonlinear conservation laws}, SIAM J. Numer. Anal. 30 (1993) 321--342.

\bibitem{Tadmor89}
E.~Tadmor, {Convergence of spectral methods for nonlinear conservation laws},
  SIAM J. Numer. Anal. 26 (1989) 30--44.

\bibitem{Ching2019}
E.~J. Ching, Y.~Lv, P.~Gnoffo, M.~Barnhardt, M.~Ihme, {Shock capturing for
  discontinuous Galerkin methods with application to predicting heat transfer
  in hypersonic flows}, Journal of Computational Physics 376 (2019) 54--75.
\newblock \href {https://doi.org/10.1016/j.jcp.2018.09.016}
  {\path{doi:10.1016/j.jcp.2018.09.016}}.

\bibitem{HartmannHoustonCompressible02}
R.~Hartmann, P.~Houston, {Adaptive discontinuous Galerkin finite element
  methods for the compressible Euler equations}, Journal of Computational
  Physics 183~(2) (2002) 508--532.
\newblock \href {https://doi.org/10.1006/jcph.2002.7206}
  {\path{doi:10.1006/jcph.2002.7206}}.

\bibitem{Bai2022a}
Y.~Bai, K.~J. Fidkowski, \href{https://doi.org/10.2514/1.J061783}{{Continuous
  Artificial-Viscosity Shock Capturing for Hybrid Discontinuous Galerkin on
  Adapted Meshes}}, AIAA Journal 60~(10) (2022) 5678--5691.
\newblock \href {https://doi.org/10.2514/1.J061783}
  {\path{doi:10.2514/1.J061783}}.
\newline\urlprefix\url{https://doi.org/10.2514/1.J061783}

\bibitem{Vila-Perez2021}
J.~Vila-P{\'{e}}rez, M.~Giacomini, R.~Sevilla, A.~Huerta,
  \href{https://doi.org/10.1007/s11831-020-09508-z}{{Hybridisable Discontinuous
  Galerkin Formulation of Compressible Flows}}, Archives of Computational
  Methods in Engineering 28~(2) (2021) 753--784.
\newblock \href {https://doi.org/10.1007/s11831-020-09508-z}
  {\path{doi:10.1007/s11831-020-09508-z}}.
\newline\urlprefix\url{https://doi.org/10.1007/s11831-020-09508-z}

\bibitem{Bhagatwala2009}
A.~Bhagatwala, S.~K. Lele,
  \href{http://linkinghub.elsevier.com/retrieve/pii/S0021999109002034}{{A
  modified artificial viscosity approach for compressible turbulence
  simulations}}, Journal of Computational Physics 228~(14) (2009) 4965--4969.
\newblock \href {https://doi.org/10.1016/j.jcp.2009.04.009}
  {\path{doi:10.1016/j.jcp.2009.04.009}}.
\newline\urlprefix\url{http://linkinghub.elsevier.com/retrieve/pii/S0021999109002034}

\bibitem{Cook2007}
A.~W. Cook,
  \href{http://link.aip.org/link/PHFLE6/v19/i5/p055103/s1{\&}Agg=doi}{{Artificial
  fluid properties for large-eddy simulation of compressible turbulent
  mixing}}, Physics of Fluids 19~(5) (2007) 055103.
\newblock \href {https://doi.org/10.1063/1.2728937}
  {\path{doi:10.1063/1.2728937}}.
\newline\urlprefix\url{http://link.aip.org/link/PHFLE6/v19/i5/p055103/s1{\&}Agg=doi}

\bibitem{Premasuthan2010b}
S.~Premasuthan, C.~Liang, A.~Jameson, {Computation Of Flows with Shocks Using
  Spectral Difference Scheme with Artificial Viscosity}, in: 48th AIAA
  Aerospace Sciences Meeting, Orlando, FL, 2010.

\bibitem{Zahr2018}
M.~J. Zahr, P.~O. Persson, {An optimization-based approach for high-order
  accurate discretization of conservation laws with discontinuous solutions},
  Journal of Computational Physics 365 (2018) 105--134.
\newblock \href {http://arxiv.org/abs/1712.03445} {\path{arXiv:1712.03445}},
  \href {https://doi.org/10.1016/j.jcp.2018.03.029}
  {\path{doi:10.1016/j.jcp.2018.03.029}}.

\bibitem{Zahr2020}
M.~J. Zahr, A.~Shi, P.~O. Persson, {Implicit shock tracking using an
  optimization-based high-order discontinuous Galerkin method}, Journal of
  Computational Physics 410 (2020) 109385.
\newblock \href {http://arxiv.org/abs/1912.11207} {\path{arXiv:1912.11207}},
  \href {https://doi.org/10.1016/j.jcp.2020.109385}
  {\path{doi:10.1016/j.jcp.2020.109385}}.

\bibitem{Shi2022}
A.~Shi, P.~O. Persson, M.~J. Zahr, {Implicit shock tracking for unsteady flows
  by the method of lines}, Journal of Computational Physics 454 (2022).
\newblock \href {https://doi.org/10.1016/j.jcp.2021.110906}
  {\path{doi:10.1016/j.jcp.2021.110906}}.

\bibitem{Corrigan2019}
A.~Corrigan, A.~D. Kercher, D.~A. Kessler, {A moving discontinuous Galerkin
  finite element method for flows with interfaces}, International Journal for
  Numerical Methods in Fluids 89~(9) (2019) 362--406.
\newblock \href {https://doi.org/10.1002/fld.4697}
  {\path{doi:10.1002/fld.4697}}.

\bibitem{Kercher2021}
A.~D. Kercher, A.~Corrigan, D.~A. Kessler, {The moving discontinuous Galerkin
  finite element method with interface condition enforcement for compressible
  viscous flows}, International Journal for Numerical Methods in Fluids 93~(5)
  (2021) 1490--1519.
\newblock \href {http://arxiv.org/abs/2002.12740} {\path{arXiv:2002.12740}},
  \href {https://doi.org/10.1002/fld.4939} {\path{doi:10.1002/fld.4939}}.

\bibitem{Kercher2021a}
A.~D. Kercher, A.~Corrigan, {A least-squares formulation of the Moving
  Discontinuous Galerkin Finite Element Method with Interface Condition
  Enforcement}, Computers and Mathematics with Applications 95 (2021) 143--171.
\newblock \href {http://arxiv.org/abs/2003.01044} {\path{arXiv:2003.01044}},
  \href {https://doi.org/10.1016/j.camwa.2020.09.012}
  {\path{doi:10.1016/j.camwa.2020.09.012}}.

\bibitem{Hicken2011}
J.~E. Hicken, H.~Buckley, M.~Osusky, D.~W. Zingg, {Dissipation-based
  continuation: A globalization for inexact-newton solvers}, in: 20th AIAA
  Computational Fluid Dynamics Conference 2011, 2011, pp. AIAA--2011--3237.
\newblock \href {https://doi.org/10.2514/6.2011-3237}
  {\path{doi:10.2514/6.2011-3237}}.

\bibitem{Nguyen2012}
N.~C. Nguyen, J.~Peraire,
  \href{http://linkinghub.elsevier.com/retrieve/pii/S0021999112001544}{{Hybridizable
  discontinuous Galerkin methods for partial differential equations in
  continuum mechanics}}, Journal of Computational Physics 231~(18) (2012)
  5955--5988.
\newblock \href {https://doi.org/10.1016/j.jcp.2012.02.033}
  {\path{doi:10.1016/j.jcp.2012.02.033}}.
\newline\urlprefix\url{http://linkinghub.elsevier.com/retrieve/pii/S0021999112001544}

\bibitem{Fernandez2018a}
P.~Fernandez, A.~Christophe, S.~Terrana, N.~C. Nguyen, J.~Peraire,
  \href{http://link.springer.com/10.1007/s10915-018-0811-x}{{Hybridized
  discontinuous Galerkin methods for wave propagation}}, Journal of Scientific
  Computing 77~(3) (2018) 1566--1604.
\newblock \href {https://doi.org/10.1007/s10915-018-0811-x}
  {\path{doi:10.1007/s10915-018-0811-x}}.
\newline\urlprefix\url{http://link.springer.com/10.1007/s10915-018-0811-x}

\bibitem{Moro2011a}
D.~Moro, N.~C. Nguyen, J.~Peraire,
  \href{http://arc.aiaa.org/doi/abs/10.2514/6.2011-3407}{{Navier-stokes
  solution using Hybridizable discontinuous Galerkin methods}}, in: 20th AIAA
  Computational Fluid Dynamics Conference 2011, American Institute of
  Aeronautics and Astronautics, Honolulu, Hawaii, 2011, pp. AIAA--2011--3407.
\newblock \href {https://doi.org/10.2514/6.2011-3407}
  {\path{doi:10.2514/6.2011-3407}}.
\newline\urlprefix\url{http://arc.aiaa.org/doi/abs/10.2514/6.2011-3407}

\bibitem{Peraire2010}
J.~Peraire, N.~C. Nguyen, B.~Cockburn, {A hybridizable discontinuous Galerkin
  method for the compressible euler and Navier-Stokes equations}, in: 48th AIAA
  Aerospace Sciences Meeting Including the New Horizons Forum and Aerospace
  Exposition, 2010, pp. AIAA 2010--363.

\bibitem{Woopen2014c}
M.~Woopen, A.~Balan, G.~May, J.~Sch{\"{u}}tz, {A comparison of hybridized and
  standard DG methods for target-based hp-adaptive simulation of compressible
  flow}, Computers and Fluids 98 (2014) 3--16.
\newblock \href {https://doi.org/10.1016/j.compfluid.2014.03.023}
  {\path{doi:10.1016/j.compfluid.2014.03.023}}.

\bibitem{Fidkowski2016}
K.~J. Fidkowski, {A hybridized discontinuous Galerkin method on mapped
  deforming domains}, Computers and Fluids 139 (2016) 80--91.
\newblock \href {https://doi.org/10.1016/j.compfluid.2016.04.004}
  {\path{doi:10.1016/j.compfluid.2016.04.004}}.

\bibitem{Fernandez2017a}
P.~Fernandez, N.~C. Nguyen, J.~Peraire, {The hybridized Discontinuous Galerkin
  method for Implicit Large-Eddy Simulation of transitional turbulent flows},
  Journal of Computational Physics 336 (2017) 308--329.
\newblock \href {https://doi.org/10.1016/j.jcp.2017.02.015}
  {\path{doi:10.1016/j.jcp.2017.02.015}}.

\bibitem{williams2018entropy}
D.~Williams, {An entropy stable, hybridizable discontinuous Galerkin method for
  the compressible Navier-Stokes equations}, Mathematics of Computation
  87~(309) (2018) 95--121.

\bibitem{Bianchini2005}
S.~Bianchini, A.~Bressan, {Vanishing viscosity solutions of nonlinear
  hyperbolic systems}, Annals of Mathematics 161~(1) (2005) 223--342.
\newblock \href {http://arxiv.org/abs/0111321} {\path{arXiv:0111321}}, \href
  {https://doi.org/10.4007/annals.2005.161.223}
  {\path{doi:10.4007/annals.2005.161.223}}.

\bibitem{Chen2015}
G.~Q.~G. Chen, M.~Perepelitsa, {Vanishing Viscosity Solutions of the
  Compressible Euler Equations with Spherical Symmetry and Large Initial Data},
  Communications in Mathematical Physics 338~(2) (2015) 771--800.
\newblock \href {https://doi.org/10.1007/s00220-015-2376-y}
  {\path{doi:10.1007/s00220-015-2376-y}}.

\bibitem{CuongNguyen2022}
N.~C. Nguyen, S.~Terrana, J.~Peraire, {Large-Eddy Simulation of Transonic
  Buffet Using Matrix-Free Discontinuous Galerkin Method}, AIAA Journal 60~(5)
  (2022) 3060--3077.
\newblock \href {https://doi.org/10.2514/1.j060459}
  {\path{doi:10.2514/1.j060459}}.

\bibitem{Nguyen2023a}
N.~C. Nguyen, S.~Terrana, J.~Peraire,
  \href{https://doi.org/10.2514/6.2023-0659}{{Implicit Large eddy simulation of
  hypersonic boundary-layer transition for a flared cone}}, in: AIAA SCITECH
  2023 Forum, AIAA SciTech Forum, American Institute of Aeronautics and
  Astronautics, 2023, pp. AIAA 2023--0659.
\newblock \href {https://doi.org/10.2514/6.2023-0659}
  {\path{doi:10.2514/6.2023-0659}}.
\newline\urlprefix\url{https://doi.org/10.2514/6.2023-0659}

\bibitem{Geuzaine2009}
C.~Geuzaine, J.~F. Remacle, {Gmsh: A 3-D finite element mesh generator with
  built-in pre- and post-processing facilities}, International Journal for
  Numerical Methods in Engineering 79~(11) (2009) 1309--1331.
\newblock \href {https://doi.org/10.1002/nme.2579}
  {\path{doi:10.1002/nme.2579}}.

\end{thebibliography}





\end{document}